\documentclass[aps,prx,twocolumn,showkeys]{revtex4-1}
\usepackage[export]{adjustbox}
\usepackage{physics}
\usepackage{dcolumn}
\usepackage{graphicx}
\usepackage{indentfirst}
\usepackage{braket}
\usepackage{float}
\usepackage{amsmath}
\usepackage{amssymb}
\usepackage{amsfonts}
\usepackage{epstopdf}
\usepackage{CJK}
\usepackage{esint}
\usepackage{color}
\usepackage[T1]{fontenc}
\usepackage{subfigure}
\usepackage{amsfonts}
\usepackage{natbib}
\usepackage{footmisc}
\usepackage{scrextend}
\usepackage{verbatim}
\usepackage{multirow}
 \usepackage{mathtools}
 \usepackage{pdfpages}
 \makeatletter
 \AtBeginDocument{\let\LS@rot\@undefined}
 \makeatother

\usepackage{hyperref}

\usepackage[english]{babel}
\usepackage{url}
\usepackage{bm}
\usepackage{hyperref}
\definecolor{darkblue}{rgb}{0,0,0.5}
\hypersetup{
colorlinks=true,
linkcolor=black,
filecolor=blue,
citecolor=darkblue,  
urlcolor=black,
}

\newlength{\fighskip} \fighskip=2pt
\newlength{\figvskip} \figvskip=3pt

\newcommand*{\figbox}[2]{{
  \def\figscale{#1}
  \def\arraystretch{0.8}
  \arraycolsep=0pt
  \begin{array}{c}
    \vbox{\vskip\figscale\figvskip
      \hbox{\hskip\figscale\fighskip
        \includegraphics[scale=\figscale]{#2}}}
  \end{array}}}
  
\usepackage{enumitem,kantlipsum}
\usepackage{refcount}

\newcounter{lemmacounter}
\newcounter{theoremcounter}
\newcounter{conjecturecounter}
\newcounter{definitioncounter}
\def\be{\begin{equation}}
\def\ee{\end{equation}}
\def\ba{\begin{eqnarray}}
\def\ea{\end{eqnarray}}
\newcommand{\calC}{{\cal C}}

\newcommand{\calE}{{\cal E}}
\newcommand{\calF}{{\cal F}}

\newcommand{\calL}{{\cal L}}
\newcommand{\calN}{{\cal N}}

\newcommand{\calH}{{\cal H}}
\newcommand{\calJ}{{\cal J}}
\renewcommand{\tr}{{\rm Tr}}

\newcommand{\1}{^{(1)}}

\newcommand{\state}[1]{\ketbra{#1}{#1}}

\newcommand{\eq}[1]{(\hyperref[eq:#1]{\ref*{eq:#1}})}

\renewcommand{\sec}[1]{\hyperref[sec:#1]{Sec.~\ref*{sec:#1}}}
\newcommand{\thrm}[1]{\hyperref[thm:#1]{Theorem~\ref*{thm:#1}}}
\newcommand{\lemm}[1]{\hyperref[lemm:#1]{Lemma~\ref*{lemm:#1}}}
\newcommand{\pro}[1]{\hyperref[pro:#1]{Proposition~\ref*{pro:#1}}}
\newcommand{\corr}[1]{\hyperref[corr:#1]{Corollary~\ref*{corr:#1}}}
\newcommand{\fig}[1]{\hyperref[fig:#1]{\ref*{fig:#1}}}
\newcommand{\tbl}[1]{\hyperref[fig:#1]{\ref*{tbl:#1}}}

\renewcommand{\b}{\boldsymbol}

\renewcommand{\ip}[2]{( #1 | #2 )}

\usepackage[toc,page]{appendix}
\newcommand{\nocontentsline}[3]{}
\newcommand{\tocless}[2]{\bgroup\let\addcontentsline=\nocontentsline#1{#2}\egroup}

\begin{document}

\title{Scrambling and Complexity in Phase Space}

\author{Quntao Zhuang$^{1}$}
\author{Thomas Schuster$^{1}$}
\author{Beni Yoshida$^{2}$ }
\author{Norman Y. Yao$^{1,3}$ }
\affiliation{$^1$Department of Physics, University of California, Berkeley, California 94720, USA
\\
$^2$Perimeter Institute for Theoretical Physics, Waterloo, Ontario N2L 2Y5, Canada
\\
$^3$Materials Science Division, Lawrence Berkeley National Laboratory, Berkeley, California 94720, USA
}
\date{\today}

\begin{abstract} 
The study of information scrambling in many-body systems has sharpened our understanding of quantum chaos, complexity and gravity. 
Here, we extend the framework for exploring information scrambling to infinite dimensional continuous variable (CV) systems.
Unlike their discrete variable cousins, continuous variable systems exhibit two complementary domains of information scrambling: i) scrambling in the phase space of a single mode and ii) scrambling across multiple modes of a many-body system. 
Moreover, for each of these domains, we identify two distinct `types' of scrambling; genuine scrambling, where an initial operator localized in phase space spreads out and quasi scrambling, where a local ensemble of operators distorts but the overall phase space volume remains fixed. 
To characterize these behaviors, we introduce a CV out-of-time-order correlation (OTOC) function based upon displacement operators and offer a number of results regarding the CV analog for unitary designs.
Finally, we investigate operator spreading and entanglement growth in random local Gaussian circuits; to explain the observed behavior, we propose a simple hydrodynamical model that relates the butterfly velocity, the growth exponent and the diffusion constant. 
Experimental realizations of continuous variable scrambling as well as its characterization using CV OTOCs will be discussed.
\end{abstract} 


\maketitle


\section{Introduction}

Scrambling refers to the dynamical delocalization of quantum information over an entire system's degrees of freedom~\cite{hayden2007black, sekino2008fast, Lashkari13, Roberts:2014isa}.
%
Recent developments in the study of quantum information scrambling have led to the discovery of novel soluble models of holography~\cite{sachdev1993,kitaev2015simple,gu2017local,kitaev2018soft}; these developments have also begun to shed light on a disparate array of fundamental questions, including: black hole information problems in quantum gravity~\cite{hayden2007black, hosur2016chaos, Yoshida:2017aa, gao2017traversable, maldacena2017diving,sekino2008fast,maldacena2016bound,shenker2014black,Lashkari13, Roberts:2014isa}, transport properties of non-Fermi liquids~\cite{banerjee_2017,patel2017quantum}, and chaotic thermalization dynamics in isolated, many-body systems~\cite{kim2013ballistic,luitz2017information,huang2017out,chen2017out,fan2017out,gopalakrishnan2018hydrodynamics}.

Prior studies of scrambling have generally focused on discrete variable (DV) systems, where the many-body Hilbert space is composed of a tensor product of local qubits. 
One of the defining features of such DV scrambling is the notion of operator growth, where the time evolution of an initially simple, local operator $\mathcal{V}$, 
yields a more complex, late-time operator, $\mathcal{V}(t)= U^\dagger\left(t\right) \mathcal{V} U\left(t\right)$~\cite{Roberts:2014isa}, whose decomposition is  dominated by non-local operator strings.
%
%
A particularly powerful quantitative diagnostic of operator growth is provided by the so-called out-of-time-order correlation (OTOC) function $\langle \mathcal{V}^\dagger(t) \mathcal{W}^\dagger(0) \mathcal{V}(t) \mathcal{W}(0) \rangle$, which measures the spreading of $\mathcal{V}(t)$ via another local probe operator $\mathcal{W}$~\cite{larkin1969quasiclassical, nahum2018operator, Keyserlingk:2018aa, Rakovszky:2018aa, Khemani:2018aa, xu2018locality, Bentsen2018}. 
In addition to its use on the theory front, OTOCs have also attracted a significant amount of experimental interest and attention~\cite{garttner2017measuring,li2017measuring,meier2017exploring,wei2018exploring,landsman2018verified}.

While discrete variable OTOCs are relatively well understood, both from an information-theoretic perspective and in terms of physical interpretation, their continuous variable (CV) cousin remains poorly explored. 
This owes to a number of intrinsic subtleties associated with CV systems, which describes systems with infinite dimensional local Hilbert space, such as harmonic oscillators.
For example, their infinite dimensional local Hilbert space leads to technical and conceptual challenges requiring the need for short- and long-distance cutoffs as well as ambiguities related to properly defining the `volume' of an operator.
Moreover, for CV systems, there naturally exists \emph{two} distinct notions of scrambling; namely, scrambling \emph{within} the phase space of a single degree of freedom and scrambling \emph{across} the phase space of many coupled degrees of freedom.

To this end, the broad goal of our manuscript is to lay out a theoretical foundation for investigating quantum information scrambling in CV systems \cite{gu2017local, caputa2016out, Roberts_2015, rozenbaum2017lyapunov,hashimoto2017out,rammensee2018many,chavez2018quantum,borgonovi2018emergence,cotler2018out}.
This is motivated in part, by an abundance of strongly-interacting, controllable physical systems, whose microscopic degrees of freedom are continuous variable; these include quantum optical systems, cavity/circuit QED, photonic networks and more abstractly, quantum field theories in general.

\emph{Analogy between DV and CV systems}---Throughout our manuscript, we find it illuminating to frame our results in analogy with well-known ideas from DV systems. 
Here, we will begin by introducing this dictionary before summarizing the organization of the remainder of the paper.

In an $N$-qudit DV system (dimension $d=q^N$ with $q$-state qudits), it is common to diagnose scrambling by observing how a single-qudit Pauli operator, $P$, evolves under unitary time evolution, $U\left(t\right)$. 
%
Since the Pauli operators form a complete basis, the time-evolved $P(t)$ can be re-expanded as,
\begin{align}
P(t) = \sum_{Q \in \text{Pauli}} f\left[Q;P(t)\right] Q
\label{Pauli_decomposition}
\end{align}
where $|f\left[Q;P(t)\right]|^2$ can be interpreted as the probability distribution of $P(t)$ over $Q$ and is normalized such that, $\sum_{Q}  |f\left[Q;P(t)\right]|^2 = 1$. 
As aforementioned, scrambling in DV systems corresponds to the fact that strongly-interacting time evolution generically leads the operator expansion in Eqn.~(1) to have significant overlaps with `high-weight' Pauli operators. Here, the weight of a Pauli operator $Q$ quantifies its non-locality and equals the number of qudits on which $Q$ acts non-trivially.

A central motif of our work is the analogy between Pauli operators in DV systems and \emph{displacement operators} in CV systems. 
In a single-mode CV system (e.g. a simple harmonic oscillator), the displacement operators (which shift a coherent state in phase space) form a complete basis:
\be
D\left(\xi_{1},\xi_{2}\right) \equiv \exp\left[i \left(\xi_{2} q-\xi_{1}p\right) \right],
\ee
where $\xi_1$ and $\xi_2$ are (respectively) the shifts of $q$ and $p$, the canonical position and momentum quadrature operators. More generally, an $N$-mode displacement operator can be written as the tensor product of local displacements.

We propose to characterize information scrambling in CV systems by considering the time-evolution of displacement operators, $D\left(\bm \xi_1;t\right) \equiv U\left(t\right)^\dagger D\left(\bm \xi_1\right) U\left(t\right)$.
To understand why, let us begin by decomposing,
\be
D\left(\bm \xi_1;t\right)=\frac{1}{\pi^N}\int d^{2N} {\bm \xi}_2 \ \chi\left[{\bm \xi}_2;D\left(\bm \xi_1;t\right)\right]D\left(-{\bm \xi}_2\right),
\label{D_decomposition_intro}
\ee
where
$
\chi\left({\bm \xi};A\right)\equiv\tr \left[A D\left({\bm \xi}\right)\right]
$ is  the Wigner characteristic function.
Initially, the characteristic function $\chi\left[{\bm \xi}_2;D\left(\bm \xi_1;0\right)\right]=\pi^N\delta \left(\bm \xi_2+\bm \xi_1\right)$ is highly localized; thus, scrambling in CV systems can be identified using the spread of $\chi\left[{\bm \xi}_2;D\left(\bm \xi_1;t\right)\right]$ in phase space for generic choices of $\bm \xi_1$, much like the spreading of $P(t)$ into high-weight Pauli operators in the DV case.

With this analogy in hand, let us now introduce the organization and summary of our results. 

\textbf{Quasi vs genuine scrambling:} In Section~\ref{sec:genuine vs quasi}, we introduce an important distinction between two types of scrambling, which we term: quasi scrambling and genuine scrambling. 
In DV systems, both types are captured by the growth of $P(t)$ into high-weight operators (i.e.~large-scale entanglement is generated). The key distinction lies in the nature of the operator \emph{distribution} as quantified by $|f\left[Q;P(t)\right]|^2$.
In particular, we refer to a system as a quasi scrambler if the operator distribution remains localized on a \emph{single} high-weight operator, and as a genuine scrambler if the distribution is spread out over a \emph{large} number of high-weight operators. 
%
This difference in operator distribution behavior can be unambiguously identified via OTOCs and roughly corresponds to whether the scrambling unitary is generated by Clifford (quasi) or non-Clifford (genuine) operators. 

Having defined two classes of scrambling in DV systems, we will find that the same distinction applies for CV systems. In particular, we propose that CV quasi scrambling corresponds to situations where the Wigner characteristic function stays localized, while genuine CV scrambling results in the delocalization of the characteristic function. 
Interestingly, in the case of CV systems, this distinction corresponds to  whether the scrambling unitary is generated by Gaussian (quasi) or non-Gaussian (genuine) operators.

\textbf{Operator spreading and OTOCs:}
	In Section~\ref{sec:genuine}, we develop theoretical tools to study operator spreading in CV systems using CV OTOCs. 
We begin by establishing an intuitive measure of genuine scrambling, in terms of the volume of a time-evolved operator's distribution in phase space. 
Then, we present a Fourier transform-like formula which relates this distribution to OTOCs of the time-evolved operator, from which we see that individual OTOCs can detect the distinction between genuine and quasi scramblers. 
This allows us to propose a physical observable --- the OTOC magnitude --- which measures the non-Gaussianity of the dynamics. 
Furthermore, we show that averages of OTOCs, weighted over ensembles of displacement operators, can probe an operator's spread in phase space with tunable short and long-distance cutoffs. 
This encompasses both operator spreading in real space, as studied in DV systems, as well as within a single mode. 
Finally, we apply our findings to study operator growth in two examples of genuine scramblers: time-evolution via cubic phase gates and the Hellon-Heiles potential. 

\textbf{Random Gaussian circuits:}
In Section~\ref{Sec_quasi}, we turn our attention to quasi scrambling. Despite the fact that quasi scramblers do not fully spread operators (i.e.~they simply map one displacement operator to another), useful insights into genuine scrambling can be obtained by studying the stochastic evolution of quasi scramblers. 
Indeed, recent progress~\cite{harrow2009random, Brandao12, Hartman13, Pastawski15b, hosur2016chaos, hayden2016holographic} on the interplay between holography and quantum information theory has revealed that quantum circuits composed of short-range random Clifford unitaries can provide useful intuition for understanding entanglement and operator propagation in DV many-body systems~\cite{nahum2018operator, Keyserlingk:2018aa, Rakovszky:2018aa, Khemani:2018aa, xu2018locality, Bentsen2018}.

This motivates us to explore local random Gaussian circuits (Sec.~\ref{Sec_circuit}) as an analytically and numerically tractable toy model of CV scrambling.
We observe a number of intriguing features.
In particular, we find that such random circuits exhibit exponential growth of displacements \emph{within} each local mode, while operators spread \emph{ballistically} to distant modes. 
In the former case, the observed exponential growth also leads to other surprising consequences; for example, entanglement in the system grows quadratically in time, in contrast to the previously found linear growth~\cite{nahum2018operator}.
In the latter case, we observe that the butterfly velocity $v_{B}$ of the ballistic spread depends on the local growth exponent in a similar fashion to large-$N$ models.
To this end, we propose a simple hydrodynamical model which relates the butterfly velocity, the growth exponent and the diffusion constant. 

\textbf{Unitary designs for CV systems:} Building upon our exploration of random Gaussian unitaries (see Section~\ref{Sec_circuit}), in Section~\ref{Sec_designs} we investigate the statistical properties of ensembles of such unitaries. 
In particular, we attempt to construct CV analogs of Haar randomness and unitary designs~\cite{footnote1}. 
To begin, we provide a plausible definition for unitary designs in CV systems. 
Using this definition, we find that Gaussian distributed displacement operators asymptotically form a CV 1-design. 
However, we find that Gaussian unitaries do not form an exact 2-design --- consistent with prior results on state designs~\cite{blume2014curious} --- but nevertheless capture many qualitative 2-design features in the limit of large squeezing. 
Finally, we provide a unique generalization of the so-called `frame potential' to CV systems, enabling the quantitative verification of designs.

\textbf{Experimental realizations of CV scrambling:}
 In Section~\ref{Sec_experiment}, we propose and analyze a concrete experimental realization of CV scrambling in a cavity-QED architecture, where non-Gaussian unitaries are generated via the SNAP (selective number-dependent arbitrary
phase) gate~\cite{heeres2015cavity,krastanov2015universal}.
Next, we present concrete protocols for the measurement of both individual and average OTOCs in CV systems.
In addition, we also present a CV analog of a teleportation-based protocol for verifying scrambling \cite{Yoshida:2017aa, yoshida2018disentangling, landsman2018verified}. 
Interestingly, while measurement uncertainties may damage the teleported quantum state, our CV teleportation protocol can be made fault-tolerant by using the Gottesman-Kitaev-Preskill (GKP) code (when the squeezing parameter is sufficiently large). 

Finally, in Section~\ref{Sec_discussion}, we conclude by offering our perspective on a number of exciting open questions and directions. Since many detailed calculations are relegated to the appendices, we provide a short description here:
\begin{itemize}
\item[i)] Appendix~\ref{App_Gaussian} summarizes key properties of basic Gaussian unitaries. 
\item[ii)]
In Appendix~\ref{sec:volume_quasi_A}, we give some details of quasi scramblers. In Appendix~\ref{App_Louiville}, we prove a quantum analog of Liouville's theorem, stating that the total volume of operator distributions is preserved under Gaussian dynamics. This suggests that quasi scrambling corresponds to an increase in a coarse-grained and/or projected volume, while the total volume in the $2N$-dimensional phase space is conserved.
\item[iii)] Appendix~\ref{App_OTO_loss} discusses the effect of photon loss on OTOCs. 
\item[iv)] Appendix~\ref{App_FP} is dedicated to further discussions on CV unitary designs from the perspective of the twice-regulated finite temperature frame potential introduced in Section~\ref{Sec_designs}.
\end{itemize}

\section{Genuine and quasi scrambling}
\label{sec:genuine vs quasi}


We begin by presenting a broad overview of scrambling in DV and CV systems. Scrambling refers to delocalization of quantum information over the entire system under unitary dynamics $U\left(t\right)\equiv \exp\left(-i\int H(t)dt\right)$, where $H(t)$ is the Hamiltonian of the system and can in principle be time dependent. In terms of operators, an initially simple operator $\mathcal{O}$ evolves to a more complex operator $\mathcal{O}(t)= U^\dagger\left(t\right) \mathcal{O} U\left(t\right)$, e.g. local operators become highly non-local~\cite{Roberts:2014isa}. In an $N$-qudit DV system ($d=q^N$ with $q$-state qudits), it is common to diagnose scrambling by observing how a single-qudit Pauli operator $P$ evolves in time. Since Pauli operators form a complete basis -- $\frac{1}{d}\Tr\left( P Q^{\dagger} \right) = \delta_{P,Q}$ for $P,Q \in \text{Pauli}$ and $\frac{1}{d}\sum_{P \in \text{Pauli}} P \mathcal{O} P^{\dagger} = \Tr(\mathcal{O}) \bm I$ -- the time-evolved $P(t)$ can be expanded as
\begin{align}
P(t) = \sum_{Q \in \text{Pauli}} f\left[Q;P(t)\right] Q, \qquad \sum_{Q}  |f\left[Q;P(t)\right]|^2 = 1,
\end{align}
where the second constraint results from the unitarity of $U(t)$. As such, $|f\left[Q;P(t)\right]|^2$ can be interpreted as a probability distribution over $Q$ for each $P(t)$. 

Scrambling in DV systems corresponds to the growth of $f\left[Q;P(t)\right]$ such that the quantity $|f\left[Q;P(t)\right]|^2$ for `high-weight' Pauli operators becomes significant. Here the weight of a Pauli operator $Q$ quantifies the non-locality of $Q$, and equals the number of qudits on which $Q$ acts non-trivially. Within this correspondence, we identify two different classes of scrambling dynamics. Time-evolution $U$ is called \emph{quasi scrambling} if the operator distribution $|f\left[Q;P(t)\right]|^2$ remains concentrated on only a few Paulis $Q$ even as the weight of these Paulis grows significantly. In contrast, time-evolution $U$ is called \emph{genuine scrambling} if $|f\left[Q;P(t)\right]|^2$ has support on many high-weight Pauli operators for a generic low-weight Pauli operator $P(0)$. 

Interestingly, these two classes of scramblers are related to Clifford and non-Clifford unitary operators. Clifford operators are unitary operators which transform Pauli operators into Pauli operators; $U P U^{\dagger} \in \text{Pauli}$ for all $P\in \text{Pauli}$. Examples include the Hadamard gate and the Control-Not gate. Under a random Clifford operator, the time-evolved operator $P(t)$ becomes a high-weight $\sim O(N)$ Pauli operator. However, by definition, the operator distribution $f\left[Q;P(t)\right]$ remains concentrated on a single Pauli $Q = P(t)$, hence it is only quasi scrambled. On the other hand, if $U$ is a non-Clifford operator, e.g. a Haar random unitary, the operator distribution becomes almost uniform,
\begin{align}
|f\left[Q;P(t)\right]|^2 \sim \frac{1}{d^2-1} \qquad Q\not= \bm I,
\end{align}
achieving genuine scrambling. 

Many previous studies in quantum information literature recognize random Clifford unitaries as scramblers since they delocalize operators and create nearly maximal entanglement when applied to arbitrary product states (see~\cite{Lashkari13, Brown15} for instance). However, Clifford operators represent only a very restricted subset of all possible time-evolutions, and one expects that many aspects of thermalization in many-body quantum systems require more complex unitaries to capture. Hence, we think it is important to distinguish two classes of scramblers. A certain plausible definition of scrambling which distinguishes two different classes was proposed in Ref.~\cite{Yoshida:2017aa} based on late-time asymptotic behavior of OTOCs. 



We now turn our attention to CV systems.
A central motif of our work will be an analogy between Pauli operators in DV systems and \emph{displacement operators} in CV systems.
In a single-mode CV system (e.g. the simple harmonic oscillator), the displacement operator shifts a coherent state by position $\xi_1$ and momentum $\xi_2$ in phase space. 
It takes the form 
\be
D\left(\xi_{1},\xi_{2}\right) \equiv \exp\left[i \left(\xi_{2} q-\xi_{1}p\right) \right],
\ee
where  $q$ and $p$ are the canonical position and momentum quadrature operators.
Similar to Paulis, the displacement operators of an $N$-mode CV system are formed as tensor products of local displacement operators, written as 
\be
D\left({\bm \xi}\right)\equiv \bigotimes_{k=1}^N D^k\left(\xi_{2k-1},\xi_{2k}\right),
\ee
where $D^k\left(\xi_{2k-1},\xi_{2k}\right)$ is the mode $k$ single-mode displacement operator, and $\bm \xi\in R^{2N}$ is a $2N$-component vector of displacements.
Like Paulis, displacement operators form a complete operator basis, obeying
$
\tr\left( D\left({\bm \xi}\right) D\left({\bm \xi}^\prime\right)\right)=\pi^N \delta\left(\bm \xi+\bm \xi^\prime \right)$ and ${1}/{\pi^N}\int d^{2N} {\bm \xi} \ D\left({\bm \xi}\right) A D^\dagger\left({\bm \xi}\right)=\tr\left(A\right){\bm I}$.

\begin{figure}
\centering
\includegraphics[width=0.4\textwidth]{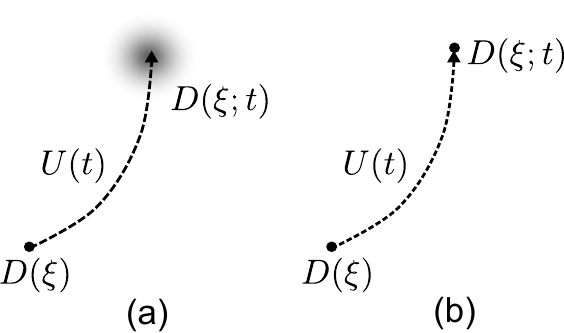}
\caption{Operator spreading in two types of CV scrambling. (a) In genuine scrambling, an initial displacement operator time-evolves into a sum of many displacements, spread throughout phase space. (b) In quasi scrambling, displacements may move around phase space, but remain localized.
\label{strong_weak}
}
\end{figure}

Inspired by the similarities between displacement operators and Paulis, we will characterize scrambling in CV systems by considering the time-evolution of displacement operators, $D\left(\bm \xi_1;t\right)\equiv U\left(t\right)^\dagger D\left(\bm \xi_1\right) U\left(t\right)$.
 The completeness of displacement operators allows the decomposition
\be
D\left(\bm \xi_1;t\right)=\frac{1}{\pi^N}\int d^{2N} {\bm \xi}_2 \ \chi\left[{\bm \xi}_2;D\left(\bm \xi_1;t\right)\right]D\left(-{\bm \xi}_2\right),
\label{D_decomposition}
\ee
where
$
\chi\left({\bm \xi};A\right)\equiv\tr \left[A D\left({\bm \xi}\right)\right]
$ is known as the Wigner characteristic function.
Initially, the characteristic function $\chi\left[{\bm \xi}_2;D\left(\bm \xi_1;0\right)\right]=\pi^N\delta \left(\bm \xi_2+\bm \xi_1\right)$ is highly localized in phase space; scrambling in CV systems can therefore be characterized by the growth of $\chi\left[{\bm \xi}_2;D\left(\bm \xi_1;t\right)\right]$. As in the DV systems, we identify two distinct classes of scrambling dynamics (visualized in Fig.~\ref{strong_weak}). Time-evolution $U\left(t\right)$ is \emph{quasi scrambling} if $\chi\left[{\bm \xi}_2;D\left(\bm \xi_1;t\right)\right]$ remains highly localized in phase space, but spreads to multiple modes. Time-evolution $U\left(t\right)$ is \emph{genuine scrambling} if $\chi\left[{\bm \xi}_2;D\left(\bm \xi_1;t\right)\right]$ spreads significantly over phase space for generic choices of $\bm \xi_1$. 

This separation of CV scramblers can be related to a common classification of CV unitaries into Gaussian and non-Gaussian operators. 
A complete introduction to Gaussian unitaries can be found in Ref.~\cite{Weedbrook_2012}; we provide a brief overview here. 
We begin by condensing notation, defining the vector of quadrature operators ${\bm x}=\left(q_1,p_1,\cdots, q_N,p_N\right)$. 
This allows us to concisely write an $N$-mode displacement operator as $D\left({\bm \xi}\right)=\exp\left(i {\bm x}^T {\bm \Omega} {\bm \xi} \right)$, defining the block diagonal matrix ${\bm \Omega} = \bigoplus_{k=1}^N \begin{psmallmatrix} 0 & 1\\ -1 & 0\end{psmallmatrix} $.
The product of displacement operators is now given by the simple addition,
\begin{align}
D\left({\bm \xi_{1}}\right)D\left({\bm \xi_{2}}\right) = e^{-i \bm \xi_{1}^{T} \bm \Omega \bm \xi_{2}} D\left({\bm \xi_1 + \bm \xi_{2}}\right). \label{displacement addition}
\end{align}

Gaussian unitaries are generated by Hamiltonians that are second order in the quadrature operators. 
One can represent a Gaussian unitary via its action on quadrature operators, which takes the form
\begin{align}
{U}_{\bm S,\bm d}^\dagger {\bm x} {U}_{\bm S,\bm d}=\bm S {\bm x}+{\bm d},
\end{align}
where the unitary is labelled by a $2N$-component displacement ${\bm d} $ and a $2N \times 2N$ symplectic matrix $\bm S$,
${\bm S}{\bm \Omega} {\bm S}^T={\bm \Omega}$. 
%
From this, one can show that Gaussian unitaries transform displacement operators into other displacement operators according to
\be 
U_{\bm S, \bm d}^\dagger D\left(\bm \xi\right)U_{\bm S, \bm d}
=  \exp\left(i{\bm d}^T {\bm \Omega} {\bm \xi}\right) D\left({\bm S}^{-1}{\bm \xi} \right),
\label{Gaussian_map}
\ee
analogous to the action of Clifford operators on Paulis in DV systems.
As such, the Wigner characteristic function of $D(\bm \xi_1;t)$ remains highly localized under time-evolution by a Gaussian unitary, and so we identify Gaussian unitaries as quasi scramblers. 
Genuine scramblers correspond to non-Gaussian unitaries, generated by Hamiltonians that are third or higher order in the quadrature operators. 
This can arise from interactions, hard boundary conditions (e.g. quantum billiard systems), or non-linear gates such as the single-mode cubic phase gate~\cite{gottesman2001encoding} or the Kerr effect~\cite{Ding2017} in quantum optical experiment.

\section{Operator spreading in genuine scrambling}
\label{sec:genuine}

In this section, we develop a basic formalism to characterize a time-evolved operator's distribution in \emph{phase space} as a probe of genuine CV scrambling. 
We begin by demonstrating that one aspect of this distribution --- its volume --- can be measured in two ways: via averages of \emph{time-ordered} correlation (TOC) functions of (generically non-local) displacement operators, and via the so-called frame potential,  previously used to study the complexity and pseudorandomness of DV unitary ensembles~\cite{roberts2017chaos}.
While the two measures are identical at infinite temperature, they differ when regularizing by a density matrix, which will be necessary in  CV systems.
%
%
To probe finer-grained aspects of operator spreading we turn to OTOCs, establishing a precise relation between a time-evolved operator's phase space distribution and its OTOC with displacement operators. 
One consequence of this relation is a constraint on OTOC decay: specifically, the magnitude of the OTOC can \emph{only} decay during genuine scrambling (non-Gaussian dynamics), and is strictly 1 in quasi scrambling (Gaussian dynamics). 
Further, this relation prompts us to consider more general averages of OTOCs over ensembles of displacement operators. 
We show that average OTOCs measure a coarse-grained density of an operator's phase space distribution; their change in time thus characterizes a `flow' of the distribution.
To conclude, we apply these tools to study two specific non-Gaussian Hamiltonians, an exactly-solvable cubic potential and the chaotic Henon-Heiles potential, and show that both models lead to operator spreading and OTOC decay.

\subsection{Phase space volume and the frame potential}

We begin by introducing our notion of an operator's `phase space volume'.
We show that in DV systems one can define such a volume for time-evolved Pauli operators, (roughly) by counting the number of Pauli strings with non-zero weight when expanding the operator in the Pauli basis. 
This can be probed by TOCs at infinite-temperature or, equivalently, the $k=1$ frame potential. 
In generalizing this to CV systems, one encounters various divergences due to the lack of an infinite-temperature limit, as the CV system's Hilbert space is not bounded. 
To regulate this, we expand our definition of phase space volume to be with respect to a normalized density matrix $\rho$. 
In sufficiently low-temperature regimes, this `coarse-grained' volume behaves qualitatively differently from the previous volume.
We demonstrate this in multiple examples for the specific case of a thermal density matrix with respect to the number operator Hamiltonian.



\subsubsection{Defining phase space volume for CV systems}

We first introduce an operator volume in DV systems. Consider a system of $N$ qudits, with total Hilbert space dimension $d=q^N$. As Pauli operators form a complete basis, the time-evolution of a Pauli operator $P$ may be decomposed as $P(t) = \sum_{Q \in \text{Pauli}} f\left[Q;P(t)\right] Q$. The coefficients $f\left[Q;P(t)\right]$ coincide with the TOC of $P$ and $Q$ with respect to the infinite-temperature density matrix $\rho_\infty = \frac{1}{d} \boldsymbol{1}$,
\be 
f\left[Q;P(t)\right] = \mathcal{C}_{1}( P(t), Q(0))  \equiv \frac{1}{d}\tr\left(P(t) Q^{\dagger}(0) \right).
\ee 
One can show that $\sum_Q|f\left[Q;P(t)\right]|^2=1$. This allows one to consider the \emph{ensemble} of Pauli operators $\calE[P(t)]=\{Q|Q\sim  |f\left[Q;P(t)\right]|^2\}$, defined by the normalized probability distribution $|f\left[Q;P(t)\right]|^2$. One can now define a volume of the ensemble $\calE[P(t)]$ corresponding to the number of $Q$ with significant $|f\left[Q;P(t)\right]|^2$. This can be made rigorous using entropies of the distribution $|f\left[Q;P(t)\right]|^2$. Specifically, the exponential of the R\'{e}nyi-$2$ entropy
\begin{align}
\text{vol}(\calE[P(t)]) \equiv 2^{S^{(2)}_{\calE[P(t)]}} = \frac{1}{\sum_{Q\in \text{Pauli}} |f\left[Q;P(t)\right]|^4}
\label{Volume_TOC}
\end{align}
provides a good measure of such a volume. The R\'{e}nyi-$2$ entropy of the ensemble is related to the frame potential~\cite{roberts2017chaos}, which we find corresponds to an inverse volume:
\begin{align}
\calF_{\calE[P(t)]} \equiv
\mathbb{E}_{U, V\sim \calE[P(t)]}
\left\{ \big|\tr\left( U^\dagger V\right)\big|^{2}\right\}
= d^2 \, 2^{-S^{(2)}_{\calE[P(t)]}},
\label{frame_potential_volume}
\end{align}
where $\mathbb{E}_{U,V\sim \calE[P(t)]}$ denotes the expectation value when $U,V$ are sampled independently from the ensemble $\calE[P(t)]$. 

In generalizing this relation to CV systems, we encounter one of our first obstacles in working with infinite-dimensional systems: the traces used in DV expressions, regulated by the infinite-temperature density matrix, are often infinite in the CV context, where infinite-temperature is not well-defined. For instance, when trying to characterize the volume of a time-evolved displacement operator $D\left(\bm \xi_1;t\right)$, the analog of $f\left[Q;P(t)\right]$ would naturally be the characteristic function $\chi\left[{\bm \xi};D\left(\bm \xi_1;t\right)\right]$. However, the volume of the characteristic function is not well-defined, as $|\chi\left[{\bm \xi};D\left(\bm \xi_1;t\right)\right]|^2$ has infinite norm [for example, at time $t = 0$, $\chi \sim \delta(\bm \xi - \bm \xi_1)$].

To remedy such divergences, it is natural to consider TOCs at a finite temperature,
\begin{align}
&\calC_1(\bm \xi_1,\bm \xi_2;t)_\rho=\tr\left[\rho D\left(\bm \xi_1;t\right)D\left(\bm \xi_2\right)\right].
\label{correlator}
\end{align}
Expressed in terms of the operators' characteristic functions, the TOC can be viewed as $\chi\left[{\bm \xi};D\left(\bm \xi_1;t\right)\right]$ `smeared' over a width in phase space determined by $\chi\left[{\bm \xi};\rho\right]$,
\begin{align}
&\calC_1(\bm \xi_1,\bm \xi_2;t)_\rho
=
\nonumber
\\
&
\frac{1}{\pi^{N}}
\int d^{2N} {\bm \xi} \ \chi\left[{\bm \xi};D\left(\bm \xi_1;t\right)\right] \, 
e^{i \bm \xi^T \bm \Omega \bm \xi_2}  \,
\chi\left({\bm \xi}-{\bm \xi_2};\rho\right).
\label{correlator_connection}
\end{align}
For the sake of illustration, throughout the manuscript we will frequently take $\rho$ to be the thermal density matrix for the number operator Hamiltonian, $\tilde{\rho}_{n_{\text{th}}} \sim \bigotimes_{k=1}^N e^{- \beta_{n_{\text{th}}} a^{\dagger}_k a_k}$, which we refer to as the 'thermal density matrix' for convenience. Here the effective temperature $\beta_{n_{\text{th}}}^{-1} = \ln(1 + \frac{1}{n_{\text{th}}})^{-1}$ is proportional to the mean photon number per mode $n_{\text{th}}$ in the CV limit $n_{\text{th}} \gg 1$. The characteristic function of the thermal density matrix has a width $\sim 1/n_{\text{th}}$ in phase space, $\chi\left(\bm \zeta; \tilde{\rho}_{n_{\text{th}}}\right)=\exp\left(-\left(n_{\text{th}}+1/2\right)|\bm \zeta|^2\right)$. In the high-temperature limit ($n_{\text{th}}\gg1$), this width becomes small, and $\calC_1(\bm \xi_1,\bm \xi_2;t)_\rho$ becomes proportional to the unregulated $\chi\left[{\bm \xi}_2;D\left(\bm \xi_1;t\right)\right]$.

Unlike their infinite temperature counterparts, finite temperature TOCs possess a well-defined norm $\calN_\rho\equiv\int d\bm \xi_2 |\calC_1(\bm \xi_1,\bm \xi_2;t)_\rho|^2 = \pi^N \tr\left(\rho^2\right)$.
This allows us to proceed similarly to the DV case and consider the ensemble of displacement operators $\calE = \{ D\left(\bm \xi_2\right)|\bm \xi_2\sim|\calC_1(\bm \xi_1,\bm \xi_2;t)_\rho|^2 / \calN_\rho\}$. The volume of such an ensemble can be defined similarly to Eq.~\eqref{Volume_TOC},
\be
{\rm vol}\left(\calE; \rho \right) = \left(\frac{1}{4\pi}\right)^N \frac{\calN_\rho^2}{\int d{\bm \xi} \, |\calC_1(\bm \xi_1,\bm \xi_2;t)_\rho|^4}.
\label{vol_general_displacement}
\ee
In Appendix~\ref{App_Louiville}, we use this definition to provide quantum analogs of Liouville's theorem and the Kolmogorov-Sinai entropy for Gaussian time-evolution. 

\subsubsection{Phase space volume and the CV frame potential}\label{Sec_volume_FP}

While the above satisfactorily defines the phase space volume of CV operators, we find it interesting to also generalize Eq.~\eqref{frame_potential_volume}, which provides a powerful interpretation of the frame potential as a phase space volume. To do so we again require regulation by a density matrix $\rho$, which we insert into the frame potential in two distinct ways, providing phase space interpretations for each. Further discussion of CV frame potentials is contained in Section~\ref{Sec_FP}, where they are used to verify unitary designs in CV systems.

\begin{figure}
\includegraphics[width=0.45\textwidth]{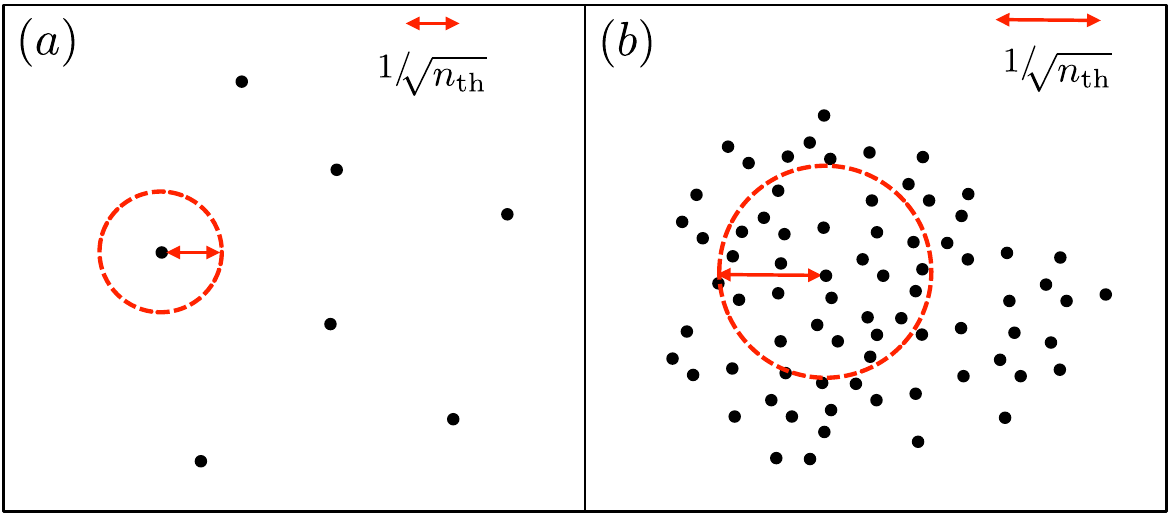}
\caption{
The finite temperature frame potential measures the volume of unitary ensembles (black dots) coarse-grained over a distance $1/\sqrt{n_{\text{th}}}$ in phase space (red), where $n_{\text{th}}$ is the number of photons in the thermal density matrix (proportional to the temperature). This leads to different behavior for (a) sparsely-distributed ensembles (high temperature) vs. (b) densely-distributed ensembles (low temperature).
\label{frame_counting}
}
\end{figure}

To begin, we consider the following finite-temperature frame potential~\cite{roberts2017chaos},
\be 
\calF_\calE \left(\rho\right)=
\mathbb{E}_{U, V\sim \calE}
\left\{ \big|\tr\left(\rho \, U^\dagger V\right)\big|^{2}\right\}.
\label{frame_simp_k1}
\ee
For the thermal density matrix, this frame potential measures the inverse volume of an ensemble `smeared' over distances $1/\sqrt{n_{\text{th}}}$, in units of $1/n_{\text{th}}^N$.
To see this, consider an individual term $|\tr\left(\tilde{\rho}_{n_{\text{th}}} U^\dagger V\right)|^{2}$ for displacements $U= D({\bm \xi}_1)$, $V=D({\bm \xi}_2)$. One computes
\begin{align}
|\tr\left(\tilde{\rho}_{n_{\text{th}}} D({\bm \xi}_1)^\dagger D({\bm \xi}_2)\right)|^{2} 
= e^{-\left(2n_{\text{th}}+1\right)|{\bm \xi}_1-{\bm \xi}_2|^2} 
\nonumber
\\
\simeq 
  \begin{cases} 
   1 & \text{if } \ |{\bm \xi_{1}} - {\bm \xi_{2}}| \lessapprox \frac{1}{\sqrt{n_{\text{th}}}}, 
   \\
   0       & |{\bm \xi_{1}} - {\bm \xi_{2}}| \gtrapprox  \frac{1}{\sqrt{n_{\text{th}}}}.
  \end{cases}
  \label{trace_DD}
\end{align}
Hence, the thermal frame potential treats two sampled displacements $U$, $V$ as identical if they are within distance $\lessapprox 1/\sqrt{n_{\text{th}}}$ of each other. In the infinite-temperature limit this distance goes to zero, and one finds a correspondence to our previous definition of volume Eq.~\eqref{vol_general_displacement},
\be
\lim_{n_{\text{th}}\to\infty}\!\! {\rm vol}\left(\calE; \tilde{\rho}_{n_{\text{th}}}\right)=\lim_{n_{\text{th}}\to\infty}\!
\left(\frac{1}{8n_{\text{th}}}\right)^N \!\!
\frac{1}{\calF_{\calE}^{(1)}\left(\tilde{\rho}_{n_{\text{th}}}\right)}.
\label{def_vol}
\ee

To further illustrate the connection between the finite-temperature frame potential and volume, we examine two example distributions, or \emph{ensembles}, of unitaries, as depicted in Fig.~\ref{frame_counting}. First, consider a discrete ensemble of $M$ displacement operators distributed sparsely in phase space compared to $1/\sqrt{n_{\text{th}}}$, shown in Fig.~\ref{frame_counting}(a). Here, the only nonzero contributions to the frame potential come from $U = V$. This occurs a fraction $1/M$ of the time, giving $\mathcal{F}_{\mathcal{E}} \approx 1/M$ --- intuitively, the smeared ensemble has a volume $(1/\sqrt{n_{\text{th}}})^{2N}$ per operator, which gives a volume $M$ in the units $(1/n_{\text{th}})^N$. Second, consider an ensemble of displacement operators densely distributed with density $\omega$ over some volume $\mathcal{V}$. Here, each displacement $U$ receives significant contributions from all operators within a $2N$-dimensional ball of radius $1/\sqrt{n_{\text{th}}}$ about $U$. There are $\sim \omega \left(1/\sqrt{n_{\text{th}}}\right)^{2N}$ such operators for each $U$, giving an inverse volume
\be
\calF_{\calE}\left(\tilde{\rho}_{n_{\text{th}}}\right) \sim \frac{ \omega \left(1/\sqrt{n_{\text{th}}}\right)^{2N}}{\mathcal{V} \omega}\sim \left(\frac{1}{n_{\text{th}}}\right)^{N}\frac{1}{\mathcal{V}}
\label{F_EC},
\ee
which is the inverse original volume, in the prescribed units. These two limits, sparsely and densely distributed, may equivalently be thought of as high and low temperature limits, as the length scale $1/\sqrt{n_{\text{th}}}$ by which the limits were defined is set by the inverse temperature.

\begin{figure}
\includegraphics[width=0.45\textwidth]{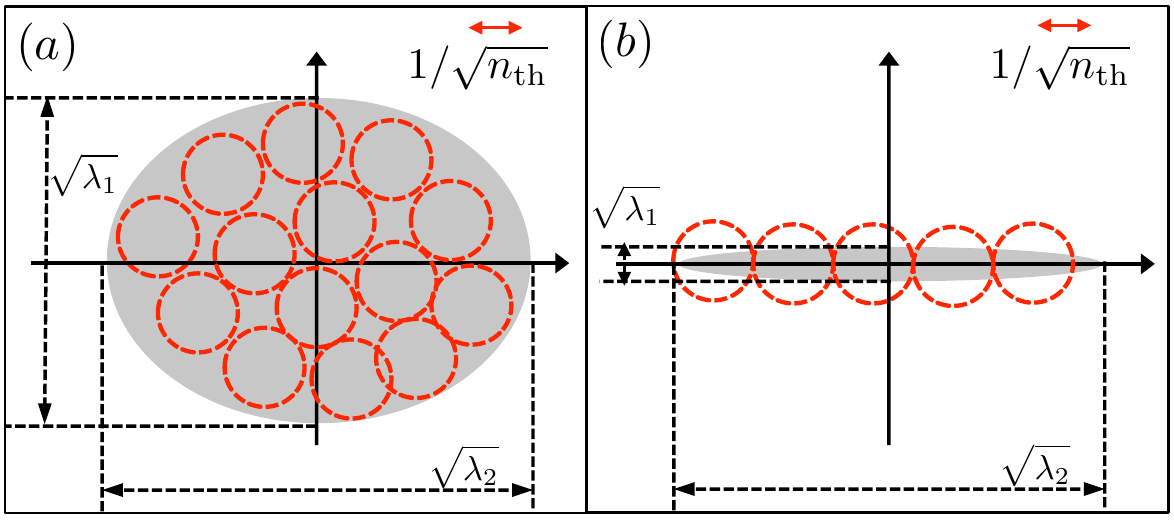}
\caption{
Schematic of the volume measured by the frame potential for two continuous ensembles of displacement operators (grey). (a) For an ensemble with large widths $\lambda_1,\lambda_2\gg 1/n_{\text{th}}$, it reproduces the traditional volume $\sqrt{\lambda_1\lambda_2}$. (b) If the ensemble is narrow in some quadrature, say $\lambda_1\ll 1/n_{\text{th}}$, it instead measures the `coarse-grained' volume $\sqrt{\lambda_2/n_{\text{th}}}$. Here $n_{\text{th}}$ is the number of photons in the thermal density matrix.
\label{frame_volume}
}
\end{figure}

The volume interpretation of the frame potential also extends to  continuous ensembles of displacement operators. We present this for the particular case of a Gaussian-distributed ensemble,
\begin{align}
\mathbb{D}_{\bm \xi_0,\bm V}=\left\{D\left(\bm \xi\right)| \bm \xi\sim 
P_D^G\left(\bm \xi;\bm \xi_0,\bm V\right)
\right\},
\label{D_Gaussian_ensemble}
\end{align}
where $P_D^G\left(\cdot;\bm \xi_0,\bm V\right)$ is a Gaussian distribution with mean $\bm \xi_0\in R^{2N}$ and covariance matrix $\bm V\in R^{2N}\times R^{2N}$. The frame potential can be computed, giving
\begin{align}
\calF_{\mathbb{D}_{\bm \xi_0,\bm V}} \left(\tilde{\rho}_{n_{\text{th}}}\right)
=\prod_{\ell=1}^{2N} \frac{1}{{\sqrt{1+4\lambda_\ell \left(2n_{\text{th}}+1\right)}}},
\label{frame_Gauss}
\end{align}
where the eigenvalues $\lambda_\ell$ of $\bm V$ give the squared width of the Gaussian along the $\ell^{\text{th}}$ eigenvector. An intuitive notion of volume would be the product of these widths, $\sqrt{{\rm det}\bm V} = \prod_\ell \sqrt{\lambda_\ell}$.
At high temperatures ($n_{\text{th}} \lambda_\ell \gg1 \, \forall \ell$, see Fig.~\ref{frame_volume}(a)), we indeed find
\be
\calF_{\mathbb{D}_{\bm \xi_0,\bm V}}\left(\tilde{\rho}_{n_{\text{th}}}\right) \simeq \left(\frac{1}{8 n_{\text{th}}}\right)^N \frac{1}{\sqrt{{\rm det}\bm V}}.
\label{frame_Gaussian_eq}
\ee
At lower temperatures this correspondence does not hold. Specifically, if $n_{\text{th}} < 1/\lambda_\ell$ for some $\ell$ (shown Fig.~\ref{frame_volume}(b)),  their contributions to the volume saturate to $O(1/n_{\text{th}}^N)$ constants. As before, this is related to the `smearing' due to the finite phase space resolution of the low temperature thermal density matrix.

An alternative route to characterizing the volume of a CV operator lies in a `twice-regularized' finite-temperature frame potential, which we introduce in detail in Sec.~\ref{Sec_FP} and Appendix~\ref{App_FP}. 
In the previous measure of volume, we inserted $\rho$ two separate times, once in the TOC and once in the finite-temperature frame potential. Here, we instead consider the ensemble of the \emph{unnormalized} characteristic function, $\calE = \{ D\left(\bm \xi_2\right)|\bm \xi_2\sim|\chi\left[{\bm \xi};D\left(\bm \xi_1;t\right)\right]|^2\}$, and measure its volume using the twice-regularized frame potential
\begin{align}
\calJ_\calE^{(1)}\left(\rho\right) \sim \mathbb{E}_{U, V\sim \calE}
\left\{ |\tr\left( \sqrt{ \rho } \, U^\dagger \sqrt{ \rho} \, V\right)|^{2}\right\}.
\end{align}
Intuitively, this frame potential both `smears' displacements over the width of $\chi[\bm \xi, \rho]$, and weights operators by their preservation of the low-energy subspace $\rho$, as detected by $\tr\left( \sqrt{ \rho } \, U^\dagger \sqrt{ \rho} \, U\right)$.
For the thermal density matrix $\tilde{\rho}_{n_{\text{th}}}$, one computes $\sqrt{ \tilde{\rho}_{n_{\text{th}}}} = \left(\frac{n_{\text{th}}' +1}{\sqrt{n_{\text{th}}+1}}\right)^N \tilde{\rho}_{n_{\text{th}}'}$
with $(n_{\text{th}}'+1)^2/n_{\text{th}}'^2 \equiv (n_{\text{th}}+1)/n_{\text{th}}$, which gives
\begin{align}
&\tr\left(\sqrt{\tilde{\rho}_{n_{\text{th}}}}D^\dagger(\bm \xi_1) \sqrt{\tilde{\rho}_{n_{\text{th}}}} D(\bm \xi_2) \right)
=
\left(\frac{\left(n_{\text{th}}^\prime+1\right)^2}{\left(n_{\text{th}}+1\right) \left(2n_{\text{th}}^\prime+1\right)}\right)^N
\nonumber
\\
&\times
\exp\left(-\frac{|\bm \xi_1|^2+|\bm \xi_2|^2}{2(2n_{\text{th}}^\prime+1)}\right) \exp\left(-\frac{|\bm \xi_1-\bm \xi_2|^2 n_{\text{th}}^\prime (n_{\text{th}}^\prime+1)}{(2n_{\text{th}}^\prime+1)}\right).
\end{align}
As expected, this contribution is significant for nearby $|\bm \xi_1 - \bm \xi_2| \lessapprox 1/\sqrt{n_{\text{th}}}$ displacements that approximately preserve the subspace of $\lessapprox n_{\text{th}}$ photons, $|\bm \xi_1|,|\bm \xi_2| \lessapprox \sqrt{n_{\text{th}}}$.

\subsection{Operator spreading and OTOCs}

As we have seen, an operators' phase space distribution can be characterized by TOCs. However, such a characterization is not always convenient, nor well-matched to physical observables. For instance, once the phase space volume of an operator becomes large, its characterization will require the measurement of a number of TOCs comparable to the dimension of the relevant many-body Hilbert space, many of which will involve highly non-local operators. 

This motivates us to turn our attention to OTOCs. In thermalizing many-body systems, the decay of OTOCs detects the spread of local operators in real space~\cite{kitaev2015simple, kitaev2018soft, harrow2009random, Brandao12, Hartman13, Pastawski15b, hosur2016chaos, hayden2016holographic, nahum2018operator, Keyserlingk:2018aa, Rakovszky:2018aa, Khemani:2018aa, xu2018locality, Bentsen2018}. In few-body CV systems, OTOCs have also found use due to their correspondence with diagnostics of classical chaos~\cite{rozenbaum2017lyapunov}.

In contrast to previous work on OTOCs in CV systems, we focus on OTOCs of displacement operators. 
We begin by establishing a precise Fourier transform-like relation between these OTOCs and operator distributions in phase space. 
We find that Gaussian and non-Gaussian unitaries, previously distinguished by their ability to spread operators in phase space, also have starkly different behavior on OTOCs. 
Gaussian time-evolution \emph{cannot} cause the OTOC magnitude to decay, and can only change the OTOC's phase---the decay of OTOCs thus serves as an indicator of non-Gaussian time-evolution.
Extending the relation between OTOCs and operator spreading, we show that the OTOC, when averaged over an ensemble of displacement operators, measures a time-evolved operator's support in a `ball', of designated width and center, in phase space.
This can probe operator spreading both in real space, by averaging over local displacement operators, as well as within the Hilbert space of a single mode, by varying the average displacement of the ensemble.
Later, in Section~\ref{Sec_experiment}, we will discuss measurement schemes for both OTOCs and average OTOCs, showing that both are efficiently measurable using only Gaussian operations. 

\subsubsection{OTOCs and operator spreading}

We begin by deriving an explicit relation between an operator's distribution in phase space and OTOCs. Again, it is helpful to first consider DV systems. OTOCs at infinite temperature are defined as
\begin{align}
\calC_2\left(
P(t),R(0)
\right) 
\equiv \frac{1}{d}\Tr[ 
P^{\dagger}(t)R^{\dagger}(0) P(t)R(0) ].
\end{align}
Expanding $P(t)=\sum_{Q}f\left[Q;P(t)\right]Q$, we have
\begin{align}
\calC_2\left(
P(t),R(0)
\right)  =  \sum_{Q \in \text{Pauli}} |f\left[Q;P(t)\right]|^2 \, 
\calC_2\left(
Q,R
\right).
\label{fourier_one_way}
\end{align}
Since Pauli operators commute up to a phase, $QR = e^{i\phi_{Q,R}}QR$, the OTOC of two Paulis $\calC_2\left(
Q,R
\right) \equiv \frac{1}{d}\Tr\left(Q^{\dagger}R^{\dagger}QR  \right)$ is an overall phase. This allows the following the inverse transformation:
\begin{align}
\frac{1}{d^2}\sum_{R\in \text{Pauli}}\calC_2\left(P, R \right)   \calC_2\left(R, Q \right)^{\star} = \delta_{P,Q},
\end{align}
which can be proven by using the orthogonality of the phase space point operators~\cite{Gross:2006aa}. Applying this to Eqs.~\eqref{fourier_one_way} we have
\begin{align}
|f\left[Q;P(t)\right]|^2  =\frac{1}{d^2} \sum_{R\left(0\right)\in \text{Pauli}} \calC_2\left(
P(t),R(0)
\right)
\calC_2\left(
R,Q
\right)^{\star}.
\label{fourier_other_way}
\end{align}
Thus, the probability distribution $|f\left[Q;P(t)\right]|^2$ can be obtained from OTOCs by a transformation similar to the discrete Fourier transform. A relation akin to this one was previously derived in Ref.\cite{roberts2017chaos}. 

In analogy to the DV case, we begin by introducing the CV OTOC with respect to displacement operators,
\begin{align}
\calC_2(\bm \xi_1,\bm \xi_2;t)_\rho=\tr\left[\rho D^\dagger\left(\bm \xi_1;t\right)D^\dagger\left(\bm \xi_2\right)D\left(\bm \xi_1;t\right)D\left(\bm \xi_2\right)\right].
\label{OTOC_general}
\end{align}
By using Eq.~\eqref{D_decomposition} this OTOC can be written in terms of the characteristic function $\chi\left[{\bm \xi};D\left(\bm \xi_1;t\right)\right]$ as
\begin{align}
&\calC_2(\bm \xi_1,\bm \xi_2;t)_\rho=\frac{1}{\pi^{2N}}
\int d^{2N}\bm \xi d^{2N} \bm \xi^\prime e^{-2i\left(\bm \xi_2-\bm \xi/2\right)^T \bm \Omega \bm \xi^\prime}
\nonumber
\\
& \ \ \ \ \ \ \ \ \times\chi^\star\left[\bm \xi; D\left(\bm \xi_1;t\right)\right]
\chi\left(\bm \xi-\bm \xi^\prime; \rho\right) 
\chi\left[\bm \xi^\prime; D\left(\bm \xi_1;t\right)\right],
\label{OTOC_connection}
\end{align}
similar to Eqs.~\eqref{fourier_one_way}, but with the characteristic function $\chi\left(\bm \zeta; \rho\right)$ of $\rho$ now acting as a transformation kernel. For the thermal density matrix in the infinite temperature limit, one has $\chi\left(\bm \xi-\bm \xi^\prime; \tilde{\rho}_{n_{\text{th}}}\right)\propto \delta\left(\bm \xi-\bm \xi^\prime\right)$ and achieves Fourier-like relations between TOCs and the OTOC:
\begin{align}
\calC_2(\bm \xi_1,\bm \xi_2;t)_\rho\propto \int d^{2N} \bm \xi \, |\calC_1(\bm \xi_1,\bm \xi;t)_\rho|^2 e^{-2i\bm\xi_2^T \bm \Omega \bm \xi},
\label{C2_highT}
\\
|\calC_1(\bm \xi_1,\bm \xi_2;t)_\rho|^2 \propto \int d^{2N} \bm \xi \,
\calC_2(\bm \xi_1,\bm \xi ;t)_\rho e^{ 2i\bm\xi^T \bm \Omega \bm \xi_2},
\label{C1_highT}
\end{align}
analogous to Eqs.~(\ref{fourier_one_way}, \ref{fourier_other_way}).
At finite temperature, the second of these becomes
\begin{align}
&\int d^{2N} \bm \xi_2 \, e^{2i\bm \xi_{2} \Omega \bm\xi^\prime} \mathcal{C}_{2}(\bm \xi_1,\bm \xi_2;t)_\rho 
= \int d^{2N} \bm \xi \, e^{i\bm \xi_{2}^T \Omega \bm \xi^\prime } \nonumber \\
& \ \ \ \ \ \ \  \ \ \times \chi^\star\left[\bm \xi; D\left(\bm \xi_1;t\right)\right] \chi\left(\bm \xi-\bm \xi^\prime; \rho\right) \chi\left[\bm \xi^\prime; D\left(\bm \xi_1;t\right)\right].
\end{align}
For the thermal density matrix, this implies that the Fourier transform of OTOCs addresses probability distributions coarse-grained over a width $\sim \frac{1}{\sqrt{n_{\text{th}}}}$.

We further illustrate this relation between OTOCs and operator spreading with a brief example. Consider an operator that spreads to have width $V$ in phase space, $\chi\left[\bm \xi; D\left(\bm \xi_1;t\right)\right]\sim \exp\left(|\bm \xi-\bm \xi_1|^2/\left(2 V\right)\right)$. This leads to an OTOC $\calC_2(\bm \xi_1,\bm \xi_2;t)_\rho\sim \exp\left(-V|\bm \xi_2|^2\right)$, where a larger phase space width implies greater decay of the OTOC.

\subsubsection{Average OTOCs}
\label{Sec_averageOTOC}

We have seen that the decay of OTOCs detects the increase in operator volume characteristic of genuine scrambling. In this section, we demonstrate that OTOCs can also probe finer-grained aspects of operator spreading. Specifically, we show that \emph{averages} of OTOCs over ensembles of displacement operators measure a coarse-grained distribution of a time-evolved operator in phase space. We begin by reviewing the use of average OTOCs to characterize real space operator spreading in DV systems. Following this, we turn to the simplest example of a CV average OTOC, which detects the amount that an operator has spread outside a ball of some radius about the identity. We then generalize this, and show that the center of the ball, as well as its widths in every direction of phase space, can be tuned by varying the ensemble over which the OTOC is averaged. This heuristic geometrical picture suggests a hydrodynamical interpretation of operator spreading, where changes of OTOCs in time characterize a flow of the operator's distribution through the surfaces of these balls. In Section~\ref{Sec_circuit}, we develop such a description for random Gaussian circuits.

In DV systems, it is well-known that averaging the OTOC over ensembles of Pauli operators can detect operator spreading in real space.
For example, consider the probability that an operator $P(t)$ has non-trivial support on the $j$-th qudit, 
\begin{align}
\mathcal{W}_{j} \equiv \sum_{Q \in \text{Pauli}\ : \  Q|_{j}\not = I } |f\left[Q;P(t)\right]|^2,
\end{align}
where $Q|_{j}$ represents the Pauli operator content at $j$-th qudit. This can be rewritten in terms of OTOCs as~\cite{hosur2016chaos}
\begin{align}
1 - \mathcal{W}_{j} =  \frac{1}{d^2} \sum_{R \in \text{Pauli} \ : \ R = I \otimes \cdots I \otimes R_{j} \otimes I \cdots \otimes I} \calC_2\left(
P(t),R
\right)
\end{align}
where $R$ are single qubit Pauli operators acting on $j$th qubit. This shows that the generic smallness of OTOCs, namely those averaged over single-qudit Pauli operator, implies higher weights in the operator spreading of $P(t)$. 



\begin{figure}
\includegraphics[width=0.35\textwidth]{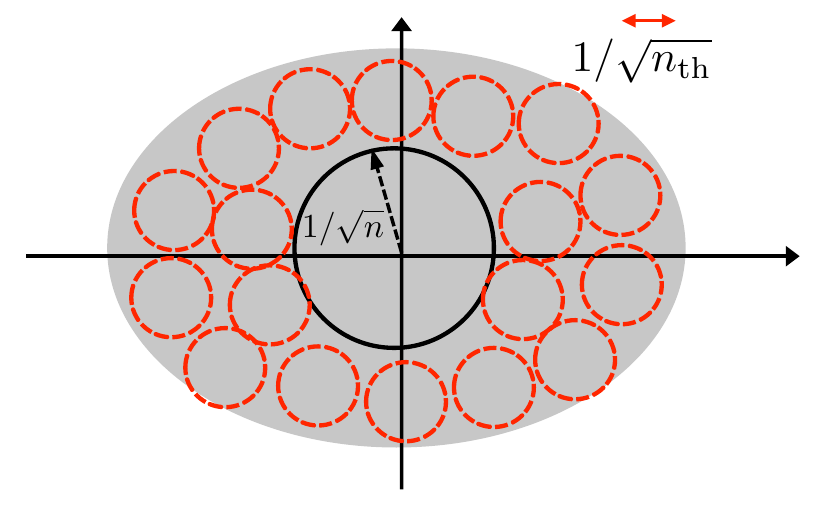}
\caption{
The average OTOC measures the extent that an operator (grey) has spread outside a ball of distance $1/\sqrt{n}$ (black) in phase space, after coarse-graining over scales $1/\sqrt{n_{\text{th}}}$ (red). Here $n$ is the width of the ensemble that the OTOC is averaged over, and $n_{\text{th}}$ is the number of photons in the thermal density matrix.
\label{OTOC_schematic}
}
\end{figure}

In CV systems, averaging the OTOC over ensembles of displacement operators can probe operator spreading not only in real space, but also phase space.
We define the average OTOC over an ensemble $\calE$ of displacement operators as: 
\be
\overline{\calC_2}\left(D\left(\bm \xi_1;t\right) , \calE \right)_\rho \equiv
\mathbb{E}_{V\sim \calE } \
[ \calC_2\left(D\left(\bm \xi_1;t\right) , V \right)_\rho].
\label{OTO_ave_general}
\ee
There are two important regularizations in this OTOC to keep in mind: the finite extent of ensemble $\calE$, and the density matrix $\rho$. To gain insight on their effects, we begin with a fully symmetric Gaussian ensemble of displacements,
\be 
\mathbb{D}_n=\left\{D\left(\bm \xi\right)| \bm \xi\sim P_D^G\left(\bm \xi; n \right)\equiv 
\exp\left(- |\bm \xi|^2/n\right)
/ \left(\pi n\right)^N  \right\}.
\label{Dn_def}
\ee
To understand the effect of the ensemble size $n$, consider the average OTOC for an unspread displacement operator $D\left(\bm \xi_1;t\right) = D\left({\bm \xi}\right)$. One finds
\be
\overline{\calC_2}\left(D\left({\bm \xi}\right), \mathbb{D}_n \right)_\rho= 
\exp\left(-n|\bm \xi|^2\right),
\label{OTOC_exp}
\ee
which is small unless $|\bm \xi| \leq {1}/{\sqrt{n}}$. This suggests that, for an arbitrary $D\left(\bm \xi_1;t\right)$ with characteristic function $\chi\left[\bm \xi; D\left(\bm \xi_1;t\right)\right]$, significant contributions to the OTOC arise only from regions $|\bm \xi| \leq {1}/{\sqrt{n}}$. This is confirmed by an explicit calculation of the average OTOC,
\begin{align}
&\overline{\calC_2}(D\left(\bm \xi_1;t\right) , \calE )_\rho=\frac{1}{\pi^{2N}}
\int d^{2N}\bm \xi d^{2N} \bm \xi^\prime  \exp\left(-n|\bm \xi |^2\right) e^{ i \bm \xi^{\prime T} \bm \Omega \bm \xi}
\nonumber
\\
& \ \ \ \ \ \ \ \ \times\chi^\star\left[\bm \xi^\prime; D\left(\bm \xi_1;t\right)\right]
\chi\left[\bm \xi^\prime-\bm \xi; \rho\right] 
\chi\left[\bm \xi; D\left(\bm \xi_1;t\right)\right],
\label{OTOC_connection2}
\end{align}
where contributions are damped by the same exponential factor $\exp\left(-n|\bm \xi |^2\right)$. On the other hand, the finite phase space resolution of $\rho$ (i.e. the width of $\chi\left[\bm \xi^\prime-\bm \xi; \rho\right]$) introduces mixing between $\chi^\star\left[\bm \xi^\prime; D\left(\bm \xi_1;t\right)\right]$ and $\chi\left[\bm \xi; D\left(\bm \xi_1;t\right)\right]$, which serves to coarse-grain the operator's phase space distribution on a scale set by $\rho$.  For a thermal state $\tilde{\rho}_{n_{\text{th}}}$, this occurs for $|\bm \xi^\prime -\bm \xi| \leq {1}/{\sqrt{n_{\text{th}}}}$. Combining the two regularizations (Fig.~\ref{OTOC_schematic}), we see that the decay of the average OTOC probes the extent that the operator distribution, coarse-grained by $\rho$, is spread outside a ball of radius $\sim 1/\sqrt{n}$ about the identity ($\bm \xi = 0$). Intuitively, one likely wants to set the coarse-graining to be on a scale smaller than the ball radius, $1/\sqrt{n_{\text{th}}} < 1/\sqrt{n}$.

From the analysis above, we see a clear distinction between volume characterization by the OTOC and the frame potential. Namely, the frame potential measures the overall volume of the coarse-grained operator spreading, whereas the average OTOC measures the extent to which the operator has spread outside the ball of radius ${1}/{\sqrt{n}}$. Broadly, this implies that the average OTOC decay earns contributions only from operators with large displacements (compared to $1/\sqrt{n}$). This feature is intuitively favorable for characterizing scrambling and operator growth in CV systems because larger displacement operators are more complex, requiring larger energy to implement.


One can also consider the average OTOC with respect to more generic Gaussian distributions $\mathbb{D}_{\bm \xi_0,\bm V}$ defined in Eq.~(\ref{D_Gaussian_ensemble}).
These detect the portion of a coarse-grained operator's distribution within a distorted ball of width(s) determined by $\bm V^{-1}$ about a center $\bm \xi_0$.
 As such, in principle, the average OTOC can measure a coarse-grained local probability density anywhere in the phase space with tunable short ($\sim 1/\sqrt{n_{\text{th}}}$) and long ($\sim 1/\sqrt{n}$) distance cut-offs. Such measures include probing operator spreading in real space, by choosing the widths of $\bm V$ to be large for modes inside some spatial region and small otherwise. Finally, we note that our focus on Gaussian ensembles is not only analytically convenient; in Sec.~\ref{Sec_experiment} we show that such average OTOCs are naturally measured in quantum optical experiments. 
 


\subsubsection{Measuring non-Gaussianity with OTOCs}
\label{Sec_nonGaussianity}

We have seen that Gaussian and non-Gaussian dynamics have different abilities to spread operators in phase space, termed genuine and quasi scrambling. Separately, we showed that operators' phase space distributions are related to OTOCs via Eq.~\eqref{OTOC_connection} (with Eqs.~(\ref{C2_highT}-\ref{C1_highT}) as limiting cases). Here we complete this triangle of relations, showing that decay of the OTOC \emph{magnitude} measures non-Gaussianity and operator spreading.

Once again, we begin by constructing a measure of non-Cliffordness for DV time-evolution. Using the commutation $PQ = e^{i\phi_{P,Q}} QP$ of Pauli operators, one can show that OTOCs with respect to Paulis satisfy ($P\neq I$)
\begin{align}
\big| \mathcal{C}_{2} ( P(t), R ) \big| 
&= 1, \ \text{ $\forall \,  R \in \text{Pauli}$ $\Leftrightarrow$ $U$ is Clifford,} 
\nonumber
\\
&< 1, \ \ \text{$\exists \, R \in \text{Pauli}$ $\Leftrightarrow$ $U$ is non-Clifford}.
\end{align}
Hence, the decay of the amplitude of OTOCs is sensitive to non-Cliffordness of $U$. This is another way to see why Clifford unitaries should not be called genuine scramblers. 
This prompts us to consider the sum of OTOC amplitudes as a faithful measure of non-Cliffordness:
\begin{align}
\mathcal{M}^{DV} &\equiv \frac{1}{d^4}\sum_{P, R\in \text{Pauli}}\big| \mathcal{C}_{2} ( P(t), R ) \big|^2 
\nonumber
\\
&= \frac{1}{d^2}\sum_{P\in \text{Pauli}} |f\left[Q;P(t)\right]|^4 
\nonumber
\\
&= \frac{1}{d^2}\sum_{P\in \text{Pauli}} \text{vol}(\calE[P(t)])^{-1},
\end{align}
where decay of $\mathcal{M}^{DV}$ from $1$ indicates non-Clifford behavior, and where we use Eqs.~\eqref{Volume_TOC} to relate $\mathcal{M}^{DV}$ to the phase space volume of time-evolved Pauli operators. This provides an explicit relation between OTOCs, operator spreadings and non-Cliffordness.

Analogous relations can be derived for CV systems. Using the commutation $D\left({\bm \xi_1} \right) D\left({\bm \xi_2} \right) = e^{-2i {\bm \xi_1}^T {\bm \Omega} {\bm \xi_2}} D\left({\bm \xi_2} \right) D\left({\bm \xi_1} \right)$, the OTOC under Gaussian time-evolution can be computed exactly, giving the overall phase
$
\calC_2(\bm \xi_1,\bm \xi_2;t)_\rho=e^{-2i{\left(\bm S^{-1}\bm \xi_1\right)}^T \bm \Omega \bm \xi_2}
$ with magnitude $1$. One can show that ($\bm \xi_1\neq \bm0$ and $\rho$ full rank)
\begin{align}
\big|\calC_2(\bm \xi_1,\bm \xi_2;t)_\rho\big|
&= 1, \ \ \text{$\forall  \, \bm \xi_2$ $\Leftrightarrow$ $U$ is Gaussian.} \\
&< 1, \ \ \text{$\exists \, \bm \xi_2$ $\Leftrightarrow$ $U$ is non-Gaussian.}
\end{align}
This suggests the following measure of non-Gaussianity
\begin{align}
\mathcal{M}^{CV} \equiv \mathbb{E}_{\bm \xi_{1}\sim \mathcal{E}_{1}, \bm \xi_{2}\sim \mathcal{E}_{2} } \big|\calC_2(\bm \xi_{1},\bm \xi_{2};t)_\rho\big|^2
\end{align}
with respect to some ensembles of displacement operators $\mathcal{E}_{1}, \mathcal{E}_{2}$, where decay of $\mathcal{M}^{CV}$ from $1$ indicates non-Gaussian behavior. In Section~\ref{Sec_experiment}, we present a measurement protocol of this quantity. Interestingly, our proposed quantity $\mathcal{M}^{DV}$ is closely related to a recent work on the stabilizer test by Gross, Nezami and Walter~\cite{gross2017schur}; it is an interesting future problem to relate $\mathcal{M}^{CV}$ to an analogous `Gaussianity test'.

\subsection{Examples of genuine scramblers}
\label{Sec_genuine_example}

Having established a foundation for characterizing scrambling and operator spreading in CV systems, we now apply these tools to two specific examples of non-Gaussian Hamiltonians: an exactly-solvable cubic phase gate in Section \ref{cubic phase}  and the quantum chaotic Henon-Heiles potential~\cite{feit1984wave,zhuang2013equilibration,zhuang2014quantum} in Section \ref{hh}. We show that both  lead to operator spreading in phase space and the decay of OTOCs. Each Hamiltonian is of the form
\be
H_{NG}=\sum_{k=1}^N p_k^2/2m+V\left(q_1,\cdots, q_N\right),
\label{HNG_general}
\ee
where the non-Gaussianity arises from a non-linear potential $V\left(q_1,\cdots, q_N\right)$. Such Hamiltonians encompass a variety of many-body phases, as well as the few-mode billiard systems often studied in quantum chaos. In addition to the examples given here, in Section~\ref{Sec_experiment} we describe genuine scrambling operations suitable to be realized in present-day quantum optics experiment.

\subsubsection{Cubic phase gate}\label{cubic phase}

We begin with an analytically tractable toy model: a single-mode Hamiltonian with a cubic potential $V=q^3/3!$, in the $m\to\infty $ limit~\cite{gottesman2001encoding}. Here we can solve exactly for the time-evolved displacement operator $D\left(\alpha;t\right)$. We use the Hadamard lemma
\be 
e^A B e^{-A}=B+\left[A,B\right]+\frac{1}{2!}\left[A,\left[A,B\right]\right]+\cdots\equiv 
\tilde{B}
\ee
and its straightforward extension $e^A f(B) e^{-A}=f(\tilde{B})$, for functions $f(B)$ with a Taylor expansion. Taking $e^A \equiv U\left(t\right)=\exp\left(-i \gamma t q^3/3!\right)$ and $f(B) = D\left(\alpha;t=0\right)$, we find
\be
D\left(\alpha;t\right)=
\exp\left[i\Big({\rm Im}\left(\alpha\right) q-{\rm Re}\left(\alpha\right)p+ \gamma t {\rm Re}\left(\alpha\right)q^2  \Big) \right].
\ee 
To compute the OTOC, we apply the Hadamard lemma again, obtaining
\begin{align}
&D^\dagger\left(\alpha;t\right) D^\dagger\left(\beta\right) D\left(\alpha;t\right)
\nonumber
\\
&=e^{i\Big(- {\rm Im}\left(\beta\right)q +{\rm Re}\left(\beta\right)p  + 2{\rm Im}\left(\alpha^\star \beta\right)+2 \gamma t {\rm Re}\left(\alpha\right) {\rm Re}\left(\beta\right) q \Big)},
\end{align}
which gives the OTOC
\begin{align}
\calC_2(\alpha,\beta;t)_\rho
= \exp\left({i \theta}\right) \chi\left(2i\gamma t{\rm Re}\left(\alpha\right) {\rm Re}\left(\beta\right);\rho\right)
.
\end{align}
Here the phase factor is given by 
\begin{align}
\theta=2{\rm Im}\left(\alpha^\star \beta\right)+ 2\gamma t {\rm R	e}\left(\alpha\right) {\rm Re}\left(\beta\right){\rm Re}\left(\alpha + \beta\right),
\end{align}
and $\chi({\rm Re}(\xi)+i{\rm Im}(\xi); \rho)$ is the characteristic function of $\rho$. For a thermal state $\tilde{\rho}_{n_{\text{th}}}$, we have
\begin{align}
\calC_2(\alpha,\beta;t)_{\tilde{\rho}_{n_{\text{th}}}}
=e^{i \theta}e^{-2\left(2n_{\text{th}}+1\right)\left({\rm Re}\left(\alpha\right) {\rm Re}\left(\beta\right) t\right)^2 \gamma^2}
\end{align}
indicating a Gaussian decay of the OTOC in time. From this, the operator's phase space distribution (diagnosed by TOCs) can be calculated via Eq.~\eqref{C1_highT}. For large $n_{\text{th}}$, we have
\begin{align}
&|\calC_1(\alpha,\beta^\prime;t)_{\tilde{\rho}_{n_{\text{th}}}}|^2
\approx  \delta \left( {\rm Re}(\alpha)-{\rm Re}(\beta') \right)
\nonumber
\\
&\times \exp\left[-  \frac{\left({\rm Im}\left(\beta'\right)-{\rm Im}\left(\alpha\right)+\gamma t {\rm Re}\left(\alpha\right)^2 \right)^2}{2(2n_{\text{th}}+1) \gamma^2 {\rm Re}(\alpha)^2  t^2 } \right] .
\end{align}
We see that the imaginary part of $D(\alpha;t)$ spreads as a Gaussian in phase space, while the real part does not spread (a consequence of the simplicity of our Hamiltonian, which commutes with $q$). The width of the operator distribution increases linearly in $t$, according to
\begin{align}
\big|{\rm Im}\left(\beta'\right) \big| \simeq \sqrt{2(2n_{\text{th}}+1)}\big|{\rm Re}\left(\alpha\right)\big| t  \gamma. 
\end{align}

\begin{figure}
\centering
\subfigure{
\includegraphics[width=0.23\textwidth,valign=t]{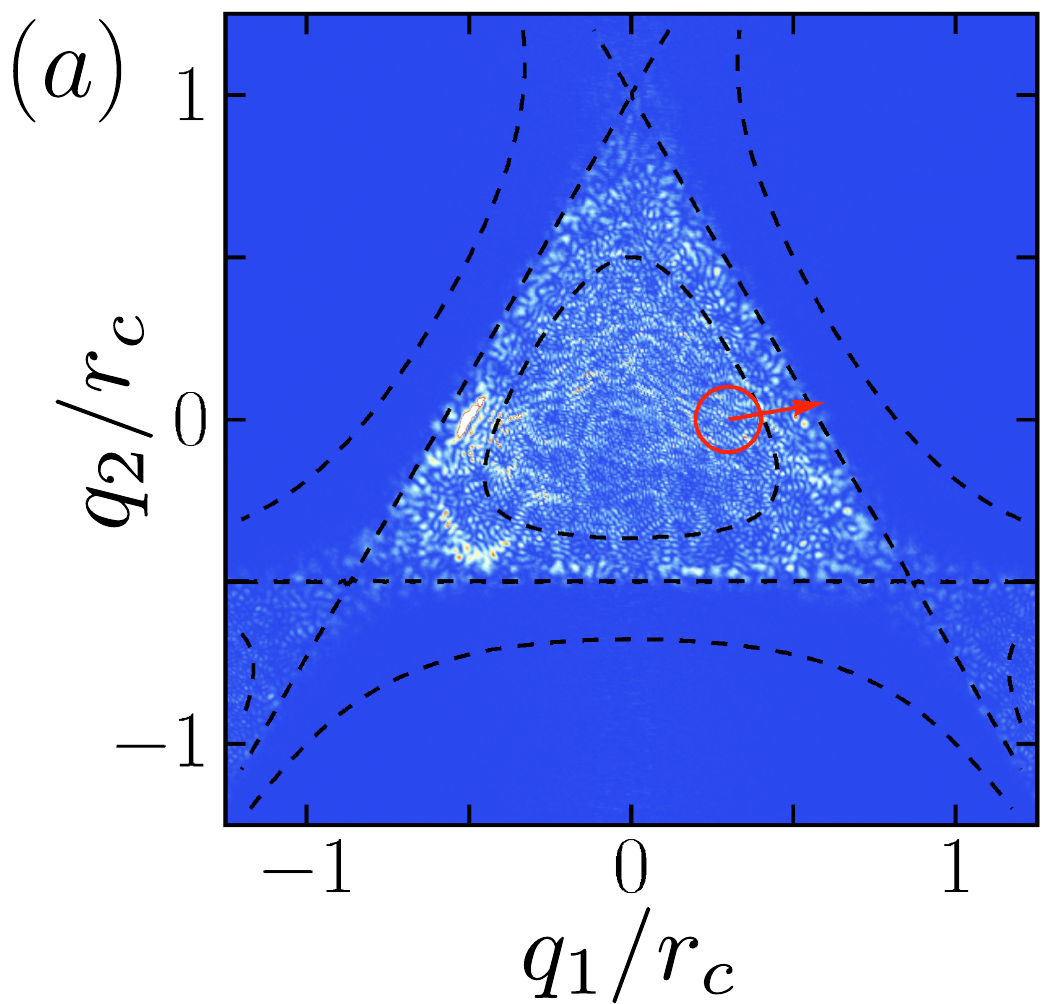}
}\hspace*{-0.85em}
\subfigure{
\includegraphics[width=0.25\textwidth,valign=t]{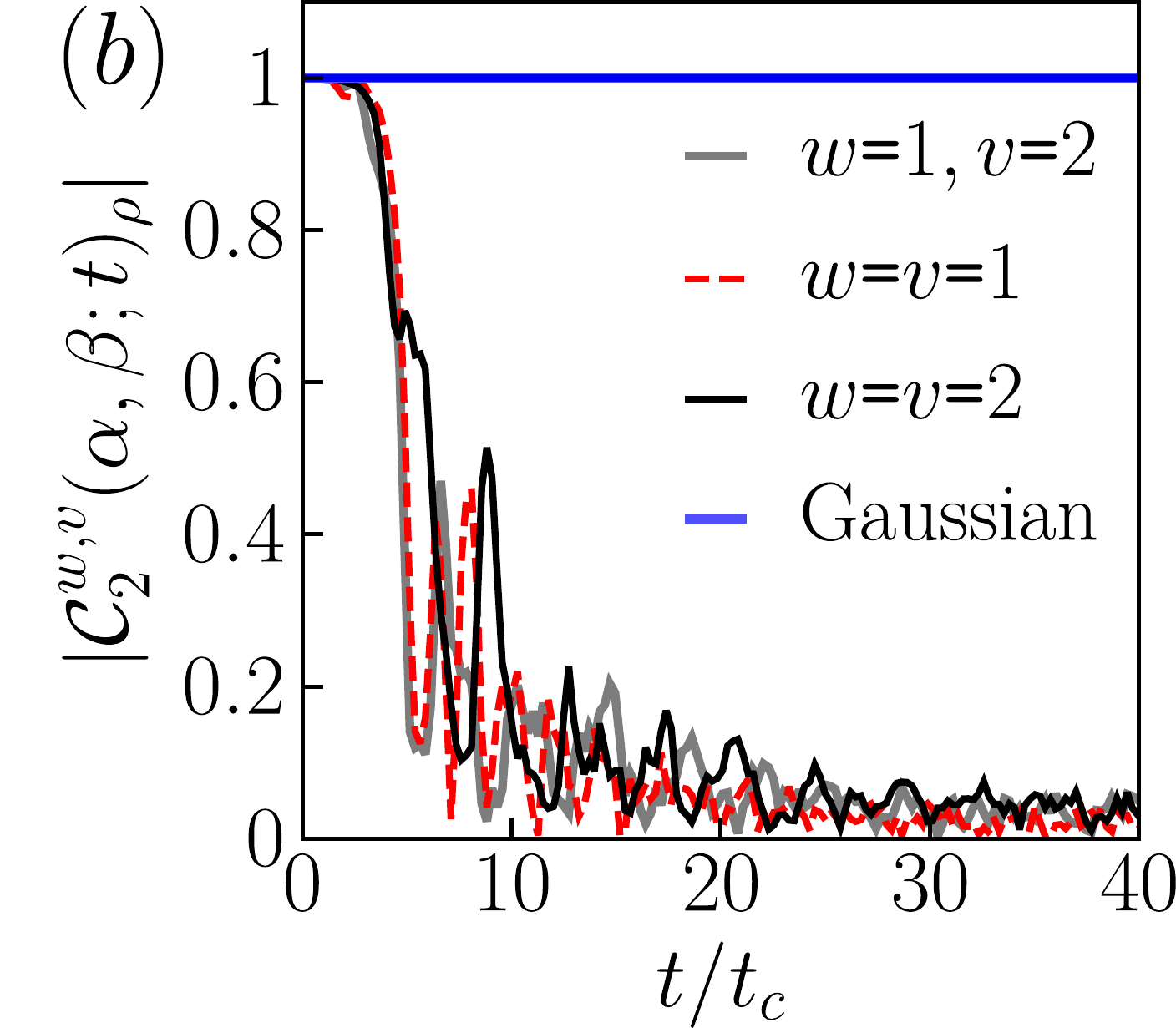}
}
\caption{Time-evolution under the Henon-Heiles potential. (a) The initial ($t=0$) wave function is localized in the red circle with momentum shown by the red arrow. Coloring displays the probability density of the final state ($t = 40t_c$), with blue indicating zero probability. Black dashed lines indicate contours $V=2V_C, V_C, V_C/2$ of the potential, and time is measured in units of $t_c=r_C/\sqrt{2V_C/m}$.  (b) The OTOC amplitude $|\calC_2^{w,v}\left(
\alpha,\beta;t
\right)_{\ket{\gamma_1}\ket{\gamma_2}}|$ for a two-mode coherent state $\gamma_1=0.15 r_C +i \sqrt{7V_C/40}\cos(10^\circ), \gamma_2=i \sqrt{7V_C/40}\sin(10^\circ)$ and displacement operators $\alpha=(1-i)/4, \beta=(1+i)/4$ on modes $v, w \in \{ 1, 2\}$. 
The OTOC decay for a different coherent state $\gamma_1=0.125 r_C +i 0.375 \sqrt{V_C}\cos(20^\circ)$, $\gamma_2=0.125r_C+i 0.375\sqrt{V_C} \sin(20^\circ)$ and displacements $\alpha=(1-i)/2, \beta=(0.4+0.3i)$ exhibits similar behavior (not shown). 
In contrast, the OTOC amplitude under Gaussian dynamics does not decay (blue).
\label{OTO_HH}
}
\end{figure}

\subsubsection{Henon-Heiles potentials}\label{hh}

The previous example of a genuine scrambler was integrable, to enable analytic treatment. To probe operator spreading in a non-integrable system, we consider the two-mode Henon-Heiles potential 
\be 
V\left(q_1,q_2\right)=U \left(q_1^2+q_2^2\right)/2+\lambda \left(q_1^2 q_2-q_2^3/3\right),
\ee 
which has a long history of study in both quantum and classical chaos~\cite{feit1984wave,zhuang2013equilibration,zhuang2014quantum}. We choose this potential partially with an eye on experiment: Hamiltonians that are low-order polynomials in $q,p$ are likely easier to realize in quantum optics experiment than those with hard-cut offs, such as billiard systems. This potential has a local minimum at the origin, and three saddle points at distance $r_C\equiv U/\lambda$ and energies $V_C= U^3/6\lambda^2$ [see Fig.~\ref{OTO_HH}(a)].
The classical orbits are chaotic for energies above $V_C/2$~\cite{hermann2016nonlinear}. In the quantum system, Ref.~\cite{zhuang2013equilibration,zhuang2014quantum} showed that initially local quantum states spread out in phase space, eventually approaching an equilibrium distribution.


We numerically study the Henon-Heiles potential with parameters $m=1/2, U=1,\lambda=0.025$. In agreement with previous works~\cite{zhuang2013equilibration,zhuang2014quantum}, we observe that an initial wavefunction localized in phase space spreads out over the entire classically allowed spatial region $V<V_C$ [Fig.~\ref{OTO_HH}(a)]. 
To study operator spreading, we numerically compute the OTOC for pure coherent states, $\rho = \ket{{\bm \gamma}}\bra{{\bm \gamma}}$. In practice, we expect such OTOCs to replicate much of the behavior of OTOCs with respect to the thermal density matrix, as, intuitively, all information about initial states other than their energy will be `forgotten' at late times due to the chaotic dynamics. Moreover, since the thermal state is a mixture of coherent states, $\tilde{\rho}_{n_{\text{th}}} \sim \int d {\bm \gamma}^{2N} e^{-\beta_{n_{\text{th}}}|\bm \gamma|^2}\ket{{\bm \gamma}}\bra{{\bm \gamma}}$, thermal OTOCs can be obtained exactly from an average over coherent state OTOCs.
As shown in  Fig.~\ref{OTO_HH}(b), the OTOC for a coherent state with energy $\sim V_C$ quickly decays on a timescale $\sim t_c =r_C/\sqrt{2V_C/m}$, roughly the time required to move from the origin to a saddle point for the classical Hamiltonian. To verify the generic nature of this decay, we computed the OTOC for a different initial state of similar energy, with respect to different displacement operators, and observe similar behavior.
Finally, as a counterpoint, we also compute the same OTOCs for a Gaussian potential, $V\left(q_1,q_2\right)=U \left(q_1^2+q_2^2\right)/2$. As expected the OTOC amplitude remains unity, indicating no genuine scrambling. 


\section{Operator distributions in quasi scrambling}\label{Sec_quasi}

In this Section, we attempt to learn more about genuine scrambling dynamics by studying \emph{quasi} scrambling systems.
Although we were able to exactly numerically simulate genuine scrambling Hamiltonians in Section \ref{Sec_genuine_example}, we were limited to few-mode Hamiltonians due to the exponential complexity of a many-mode Hilbert space.
%
%
%
In contrast, quasi scrambling (Gaussian) time-evolution can be efficiently simulated---with $N$ modes, one only needs to keep track of the $2N \times 2N$ symplectic matrix $\bm S$.

It is initially surprising that quasi scrambling can teach us anything about general scrambling systems, and we begin Section \ref{Sec_circuit} by briefly elaborating on the sense in which quasi scrambling unitaries are, and are not, capable of scrambling.
Following this, we explore random circuits of Gaussian unitaries on both single (Section \ref{Sec_single_mode}) and many (Section \ref{Sec_many_mode}) mode systems.
Unlike DV random circuits~\cite{harrow2009random, Brandao12, nahum2018operator, Keyserlingk:2018aa, Rakovszky:2018aa, Khemani:2018aa, xu2018locality, Bentsen2018}, we find that the accessible single-mode CV Hilbert space \emph{grows exponentially} in time, related to a tunable parameter, the \emph{squeezing}, of the Gaussian operation.
This in turn leads a squeezing-dependent ballistic spreading of operators in many-mode circuits, as well as an unusual quadratic growth of entanglement entropy.
To theoretically capture these results, we introduce a hydrodynamical model of operator spreading that is accurate in the low-squeezing regime.
%

\begin{figure}
\centering
\includegraphics[width=0.3\textwidth]{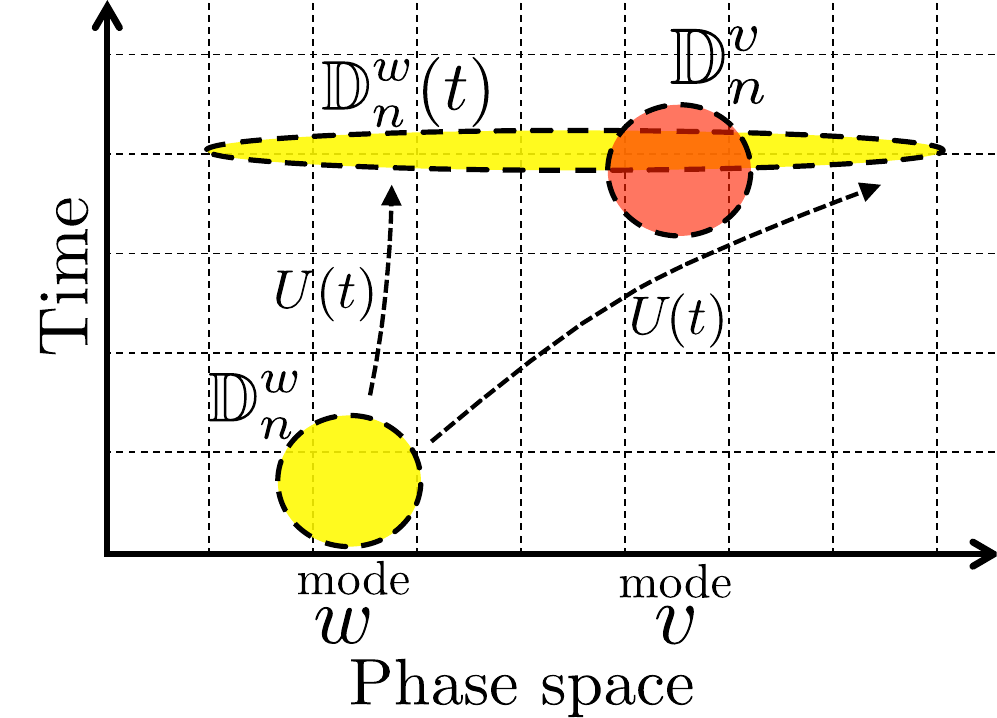}
\caption{Rough illustration of quasi scrambling and the quantum Liouville's theorem. While the global volume of an ensemble $\mathbb{D}_{n}^{w}(t)$ of time-evolved displacement operators (yellow) remains fixed under time-evolution by a Gaussian unitary $U(t)$, the projected volume on the mode $v$ (measured via an average over local displacements; red), and the coarse-grained volume (roughly, the number of boxes the ensemble occupies) may increase. 
\label{weak_scrambling} 
}
\end{figure}

%
%
In DV systems, the expectation that quasi scrambling random circuits can mimic aspects of actual physical systems is justified by the notion of unitary designs. 
In Section \ref{Sec_designs}, we adapt the definition of unitary designs to CV systems.
We provide explicit results for CV 1-designs formed by displacement operators, and analogs of 2-designs formed by Gaussian unitaries.
CV designs necessarily involve `cutting-off' an ensemble of unitaries at some finite extent, which we show can lead to ensembles that mimic design behavior on certain subspaces of a Hilbert space, but not the entire space.

\subsection{Random Gaussian circuits}\label{Sec_circuit}

Before addressing operator spreading in random Gaussian circuits, we find it useful to discuss the extent that such systems can scramble.
As we have seen, quasi scrambling unitaries cannot increase the volume of displacement operators in phase space.
In Appendix~\ref{App_Louiville}, we prove a quantum Louiville theorem that expands this non-increase to volumes of \emph{ensembles} of displacement operators.
Interestingly however, Gaussian dynamics can `squeeze' such ensembles so that their phase space volume \emph{appears} to increase after coarse-graining by a density matrix $\rho$, as depicted in Figs.~\ref{frame_volume}(b),~\ref{weak_scrambling}. 

Hints of this scrambling power were in fact already present when we considered average OTOCs. 
While quasi scrambling unitaries cannot cause any individual OTOC to decay, they can randomize the OTOC phase; when averaging many OTOCs, this leads to decay just as does genuine scrambling~\cite{footnote2}.
In what follows, we will be interested in averaging over the time-evolution \emph{itself} (i.e. the random circuits), but similar themes hold.

\subsubsection{Single-mode: growth of the accessible local Hilbert space}\label{Sec_single_mode}

We begin by studying random Gaussian circuits on a single mode. 
In DV systems, a single-qudit circuit would seem trivial --- the product of Haar random unitaries is also Haar random, and so sequential applications of them would have no interesting dynamics. 
In contrast, the set of CV Gaussian unitaries is unbounded, and to choose a random unitary we must cut-off this ensemble using a finite `squeezing' parameter.
The squeezing is \emph{not} invariant under composition of Gaussians; we will study its increase in time, and this increase's effect on the size of the accessible Hilbert space.

Borrowing terminology from quantum optics, we decompose a general $N$-mode Gaussian operation into a product of `passive linear optics' operations and `squeezing' operations.
This is known as the Euler decomposition, and takes the form
\be 
{U}_{\bm S}={U}_{\bm K}  U_{ \bm S\left(\{r_k\}\right)} U_{\bm L}.
\label{Euler_decomposition_main}
\ee
The passive linear optics operations preserve photon number, and are described by symplectic orthogonal matrices $\bm K, \bm L \in SpO(N)$.  
Single-mode squeezing operations, which increase/decrease photon number by mixing creation and annihilation operators of each mode $k$, are characterized by their strengths $r_k$ and represented by the diagonal matrix $\bm S(\{r_k\}) = \bigoplus_{k=1}^N \text{Diag}\left( e^{r_k}, e^{-r_k} \right)$.
This squeezing operation multiplies a states' width in the $q_k$-quadratures by $e^{r_k}$, and in the $p_k$-quadratures by $e^{-r_k}$. (See Appendix~\ref{App_Gaussian} for a more in-depth introduction to Gaussian operations.)

\begin{figure}
\centering
\subfigure{
\includegraphics[width=0.23\textwidth,valign=t]{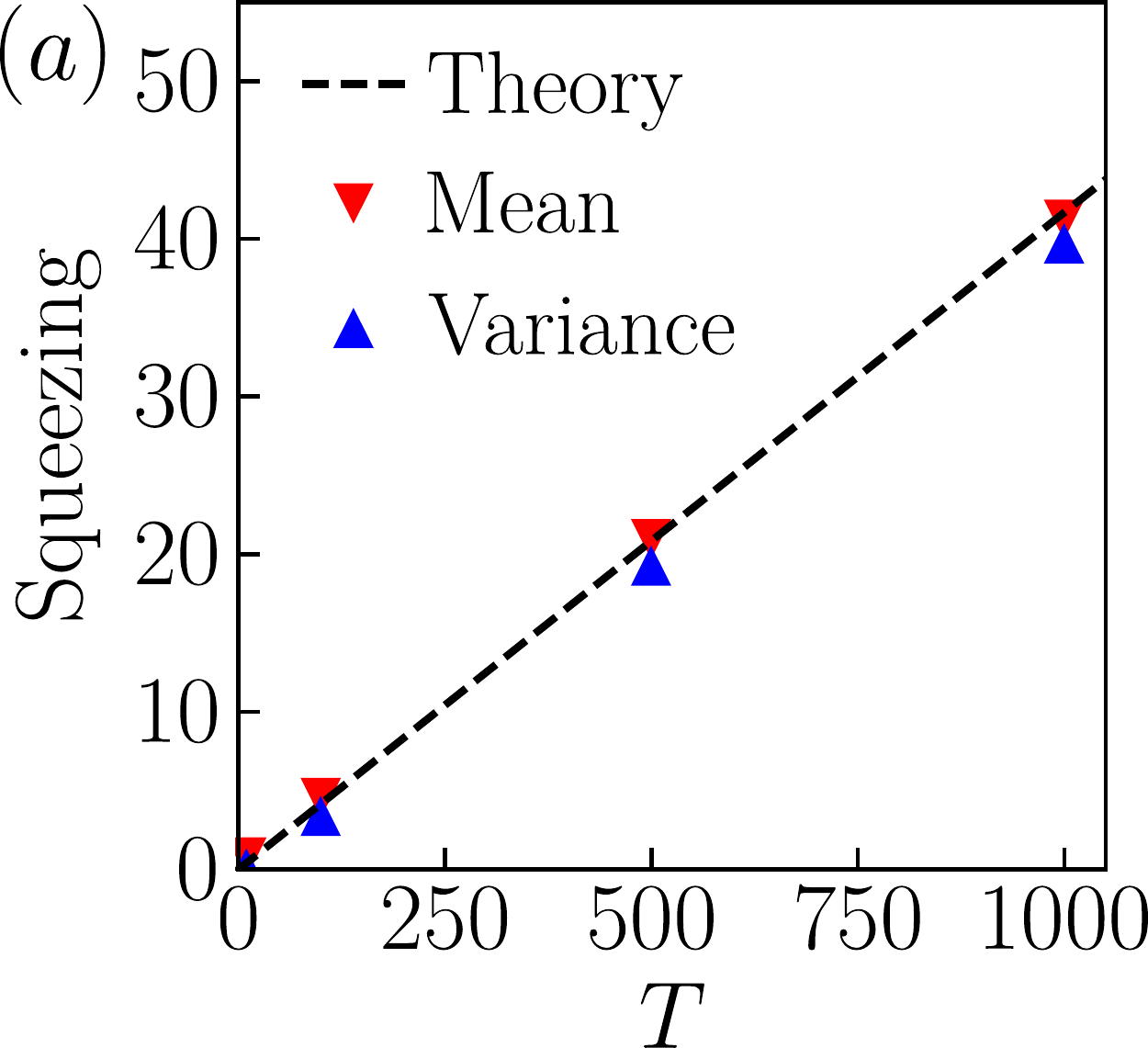}
}\hspace*{-1em}
\subfigure{
\includegraphics[width=0.23\textwidth,valign=t]{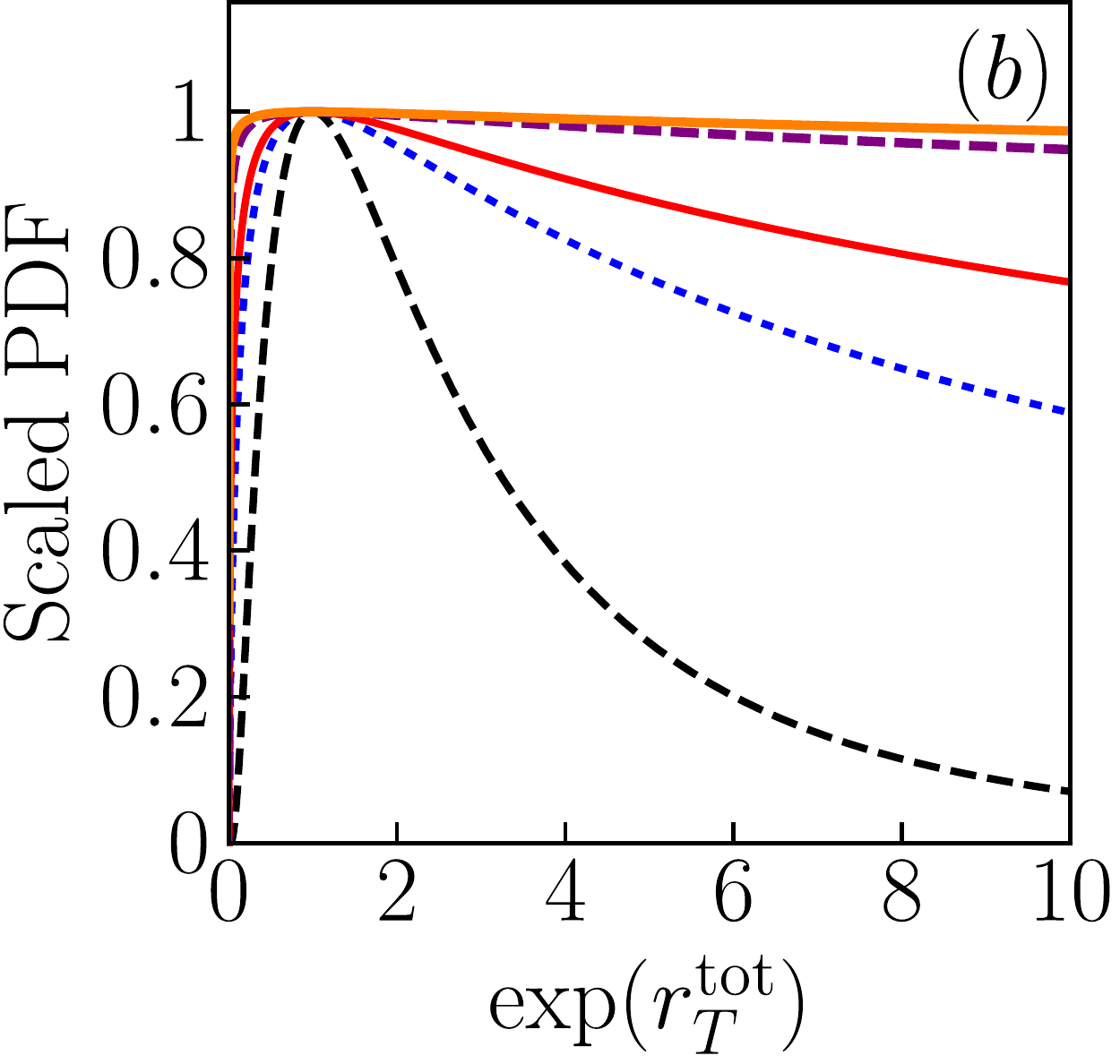}
}
\caption{(a) Increase of the mean and variance of the total squeezing $r^{\text{tot}}_T$ for a single-mode Gaussian circuit, with individual squeezings drawn from a uniform distribution in $[0,1/2]$ (other distributions exhibit similar behavior). Each data point is calculated from 2000 samples. At large $T$, the mean and variance are equal and close to the theoretical prediction $\frac{1}{2}\sum_{t=1}^{T} \braket{r_t^2}=(1/24) T $ (black dashed line).  (b) Scaled probability density function of the log-normal distribution for $e^{r^{\text{tot}}_T}$. The mean/variance ratio equals $1$(black,dashed), $5$ (blue dotted), $10$ (red), $50$ (purple thick), $100$ (orange thick), from bottom to top.
\label{single_mode_Pr}
}
\end{figure}

In our single-mode Gaussian circuit, we consider a sequence $U_1, U_2, \ldots, U_T$ of random Gaussian unitaries. We take the passive linear optics operations $\bm K, \bm L$ of each unitary to be distributed uniformly according to the Haar measure, and the squeezing strengths $r_t$ (at time step $t$) to be small, drawn from some probability distribution $P(r_t)$.
We study the growth of the total squeezing $r^{\text{tot}}_T$ of the compounded unitary $U_{T} U_{T-1} \ldots U_1$.
Averaging over angles, we find that the mean and variance of $r^{\text{tot}}_T$ increase as
\ba
&&\braket{r^{\text{tot}}_{T+1}}=\braket{r^{\text{tot}}_T}+\frac{1}{2}\braket{r_{T+1}^2},
\\
&&{\rm var}\left({r^{\text{tot}}_{T+1}}\right)={\rm var}\left({r^{\text{tot}}_{T}}\right)+\frac{1}{2}\braket{r_{T+1}^2}.
\ea
After many time steps, this approaches a normal distribution $r^{\text{tot}}_{T} \sim  \calN\left(\mu,\sigma^2\right)$ with equal mean and variance, $\mu=\sigma^2=\sum_{t=1}^{T} \braket{r_t^2/2} = T \braket{r_t^2/2}$. We verify this numerically in Fig.~\ref{single_mode_Pr}(a).

We can also track the time-evolution of a displacement operator $D(\alpha(t))$ under the random Gaussian circuit. At large squeezing, the amplitude of a typical evolved displacement is dominated by the displacement's component on the axis amplified by $e^{r^{\text{tot}}_T}$. From our previous results, the factor $e^{r^{\text{tot}}_T}$ obeys a log-normal distribution $\calL\calN\left(T \braket{r_t^2/2},T \braket{r_t^2/2}\right)$, and so the amplitude obeys
\be
|\alpha(t)| \sim \calL\calN\left(\mu=T \braket{r_t^2/2} |\alpha(0)|,\sigma^2 = T \braket{r_t^2/2}|\alpha(0)|^2\right).
\ee
This resembles a constant distribution in the range $|\alpha(0)| e^{-T \braket{r_t^2/2}} < |\alpha(t)| < |\alpha(0)|e^{T \braket{r_t^2/2}}$ [shown in Fig.~\ref{single_mode_Pr}(b)]. The phase space volume $\sim (e^{T \braket{r_t^2/2}})^2$ available to the time-evolved displacement operator increases exponentially in time.

Squeezing's effect on states parallels its effect on displacement operators. Note that, for large squeezing, the number of photons in a state typically increases as $n \rightarrow e^{2r^{\text{tot}}_T} n$ (as seen by writing  $n = a^{\dagger}a \sim q^2 + p^2 \rightarrow e^{2r}q^2 + e^{-2r}p^2 \sim e^{2r} n$). Therefore, over different circuit realizations, the time-evolved state will have some chance to be in any of the $\sim (e^{T \braket{r_t^2/2}})^2 \, n$ states of photon number $\lesssim (e^{T \braket{r_t^2/2}})^2 \, n$; hence our claim that the size of the Hilbert space `accessible' to the system grows exponentially.



\begin{figure}
\includegraphics[width=0.45\textwidth]{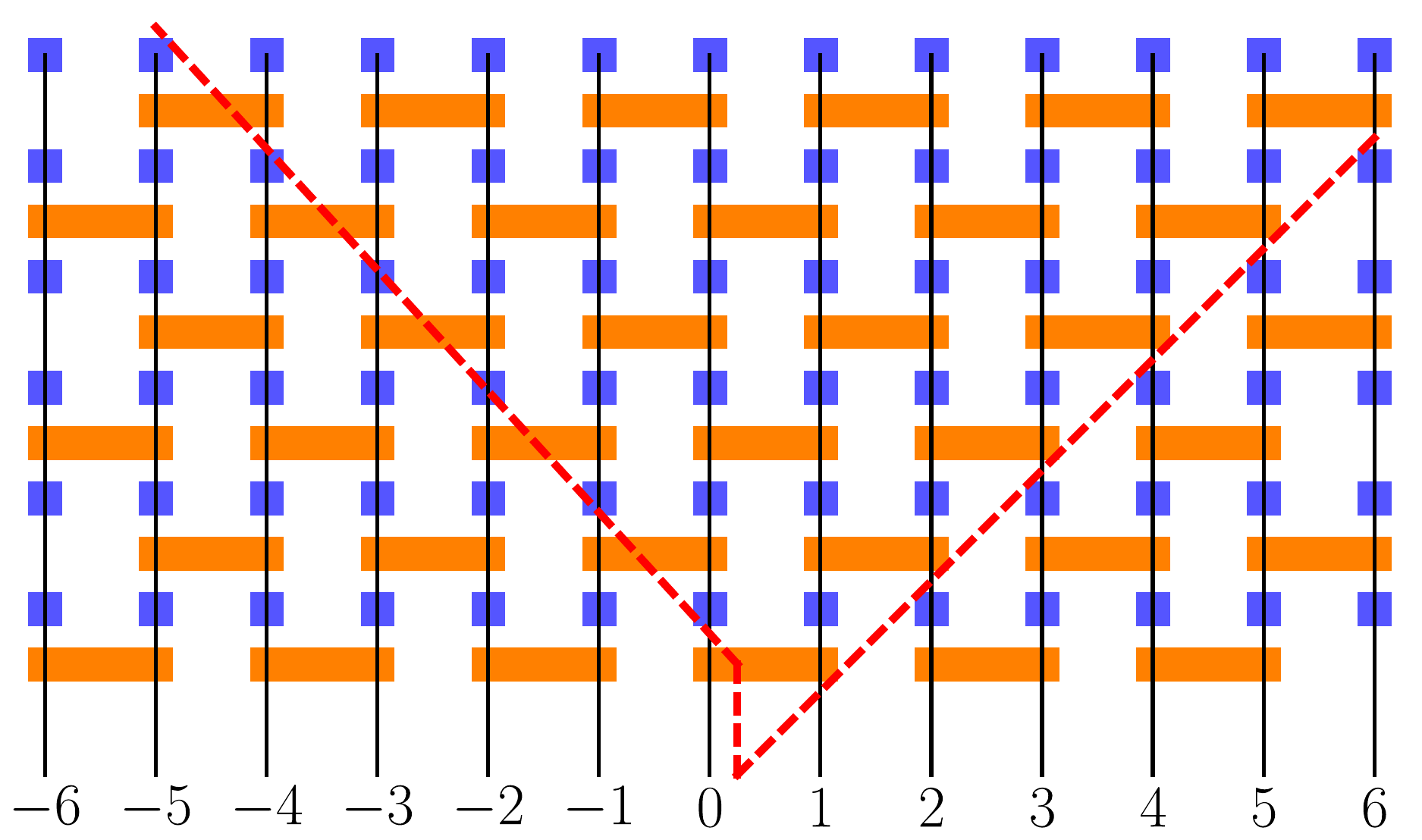}
\caption{Schematic of the local random Gaussian circuit for $13$ modes. The orange rectangles are random two-mode passive linear optics (beamsplitters and phase shifters) and the blue squares are random single-mode squeezers. The red dashed line shows the lightcone of mode $0$.
\label{random_Gaussian}
}
\end{figure}

\subsubsection{Many-modes: ballistic propagation}
\label{Sec_many_mode}

We now turn to the effect of squeezing in many-mode random Gaussian circuits. Again owing to the Euler decomposition of Gaussian unitaries in Eqs.~\eqref{Euler_decomposition_main}, we construct a general many-mode circuit by interleaving layers of single-mode squeezing and layers of multi-mode passive linear optics (i.e. beamsplitters and phase phase-shifters), as shown in Fig.~\ref{random_Gaussian}. To capture the behavior of locally-interacting physical systems, we take the passive linear optics operations to be nearest-neighbor (decomposable into nearest-neighbor beamsplitters and single-mode phase shifting operations). Each of these is described by a two-mode Gaussian unitary $U_{\bm L_{2x,t}}$, where $\bm L_{2x,t}$ is a random symplectic orthogonal matrix. Each time step also includes single-mode squeezing of amplitudes $r_{x,t}$ drawn uniformly from the interval $[0,R]$.


We characterize operator spreading in these circuits using the average OTOC. Specifically, we consider the time-evolution of an initial displacement $D^0\left(\alpha\right)$ (localized on mode $0$) under an ensemble $\mathbb{C}_{R}$ of random Gaussian circuits. This gives rise to an ensemble of displacements $\mathbb{D}^{0}\left(\alpha;t\right)=\{ U^\dagger\left(t\right)D^0\left(\alpha\right) U\left(t\right)|U\left(t\right)\sim \mathbb{C}_{R} \}$.
We measure the support of this ensemble on mode $x$ with the OTOC
\be 
\overline{\calC_2}\left(\mathbb{D}^{0}\left(\alpha;t\right),\mathbb{D}_n^x\right)_\rho=\mathbb{E}_{\bm \xi \sim \mathbb{D}^{0}\left(\alpha;t\right)} \left[ \exp\left(-n\left(\bm \xi_{2x}^2+\bm \xi_{2x+1}^2\right)\right) \right],
\label{OTOC_diff}
\ee
averaged over both $\mathbb{D}^{0}\left(\alpha;t\right)$ and a local displacement ensemble $\mathbb{D}_n^x$ of width $n$. Per Section~\ref{Sec_averageOTOC}, decay of the average OTOC indicates that significant portions of time-evolved displacements are distributed outside the ball of radius $\frac{1}{\sqrt{n}}$ about the identity on mode $x$.

\begin{figure}
\centering
\includegraphics[width=0.5\textwidth]{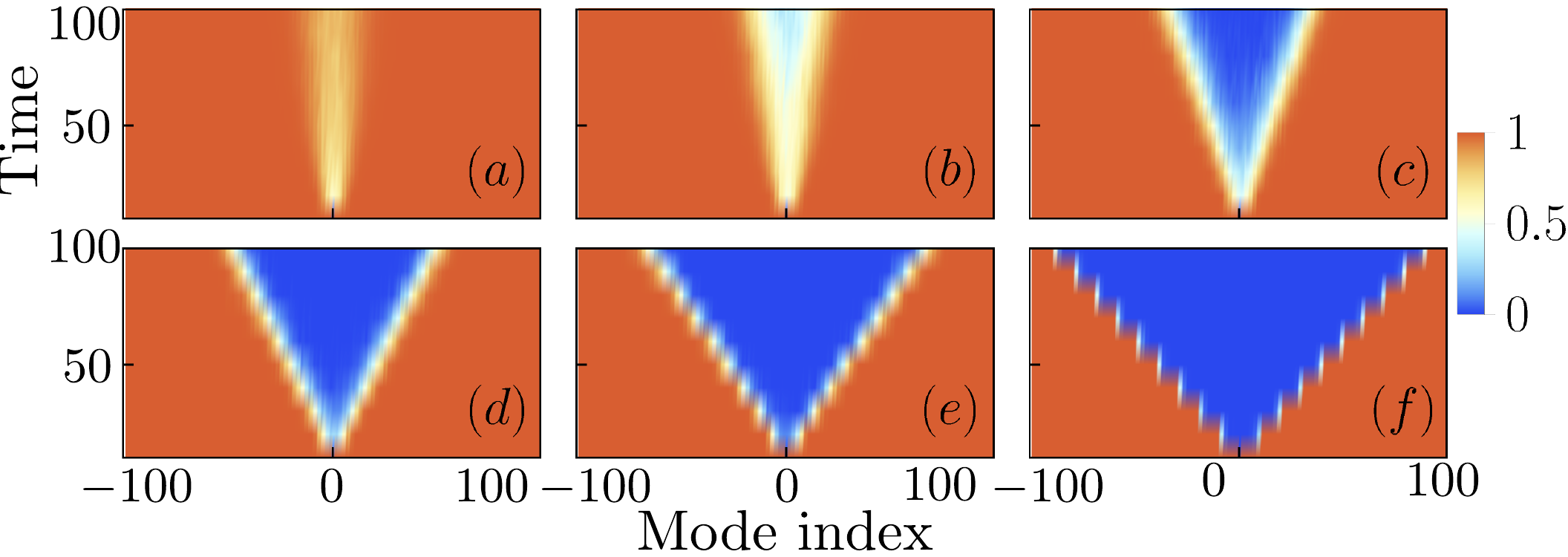}
\caption{
Average OTOC in the random Gaussian circuit, for a thermal density matrix $n_{\text{th}}=5$, as a function of both space (x-axis) and time (in units of 10, y-axis). The butterfly velocity increases to its upper bound of 1 as the squeezing is increased [$R=0,0.2,0.4,0.6,0.9,2$, from (a)-(f)]. Each average is obtained from $100$ samples.
\label{OTO_Brownian_all}
}
\end{figure}

For low squeezing $R$, operator spreading in the random Gaussian circuit can be captured by a simple hydrodynamical equation. The central object of this equation is the amplitude squared of the operator on the mode $x$, $f\left(x,t\right)=\bm \xi_{2x}^2+\bm \xi_{2x+1}^2$. To motivate the hydrodynamical description, note that, in the absence of squeezing, the total amplitude $F(t)=\int dx \, f\left(x,t\right)$ is conserved due to the orthogonality of the matrix $\bm S$ [see Eq.\eqref{Euler_decomposition_main}].  In this regime, random beamsplitters lead to diffusion of $f$. Introducing single-mode squeezing breaks conservation of $F$ and leads, on average, to its exponential growth. Together, these suggest the hydrodynamical equation
\be
\partial_t \overline{f}\left(x,t\right)=D\partial_x^2 \overline{f}\left(x,t\right)+c_R  \overline{f}\left(x,t\right),
\ee
where we denote the average of $f$ over the circuit ensemble as $\overline{f}\left(x,t\right)$, and introduce the diffusion constant $D$ and the growth exponent $c_R$. Solving this equation, we find that an initially local operator $\overline{f}\left(x,0\right)=\delta(x)$ spreads according to
\be
\overline{f}\left(x,t\right)=\frac{1}{\sqrt{4\pi D t}}\exp\left({-\frac{x^2}{4Dt}+c_R t}\right).
\label{diff_solution}
\ee
From Eq.~\eqref{OTOC_diff}, the growth of $\overline{f}$ leads to decay of the OTOC, which becomes sizable when $\overline{f} \sim 1$. This decay spreads ballistically with a wavefront $x_f\left(t\right) = v_B t $, where we define the butterfly velocity
\be
v_B=\sqrt{4D c_R}. \label{eq:Lyapunov_Diffusion}
\ee
Intriguingly, this relation between the butterfly velocity and the diffusion constant closely resembles that found for a coupled SYK chain~\cite{xu2018locality,gu2017local} and weakly interacting diffusive metal~\cite{patel2017quantum}, if one identifies the single-mode growth exponent $c_{R}$ with the Lyapunov exponent.

\begin{figure}
\centering
\includegraphics[width=0.45\textwidth]{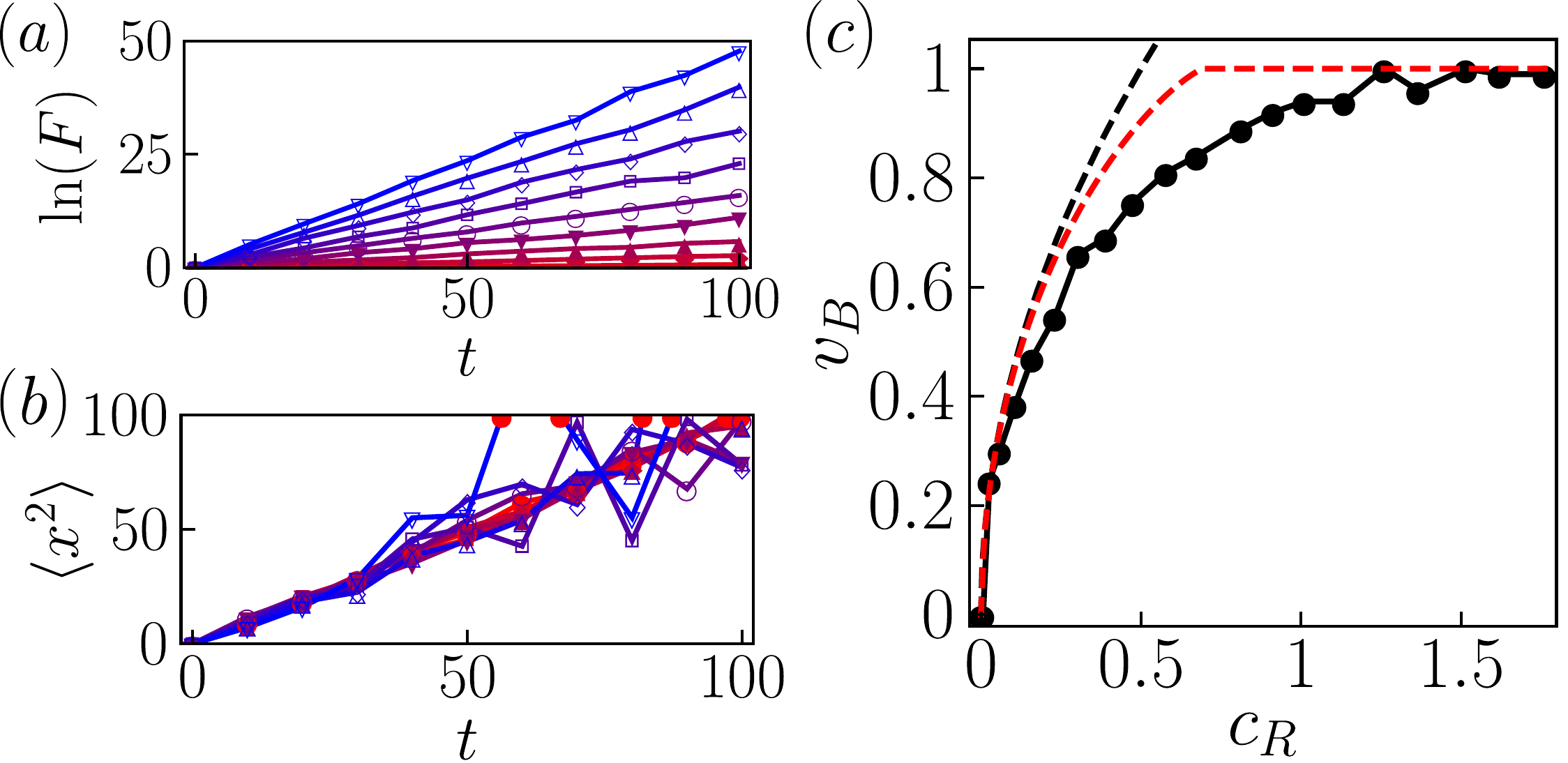}
\caption{Hydrodynamical description of the many-mode random Gaussian circuit. (a) The total operator amplitude $F$ increases exponentially in time, with growth exponent proportional to the squeezing ($R=0,0.1,\cdots, 0.9$, from red on bottom to blue on top). (b) The variance in position $\braket{x^2}$ grows linearly in time, indicating diffusive behavior. (c) These combine to give a ballistic spread of the OTOC decay, with a squeezing-dependent butterfly velocity $v_B$. The black dashed curve is the hydrodynamical prediction $v_B = \sqrt{4Dc_R}$, the red dashed curve is the binomial analysis, and black dots are numerics. Each average is obtained from $100$ samples. 
\label{verification}
}
\end{figure}


We verify our hydrodynamic model numerically on a system of $2L + 1$ modes, indexed by integers from $[-L,L]$. The total operator amplitude $n$ indeed grows exponentially, with a growth exponent  proportional to the squeezing $c_{R}\sim R$, see Fig.~\ref{verification}(a). The variance in position $\braket{x^2}$ increases linearly in time, consistent with diffusive behavior at $D=1/2$. We numerically extract the wavefront $x_f(t)$ by finding the farthest mode with average OTOC $< 0.5$. The wavefront spreads ballistically with a squeezing-dependent butterfly velocity, as shown in Figs.~\ref{OTO_Brownian_all},~\ref{verification}(c). For small $c_R$, this velocity agrees with the hydrodynamical relation Eq.~\eqref{eq:Lyapunov_Diffusion}.

At larger squeezing $c_R \sim 1$, the butterfly velocity saturates to a maximum value of $1$, and our hydrodynamical description does not apply. This maximum velocity is set by the nearest-neighbor coupling of the circuit, and we can capture this saturation by taking this discreteness into account. Note that, averaged over circuit realizations, a beam-splitter between modes $x$ and $x+1$ acts to average the values of $f$ on each mode: $\overline{f}(x,t+1) = \overline{f}(x+1,t+1) = [\overline{f}(x,t) + \overline{f}(x+1,t)]/2$. Under this process, an initially local $f$ will spread as a binomial distribution. Combining with squeezing, we predict
\begin{equation}
\overline{f}(x,t) = \text{Bi}(t,x)\, e^{c_R t} = {t \choose {\frac{x}{2} + \frac{t}{2}}} \frac{1}{2^t} \, e^{c_R t}.
\end{equation}
From this, we can solve for the butterfly velocity using only the approximation $t \gg 1$ [but not the further approximation $x_f(t) \ll t$, which would reproduce the Gaussian of Eq.~\eqref{diff_solution}]. As shown in Fig.~\ref{verification}(c), this indeed more accurately captures the squeezing dependence of the butterfly velocity~\cite{footnote3}.

\begin{figure}
\centering
\includegraphics[width=0.375\textwidth]{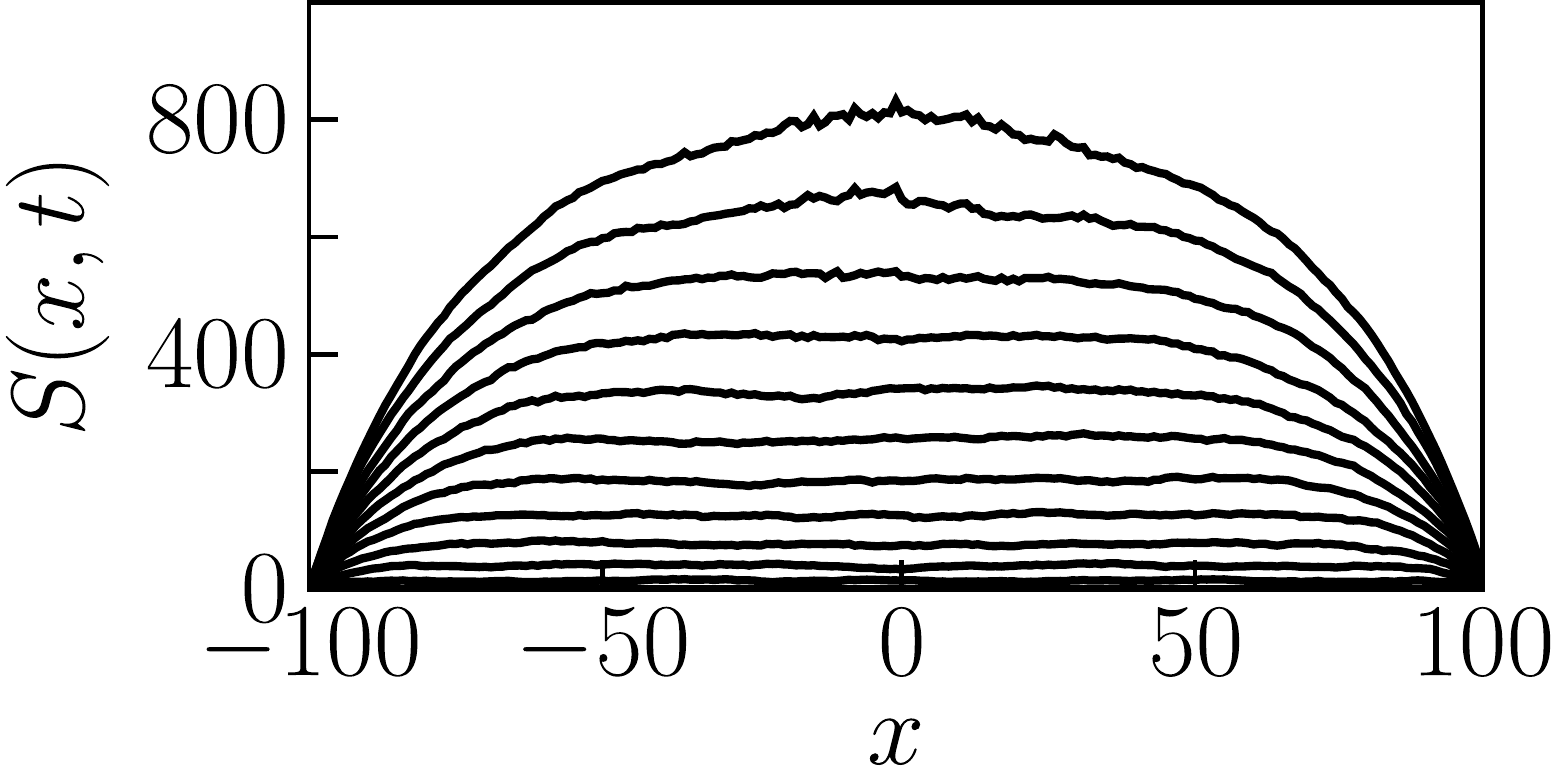}
\caption{The entanglement entropy $S(x,t)$ for a random Gaussian circuit with $L = 201$ modes and squeezing $R=0.2$ ($t=0,80, 160,\cdots,800,880$, from bottom curve to top). The entropy initially increases quadratically in time, then begins to saturate to $S(x,t) \sim t (L -|x|)$.
\label{Sxt}
}
\end{figure}

In addition to operator spreading, entanglement growth is a key diagnostic of scrambling in many-body systems. To study it in our model, we bipartition the system at position $x$ and calculate the entanglement entropy $S\left(x,t\right)$ between the left and right subsystems as a function of time. As shown in Fig.~\ref{Sxt}, for a system initially in the vacuum product state, we find that the average entanglement growth across the center cut $h\left(t\right)=\overline{S\left(0,t\right)}$ is \emph{quadratic} in time $\sim t^2$, in contrast to the linear growth characteristic of DV systems~\cite{nahum2017quantum}. We can understand this in terms of the exponential growth of the accessible Hilbert space found in Section \ref{Sec_single_mode}. In DV systems, the hard cut-off of the local Hilbert space means that qudits near the cut quickly become maximally entangled across the cut; the linear growth $\sim t$ arises from a ballistic `spread' of entanglement with faraway modes. In the CV case, one still receives this ballistic factor of $t$, \emph{in addition} to a factor $t \sim \log(e^{c_R t})$ from the growth of already-entangled modes. In a finite-size system, at some time $T \sim L$ all modes will contribute to entanglement across the cut, and we expect this growth to saturate to a linear behavior $S(x,t) \sim t (L -|x|)$. This is seen in Fig.~\ref{Sxt}, although we are limited in system size and evolution time due to the increased ill-conditioning of the state's covariance matrix  (used for efficient numerical simulation of Gaussian evolution) under squeezing.

\begin{figure}
\centering
\includegraphics[width=0.45\textwidth]{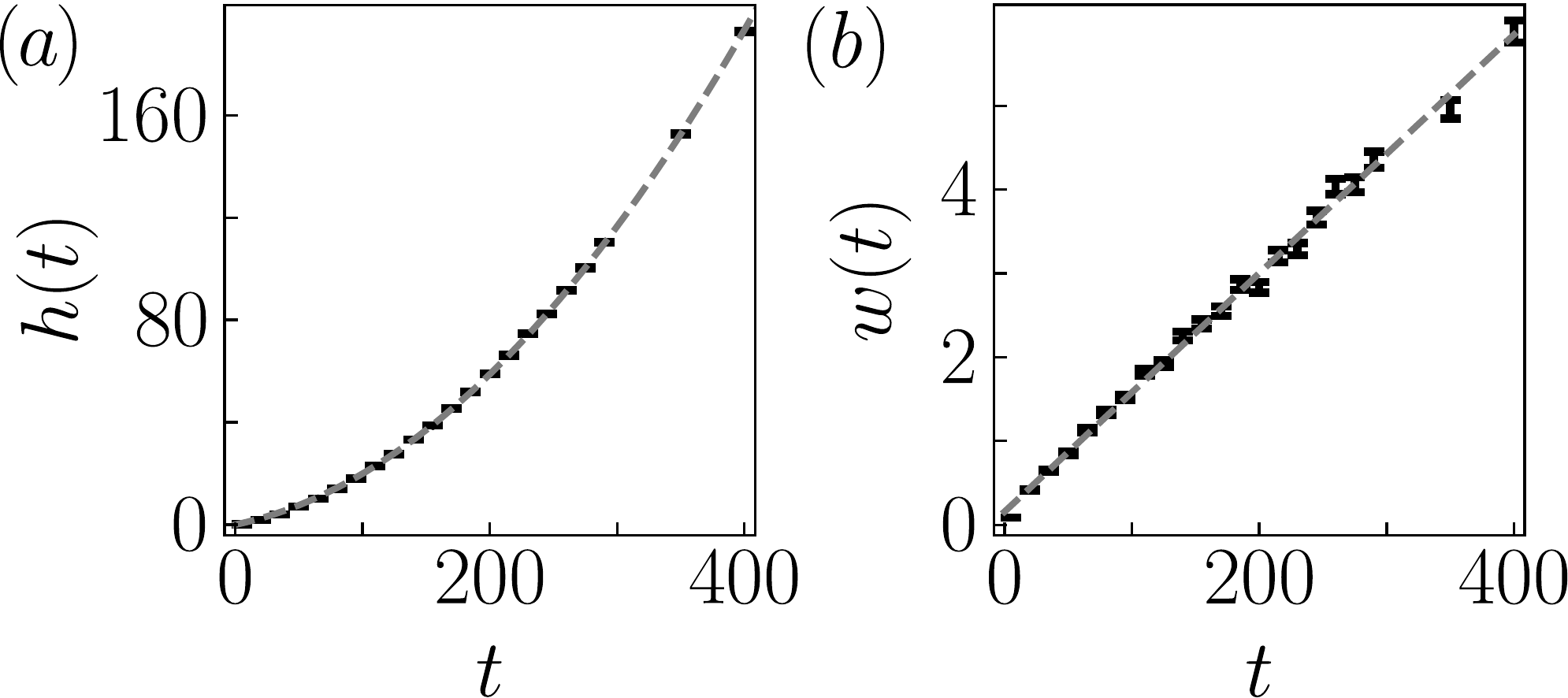}
\caption{(a) Average entanglement entropy $h(t)$ across the center cut (black dots), and a quadratic polynomial fit $0.1 t+0.001 t^2$ (grey line). (b) Fluctuation $w(t)$ of the entanglement entropy (black dots), and a linear polynomial fit $0.014 t$ (grey line). Each point is obtained from 1000 independent samples for $R=0.2$ and $L=400$. Error bars indicate standard-deviations.
\label{hw_fit}
}
\end{figure}

In addition to the average entanglement, we study its fluctuations across circuit realizations, as measured by the standard deviation $w\left(t\right)=\sqrt{\overline{\left(S\left(0,t\right)-h\left(t\right)\right)^2}}$. In DV systems, such fluctuations are predicted to lie in the Kardar-Parisi-Zhang (KPZ) universality class~\cite{nahum2017quantum}, scaling with time as $\sim t^{1/3}$. In contrast,  in Fig.~\ref{hw_fit} we observe fluctuations scaling linearly with time $\sim t$. We suspect that this arises from a dominance of fluctuations in squeezing over KPZ fluctuations, but postpone a full theoretical model to future work.

\subsection{CV unitary designs}
\label{Sec_designs}


In DV systems, the utility of quasi scrambling random circuit models is justified by an assumption that aspects of physical systems can be modeled by local, Haar random unitaries~\cite{Page93, Susskind:2015toa, Almheiri13, hayden2016holographic, Cotler:2017ab, Cotler:2017aa}.
When applicable, this assumption has incredible power --- it can be rigorously shown that averages over Haar random unitaries can be replicated by much simpler unitary ensembles, known as unitary $k$-designs.
By simulating the system using the simpler $k$-designs (or computing the Haar averages explicitly), one encounters a rare opportunity to study quantum chaotic behavior in an analytically and numerically tractable setting~\cite{harrow2009random, Brandao12, nahum2018operator, Keyserlingk:2018aa, Rakovszky:2018aa, Khemani:2018aa, xu2018locality, Bentsen2018}.
%
%
%
%
%

In DV systems, a unitary $k$-design is defined as distribution of unitaries that replicates the $k^{\text{th}}$ moments of the Haar ensemble~\cite{Ambainis07, dankert2009, roy2009unitary},
\begin{align}\label{design_def_1}
\mathbb{E}_\calE \left[f(U,U^{\dagger})\right]=\mathbb{E}_{\rm Haar} \left[f(U,U^{\dagger})\right]
\end{align}
for any polynomial $f(U,U^{\dagger})$ of order $\leq k$ in $U, U^{\dagger}$.
Designs represent a hierarchy of increasingly random behavior, which has been tied to the complexity of unitary ensembles~\cite{roberts2017chaos}.
They can also inform our understanding of scrambling in systems with a notion of locality: local randomization of the system, as diagnosed by TOCs, is captured by 1-designs, while entanglement generation and operator delocalization are captured by 2-designs~\cite{roberts2017chaos}.
In DV systems, it is well-known that the set of all Pauli operators form a 1-design, and Clifford operators a 2-design.

The extension of unitary designs to CV systems is initially unclear, as there is no Haar ensemble in an infinite-dimensional system.
Despite this, we notice that \emph{averages} over Haar unitaries can remain well-defined in the CV limit.
To see this, note that the Haar expectation of $f(U,U^{\dagger})$ can be computed power by power, which leads to an equivalent definition of a unitary $k$-design:
\begin{align}\label{design_def_2}
\mathbb{E}_\calE \left[ U^{\otimes k}\otimes {(U^\dagger)}^{\otimes k}\right]=\mathbb{E}_{\rm Haar} \left[ U^{\otimes k}\otimes {(U^\dagger)}^{\otimes k}\right],
\end{align}
where the operator $U^{\otimes k}\otimes {(U^\dagger)}^{\otimes k}$ acts on $2k$-copies of the original system.
This reformulation is convenient because the RHS can be computed explicitly; it is found to be a sum of permutation operators between the $2k$ system copies, with coefficients that depend on the dimension $d$~\cite{roberts2017chaos}.
We use this to \emph{define} a CV $k$-design as an ensemble which satisfies Eqs.~\eqref{design_def_2} in the limit $d \rightarrow \infty$, keeping only lowest order terms in $1/d$.  

In Sections \ref{Sec_1design}, we use this definition to show that displacement operators form a CV 1-design.
Intriguingly, previous work has indicated that Gaussian \emph{states} cannot form a CV state 2-design~\cite{blume2014curious}; in Section \ref{Sec_2design}, we show an analogous result for Gaussian unitaries, arising from the need to regulate, or cut-off, CV unitary ensembles.
%
%
For 1-designs, regularization naturally leads to ensembles that replicate `energy-constrained' random behavior.
For 2-designs, regularization relates to the squeezing of the Gaussian unitaries, and we find that scrambling by Gaussian unitaries \emph{necessarily} increases the `energy' (defined roughly, by the choice of regularization) of the system.
Intriguingly, large-squeezing Gaussian unitaries do exhibit some similar behavior to 2-designs in DV systems. 
%
%

To verify CV designs, in Section~\ref{Sec_FP} we adapt a DV quantity which measures closeness to Haar randomness, the \emph{frame potential}, to CV unitary ensembles.
%
%
We find that the DV finite-temperature frame potential~\cite{roberts2017chaos} has no nontrivial lower bound for CV systems, owing to the potentially infinite size of CV ensembles. 
To remedy this, we define a `twice-regulated' frame potential, which only receives contributions from unitaries that approximately preserve a `low-energy' subspace.
%

Although higher designs remain largely unknown in both DV~\cite{zhu2017multiqubit,webb2015clifford,Kueng15} and CV systems, we note that our differentiation between quasi and genuine scrambling matches that between 2-designs and higher designs.
Specifically, the volume of operators, measured by either the fourth power of the TOC or the OTOC squared, consists of four copies each of $U$ and $U^{\dagger}$.
Its `Haar average', corresponding to a large phase space volume of unitary time-evolved operators, is therefore replicated by $(k\geq4)$-designs.

Before proceeding, we contrast our work with previous results on unitary designs in CV systems. Ref.~\cite{sanders2002qubits} proposes to construct a CV 1-design by using the number and phase operators as generators; however, as they point out, their construction relies on a finite-dimension Hilbert space cut-off. Another proposal, Ref.~\cite{gross2007evenly}, defines Haar randomness and $k$-designs in CV systems using the isomorphism between unitary matrices $U(N)$ and orthogonal symplectic matrices $SpO(2N)$, the latter describing operations in passive linear optics. This approach only encompasses Gaussian operations without squeezing, and is not suited to capture chaotic behavior in interacting CV systems.

\subsubsection{CV 1-design}\label{Sec_1design}


We begin with a brief review of 1-designs in DV systems. For $k=1$, Eqs.~\eqref{design_def_2} becomes
\begin{align}
\mathbb{E}_\calE \left[ U\otimes U^\dagger\right] & = \frac{1}{d} \, S_{\leftrightarrow}
\label{1d_1}
\end{align}
where the swap operator $S_{\leftrightarrow}$ interchanges the copies of the system acted on by $U$ and $U^{\dagger}$. By rearranging indices, one can also define 1-designs by any of the equivalent conditions:
\begin{align}
\mathbb{E}_\calE \left[ U \, A \, U^\dagger\right] & = \frac{1}{d} \, \tr\left(A\right){\bm I},
\label{1d_2}
\\
\mathbb{E}_\calE \left[ \, \tr(U^{\dagger} \, A) \, U\right] & = \frac{1}{d} \, A,
\label{1d_3}
\\
\mathbb{E}_\calE \left[ U\otimes U^*\right] & =  \ket{\rm EPR}\bra{\rm EPR},
\label{1d_4}
\end{align}
where $A$ is an arbitrary operator on the system, and $\ket{\rm EPR} \equiv \frac{1}{\sqrt{d}} \sum_i \ket{i} \ket{i}^*$, with $\{ \ket{i} \}$ a complete basis of states, is an EPR pair between two copies of the system.
Moreover, as seen from Eqs.~\eqref{1d_3}, all four definitions are equivalent to requiring that $\calE$ forms a complete operator basis.
Pauli operators, as a complete operator basis, thus form a DV 1-design.

The equivalence of 1-designs and complete operator bases immediately suggests a uniform ensemble of displacement operators as a candidate CV 1-design.
Indeed, displacement operators satisfy similar relations
\begin{align}
\left( \frac{1}{\pi} \right)^N\int d^{2N}\bm \xi  D\left(\bm \xi\right)\otimes D^{\dagger}\left(\bm \xi\right)&=S_{\leftrightarrow}
\\
\left( \frac{1}{\pi} \right)^N \int d^{2N}\bm \xi \, D\left(\bm \xi\right) \, A \, D^{\dagger}\left(\bm \xi\right)&=\tr\left(A\right){\bm I}.
\end{align}
However, this ensemble is not normalized, due to the infinite volume of phase space.
This compensates for the factors of $\frac{1}{d}$ in Eqs.~(\ref{1d_1}-\ref{1d_3}), which go to zero in the CV limit.
Regularizing this ensemble, in addition to being convenient, is also physically motivated.
For example, the dynamics of a Hamiltonian system will be constrained by energy, and we shouldn't expect an unbounded ensemble of displacements to mimic typical scrambling behavior.
To regularize this, we consider the Gaussian ensemble
$
\mathbb{D}_n
$ of displacements, defined in Eq.~\eqref{Dn_def}. 
Gaussians are natural due to their stability under addition, which suggests that sequential applications of random displacements will asymptotically approach a Gaussian distribution. 
The ensemble $\mathbb{D}_n$ satisfies
\begin{align}
\lim_{n\to\infty} \! n^N \!\!\! \int d^{2N}\bm \xi \, P_D^G\left(\bm \xi;n \right) D\left(\bm \xi\right) \, A \, D^{\dagger}\left(\bm \xi\right)&=\tr\left(A\right){\bm I}
\label{Gaussian displacement 1}
\\
\lim_{n\to\infty}  \! n^N \!\!\! \int d^{2N}\bm \xi \, P_D^G\left(\bm \xi;n \right) D\left(\bm \xi\right)\otimes D^{\dagger}\left(\bm \xi\right)&=S_{\leftrightarrow}
\label{Gaussian displacement 2}
\\
\lim_{n\to\infty} \! \int d^{2N}\bm \xi \, P_D^G \left(\bm \xi;n\right) D\left(\bm \xi\right)\otimes D^*\left(\bm \xi\right) &= \ket{\rm EPR}\bra{\rm EPR},
\label{Gaussian displacement 3}
\end{align}
where the operator $A$ has a well-defined characteristic function, the conjugate $D^*\left(\bm \xi\right)$ flips the signs of the displacement's momentum quadratures~\cite{footnote4}, and the CV EPR pair is defined as $\ket{\rm EPR} = \lim_{\beta\to\infty} \sum_n e^{-\beta n} \ket{n} \ket{n}/\calN_\beta$ ($\calN_\beta$ chosen for normalization)~\cite{ban2002continuous}.


Intriguingly, at finite $n$, analogs of Eqs.(\ref{Gaussian displacement 1}-\ref{Gaussian displacement 3}) still hold for \emph{states with mean photon number} $n_{\text{th}} < n$.
For example, $n^N \int d^{2N}\bm \xi \, P_D^G\left(\bm \xi;n \right) D\left(\bm \xi\right)\otimes D^{\dagger}\left(\bm \xi\right)$ will act as the swap operator on the subspace of $\lesssim n$ photons, but not for higher photon number (we verify this in Section \ref{Sec_FP} and Appendix~\ref{App_FP}).
This can be understood intuitively. 
As we have seen, a thermal density matrix $\tilde{\rho}_{n_{\text{th}}}$ can resolve distances $1/\sqrt{n_{\text{th}}}$ in phase space. 
To swap such states, the ensemble needs to have nontrivial commutation $\sim e^{i \bm \xi^T \bm \Omega \bm \xi_0}$ [see, e.g. Eq. \eqref{displacement addition}] with displacements of this minimum distance $|\bm \xi_0| \sim 1/\sqrt{n_{\text{th}}}$. 
This occurs when $|\bm \xi| \gtrsim 1/|\bm \xi_0|$, or $\sqrt{n} \gtrsim \sqrt{n_{\text{th}}}$.
We speculate that these regulated designs may arise naturally when approximating scrambling quantum dynamics.
For example, time-evolution under a static Hamiltonian will generally be energetically restricted to some subspace of the total Hilbert space.
Approximating scrambling behavior in such systems would require operators which respect this subspace.

The regulated CV 1-design arises naturally in physical contexts. We briefly discuss two such examples. In CV state tomography~\cite{d2007homodyne,lvovsky2009continuous}, one aims to estimate the density matrix $\rho$ of an unknown quantum state through its characteristic function $\chi\left(\bm \xi; \rho\right)\equiv \tr \left[\rho D\left({\bm \xi}\right)\right]$. In reality, one can only perform a finite number of measurements, and can therefore only estimate $\chi\left(\bm \xi; \rho\right)$ in a certain region of phase space. Conventionally, one chooses to sample $\bm \xi$ according to a Gaussian distribution, obtaining the reconstruction
$\rho^\prime \sim \int d^{2N} {\bm \xi} \ P_D^G\left(\bm \xi; n\right)  \tr \left[\rho D\left({\bm \xi}\right)\right] D\left(-{\bm \xi}\right),
$
which suffers from Gaussian additive noise of strength $1/n$~\cite{GiovannettiV2014}. 
A second application of CV 1-designs concerns designs for quantum \emph{states}. In DV systems, a state 1-design is obtained by applying a unitary 1-design to a computational basis state. For CV systems, an analogous procedure gives a Gaussian distributed ensemble of coherent states. This ensemble has important applications in quantum information processing. For instance, it can be used as a basis of encoding states to achieve the classical capacity of one-mode bosonic Gaussian channels~\cite{GiovannettiV2014}, or in CV quantum key distribution protocols~\cite{grosshans2002continuous}.

\subsubsection{CV 2-design}\label{Sec_2design}

In DV systems, the set of Clifford unitaries forms a 2-design~\cite{roberts2017chaos}. 
They obey a defining equation analogous to Eq. \eqref{1d_1},
\begin{align}\label{DV2design}
&\mathbb{E}_\calE \left[ (U\otimes U) \, A \, (U^\dagger \otimes U^\dagger)\right] 
\nonumber
= \frac{1}{d^2-1} \, \left[ {\bm I} \tr\left(A\right) +
\right.
\\
&
\left. S_{\leftrightarrow} \tr\left(S_{\leftrightarrow} A\right) - \frac{1}{d} {\bm I} \tr\left(S_{\leftrightarrow} A\right) - \frac{1}{d} {\bm I} \tr\left(S_{\leftrightarrow} A\right) \right].
\end{align} 
It is insightful to observe the action of this quantum channel on Pauli matrices $A = P_1 \otimes P_2$:
\begin{align}\label{DVP2P}
&\mathbb{E}_\calE \left[ (U\otimes U) \, P_1 \otimes P_2 \, (U^\dagger \otimes U^\dagger)\right] 
\nonumber
\\
&=\left\{ \begin{array}{ll}
         {\bm I} \otimes {\bm I} & \mbox{if $P_1 = P_2 = {\bm I}$};\\
        \frac{1}{d^2-1}\sum_{P \neq {\bm I}} P \otimes P^{\dagger}  & \mbox{if $P_1 = P_2 \neq {\bm I}$};\\
        0&\mbox{if $P_1 \neq P_2$.} 
        \end{array} \right. 
\end{align}     
Intuitively, a random Clifford unitary transforms a non-identity Pauli matrix to any other non-identity Pauli with equal probability (along with the constraint $P_1 = P_2$).


In the CV case, we take the $d \rightarrow \infty$ limit and keep the leading order of Eq. \eqref{DV2design} as the definition of a 2-design~\cite{footnote5},
\begin{align}\label{CV2d1}
&\mathbb{E}_\calE \left[ (U\otimes U) \, A \, (U^\dagger \otimes U^\dagger)\right]
\nonumber
\\
& \propto \, \frac{1}{d^2} \left[ {\bm I} \tr\left(A\right) + S_{\leftrightarrow} \tr\left(S_{\leftrightarrow} A\right) \right].
\end{align}
%
%
%
We begin by exploring a particular ensemble $\calE_r$ of Gaussian unitaries that comes close to satisfying this definition.
%
%
To define this ensemble, it is helpful to decompose a given Gaussian into the product of a quadratic operation and a displacement,  $U = D(\bm d/2) U_{\bm S}$, and Euler decompose the former as $\bm S = \bm K \bm S(\{r_i\}) \bm L$, with single-mode squeezings of strength $r_i$ (see Section \ref{Sec_single_mode}). 
%
%
When acted on a displacement operator $A = D(\b{\xi}_1) \otimes D(\b{\xi}_2)$, Eq.~(\ref{CV2d1}) leads to a sum over a distribution of transformed displacements $D(\b{S} \b{\xi}_1) \otimes D(\b{S} \b{\xi}_2)$, similar to Eqs.~(\ref{DV2design}-\ref{DVP2P}).
A Gaussian distribution over $\b{d}$ gives this distribution a factor of $\delta(\b{\xi}_1-\b{\xi}_2)$, analogous to the DV requirement $P_1 = P_2$.
The distribution is rotationally symmetric if $\b{K}, \b{L}$ are Haar distributed.
Finally, we take the squeezings $r_i = r$ to be large.
In the many-mode limit, one can show (via the Central Limit theorem) that this gives displacements approximately Gaussian distributed, with a width $e^{r} |\b{\xi}_1|$ proportional to the initial displacement $\b{\xi_1}$.
Concretely, we find
\begin{align}\label{CVD2D}
&\mathbb{E}_{\calE_r} \left[ (U^{\otimes 2}) \, D(\b{\xi}_1) \otimes D(\b{\xi}_2) \, (U^{\dagger \otimes 2})\right]
\nonumber
\\
&\approx\delta({\bm \xi}_1+{\bm \xi}_2)
\int d^{2N}{\bm \xi} \,
P_D^G\left(\bm \xi; e^r |\bm \xi_1| \right)
\left[
D\left({\bm \xi} \right) \otimes D^{\dagger}\left({\bm \xi} \right)
\right],
\end{align}
 a CV analog of Eq.~(\ref{DVP2P}).

The relation Eq.~(\ref{CVD2D}) is close to the 2-design definition Eq.~(\ref{CV2d1}).
When $\b{\xi}_1 = 0$, both have a RHS proportional to the identity.
When $\b{\xi}_1 \neq 0$, the RHS of Eq.~(\ref{CVD2D}) is proportional to a SWAP operator on subspaces with less than $\sim e^r |\bm \xi_1|$ photons (see our discussion in Section \ref{Sec_1design}).
However, the constant of proportionality for the latter is off by a displacement-dependent factor of $1/(e^r |\b{\xi}_1|)^N$.
This arises from the fact that, wherever a displacement $D(\b{\xi})$ is transformed to, the displacement $D(c \, \b{\xi})$ is transformed to a displacement $c$ times as large (because Gaussian unitaries act linearly on $\b{\xi}$).
The width of the transformed displacements' distribution must then be proportional to $|\b{\xi}_1|$, leading to the prefactor $1/|\b{\xi}_1|^N$.
This prefactor is similar to that found in Ref.~\cite{blume2014curious}, when considering an ensemble of single-mode Gaussian states.

It is unclear how physically fundamental this discrepancy is.
Arguing for its physicality is that it seems as if it would arise in any soft regularization of CV systems.
For instance, when regularized by the thermal density matrix $\tilde{\rho}_{n_{\text{th}}}$, information about the state is contained in displacements as small (`low-energy') as $\b{\xi} \sim 1/\sqrt{n_{\text{th}}}$ and as large (`high-energy') as $\b{\xi} \sim \sqrt{n_{\text{th}}}$.
If the lowest-energy operator is evolved to a high-energy operator by some elements of the 2-design, then the product of many low-energy operators (which form a larger displacement, i.e. a higher energy operator) must be taken to an operator of even higher energy, outside of the subspace defined by $\tilde{\rho}_{n_{\text{th}}}$.
This seems to suggest that it is necessary for a 2-design to increase the energy/dimension of the effective CV Hilbert space, leading to the `energy-dependent'  prefactor $1/|\b{\xi}_1|^N$.
On the other hand, this restriction may be unique to the linear action of Gaussian unitaries on displacements, and it is an open question whether an ensemble of non-Gaussian unitaries can satisfy Eq.~(\ref{CV2d1}).

\begin{figure}
\centering
\includegraphics[width=0.275\textwidth]{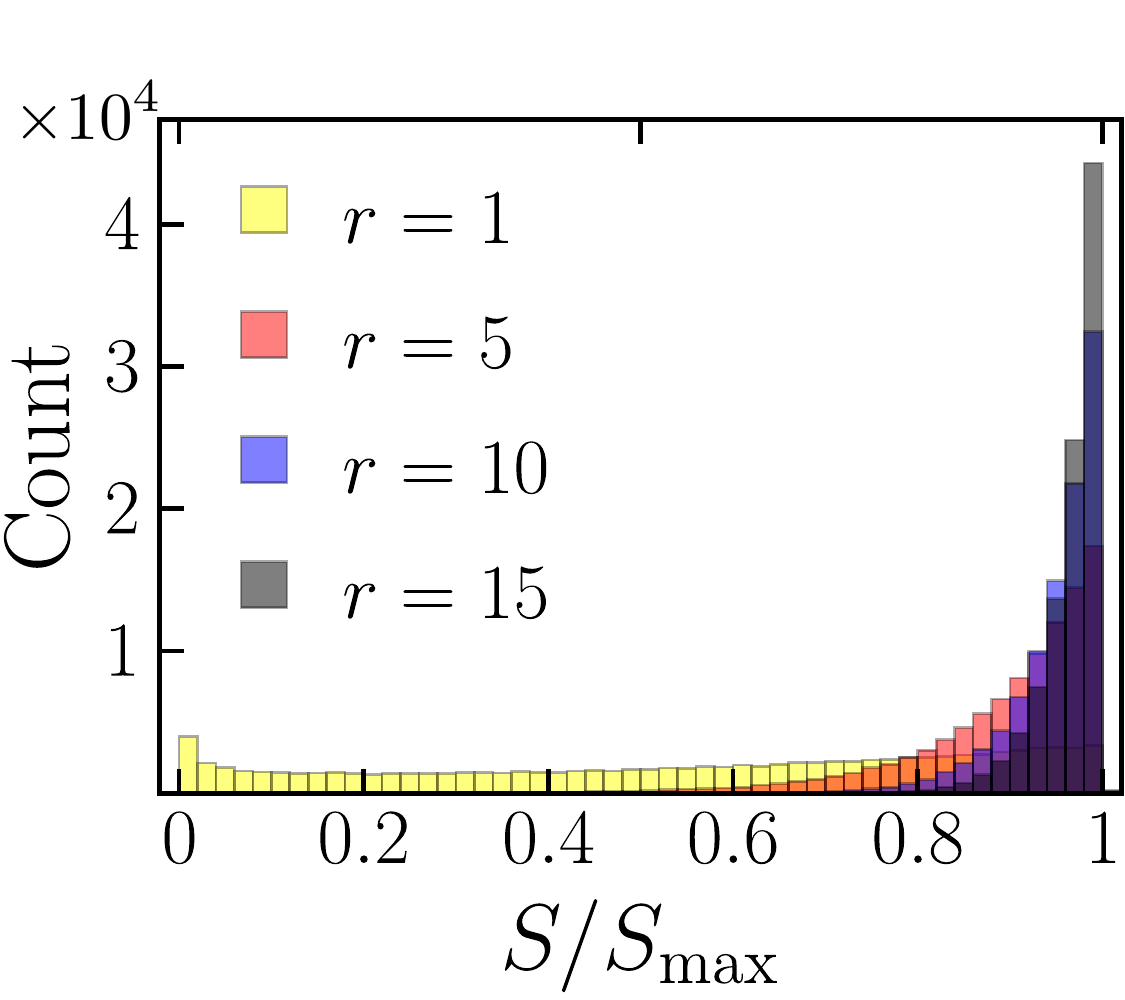}
\caption{Distribution of entanglement entropy $S$ after applying random unitaries drawn from $\calE_r$ on the two-mode vacuum state, for various squeezing strengths $r$, scaled by its (squeezing-dependent) maximum $S_{\text{max}}$, which equals the entropy of a thermal state with mean occupation number $\cosh(2r)-1$. Each distribution contains $10^4$ sampled unitaries. Light to dark color corresponds to increasing squeezing: $r=1$(yellow), $r=5$(red), $r=10$(blue), $r=15$(black).
\label{entanglement_entropy}
}
\end{figure}

Nevertheless, 2-designs are interesting in many-body physics because they can model physical processes, and the ensemble $\calE_r$ \emph{does} possess these more qualitative properties.
%
%
For instance, a key characteristic of a 2-design is the ability to generate entanglement.
In DV systems, a typical Clifford unitary applied to a product state will generate near maximal entanglement between any two subsystems. 
In CV systems, we numerically investigate the entanglement generated by the ensemble $\calE_r$ by applying a randomly sampled Gaussian on an initially unentangled two-mode system in a pure state~\cite{footnote6}.
We find that the entanglement generated grows linearly with the squeezing $r$ and concentrates around the maximum value, as shown in Fig.~\ref{entanglement_entropy}. 
This is expected, as entanglement is proportional to the logarithm of the number of states accessible to each subsystem, which increases $\sim e^{2r}$ under squeezing.

We also find that time-evolution in the many-mode random Gaussian circuit in Section \ref{Sec_many_mode} converges to $\calE_r$ (with asymptotically increasing squeezing) at long times, demonstrating that $\calE_r$ can capture operator growth and OTOC decay as observed in those circuits. 
The displacement amplitudes can be seen to be Gaussian distributed locally due to the beamsplitters, with typical width $e^{c_Rt}/\sqrt{2L+1}$ as $t \rightarrow \infty$.
The single-mode circuit in Section \ref{Sec_single_mode} does \emph{not} asymptotically converge to $\calE_r$, although this is less surprising since our arguments for $\calE_r$ were valid only in the many-mode limit. 
Instead, we observe that a displacement $D(\alpha(t))$ is time-evolved such that its \emph{amplitude} obeys a uniform distribution, which gives a distribution $P(\alpha) \sim 1/|\alpha|$ for the displacement.




\subsubsection{Finite temperature frame potential}\label{Sec_FP}
 
In DV systems, the closeness of a unitary ensemble to a $k$-design is measured by the $k^{\text{th}}$ frame potential~\cite{roberts2017chaos},
\be\label{DV frame potential k}
\calF_\calE \equiv
\mathbb{E}_{U, V\sim \calE}
\left\{ \big|\tr\left( U^\dagger V\right)\big|^{2k}\right\}.
\ee
The frame potential is 1 for a trivial ensemble, and decays to a minimum $k!/d^{2k}$ when $\calE$ is a $k$-design.
%
%
A `finite-temperature' generalization of the frame potential also exists~\cite{roberts2017chaos}, which takes the form
\be
\calF_\calE^{(k)}\left(\rho\right)=
\mathbb{E}_{U, V\in \calE}
\left\{ |\tr\left((\rho)^{\frac{1}{k}} U^\dagger V\right)|^{2k}\right\},
\label{frame_simp}
\ee
and decays from $1$ to a $\rho$-dependent constant $\sim 1/d^{2k}$.

We begin our discussion of CV frame potentials by demonstrating the obstacles encountered when applying DV frame potentials to CV unitary ensembles.
A naive application of Eq.~\eqref{DV frame potential k} to CV unitaries, say displacements $U = D(\bm \xi_U), V = D(\bm \xi_V)$, is ill-defined due to traces diverging like $\sim [\delta(\bm \xi_U - \bm \xi_V)]^{2k}$.
When discussing operator volumes in Section \ref{Sec_volume_FP}, we regulated these divergences with a density matrix $\rho$, via the finite-temperature frame potential Eq.~(\ref{frame_simp}). 
%
%
Unfortunately, since the frame potential is only lower-bounded by $0$ in the CV case, it is not clear how to use its decay to make sharp statements about an ensemble's validity as a $k$-design.
For instance, for our candidate 1-design $\mathbb{D}_n$ we find
\begin{align}
&\calF_{\mathbb{D}_n}^{(k)}\left(\tilde{\rho}_{n_{\text{th}}}\right)
\sim\left(1/k^2 nn_{\text{th}}\right)^N,
\end{align}
which indeed approaches $0$ as $n \rightarrow \infty$.
However, ensembles which are \emph{not} 1-designs lead to similar decay.
Consider two such ensembles: the sphere of displacements with fixed amplitude $\tilde{\mathbb{D}}_n=\left\{\otimes_{k=1}^N D^k\left( \sqrt{n}e^{i\theta_k}\right)|\theta_k\sim [0,2\pi]\right\}$, and the ensemble of phase shifts $\mathbb{R}=\left\{\otimes_{j=1}^N U_{\bm R\left(\theta_j\right)}| \theta_j\sim [0,2\pi]\right\}$. We find
\begin{align}
&\calF_{\tilde{\mathbb{D}}_n}^{(k)}\left(\tilde{\rho}_{n_{\text{th}}}\right)
\sim \left(1/k\sqrt{ nn_{\text{th}}}\right)^N
\\
&\calF_\mathbb{R}^{(k)}\left(\tilde{\rho}_{n_{\text{th}}}\right)\sim  \left({1}/{k n_{\text{th}}}\right)^N.
\end{align}
In particular, the frame potential for $\tilde{\mathbb{D}}_n$ decays to $0$ as $n \rightarrow \infty$, just as it does for the 1-design $\mathbb{D}_n$.
For $k=1$, this can be understood through Section \ref{Sec_volume_FP}.
The finite-temperature frame potential measures the ensemble's inverse volume in phase space --- it will decay to $0$ as long as this volume increases to infinity, regardless of whether the ensemble approaches a 1-design. 
This suggests that we may be asking the wrong question: to be a 1-design on the entire Hilbert space the ensemble must have infinite phase space volume, and accurately capturing this volume will necessarily be difficult.




Motivated by our regularization of 1- and 2-designs, we instead seek to characterize whether a unitary ensemble can form a design \emph{on a particular, `low-energy' subspace}, defined by some density matrix $\rho$.
To do so, we introduce a new `twice-regulated' frame potential that weights unitaries based on their preservation of this subspace.
This takes the form (see Appendix~\ref{App_FP}),
\begin{align}
\calJ_\calE^{(k)}\left(\rho\right) &= \frac{\int_{\mathcal{E}} dU \int_{\mathcal{E}} dV \, \big| \Tr{ \, U^{\dagger} \, \rho^{\frac{1}{2k}} \, V \rho^{\frac{1}{2k}}} \big|^{2k}}{\bigg[ k! \int_{\mathcal{E}} dU \, \big| \Tr{ \,U^{\dagger}\, \rho^{\frac{1}{k}}  \,U\, \rho^{\frac{1}{k}}}\, \big|^k \bigg]^2} 
\nonumber
\\
&
\geq 
\frac{1}{\mathcal{H}^{(k)}(\{\rho_i\})}.
\label{frame_potential_new}
\end{align}
It has no upper bound, but decays to a strict nonzero lower bound determined by the entropies of $\rho$, $\mathcal{H}^{(k)}(\{\rho_i\}) \approx  k! \, \Tr{ \rho^{\frac{1}{k}} }^{2k}$.
As we saw in Section~\ref{Sec_volume_FP}, at $k=1$, this frame potential measures the inverse coarse-grained volume of displacements obeying $|\bm \xi| < \sqrt{n_{\text{th}}}$. 
This volume is upper bounded by $\sim n_{\text{th}}^N$, leading to the nonzero lower bound of the frame potential.
In Appendix~\ref{App_FP}, we show that the frame potential measures an operator distance between the LHS and RHS of Eq. \eqref{design_def_2}, with respect to the density matrix $\rho$.

We can use the twice-regulated frame potential to verify our regularized 1-design. We find:
\be
\calJ_{\mathbb{D}_n}^{(1)}\left(\tilde{\rho}_{n_{\text{th}}}\right) 
\approx \left(\frac{(1+2\,(n_{\text{th}}/n))^2}{1+4\,(n_{\text{th}}/n)}\right)^N
\ge 1,
\ee
which indeed decays to its lower bound for $n \gg n_{\text{th}}$. We postpone evaluation of the $k=2$ frame potential for the large-squeezing Gaussian ensemble $\calE_r$ to future work.

\section{Experimental verifying scrambling}
\label{Sec_experiment}

Here, we consider the experimental detection of scrambling in CV systems. 
We begin in Section~\ref{Sec_experiment_SNAP} with the implementation of the scrambling dynamics themselves, providing concrete, precisely-controllable schemes to realize genuine scrambling dynamics using the so-called SNAP gate in a cavity-QED architecture.
We numerically simulate these dynamics when possible, and mention open questions which may be addressed by experiment.
Turning towards detection, in Section~\ref{Sec_experiment_measurement} we introduce concise measurement schemes for TOCs, individual OTOCs, and average OTOCs. 
These schemes rely only on Gaussian operations, as well as the ability to experimentally implement the (possibly non-Gaussian) scrambling operations $U(t)$ and $U^{\dagger}(t)$. 
Unfortunately, all of these schemes are prone to confusing scrambling with decoherence and experimental error, a problem well-known in DV systems. 
We address this in Section~\ref{Sec_experiment_teleportation} by introducing a robust teleportation-based measurement scheme, adapted from that in Ref.~\cite{yoshida2018disentangling}.

\subsection{Experimental realization of scramblers}
\label{Sec_experiment_SNAP}


In this Section, we present concrete proposals for the experimental realization of CV genuine scrambling dynamics. 
While we can numerically simulate these models in single- and few-mode systems, many questions arise that are beyond the scope of exact numerics, and are ripe for experimental input.
For instance, the implementation of genuine scrambling random circuits could probe the accuracy of  our conjecture in Section \ref{Sec_circuit}, that quasi scrambling circuits can replicate aspects of OTOC decay and operate spreading in genuine scrambling. 
Additionally, our proposal will also prove apt for probing scrambling in \emph{number-conserving} CV systems, which may behave qualitatively different from the non-conserving models of Section \ref{Sec_circuit}.

Measuring scrambling behavior (as discussed in the following Sections \ref{Sec_experiment_measurement} and \ref{Sec_experiment_teleportation}) will necessarily entail the precise implementation of not only the scrambling unitary $U$, but also either its inverse $U^{\dagger}$ or its conjugate $U^*$, requiring a high amount of experimental control.
Additionally, realizing genuine scrambling requires strong non-Gaussian operations.
Candidate experimental platforms for detecting CV scrambling should feature both of these properties, and might include non-linear crystals, cavities~\cite{vlastakis2013deterministically,heeres2015cavity,joshi2017qubit}, and optical Floquet systems~\cite{lemos2012experimental}. 

\begin{figure}
\centering
\subfigure{
\includegraphics[width=0.475\textwidth]{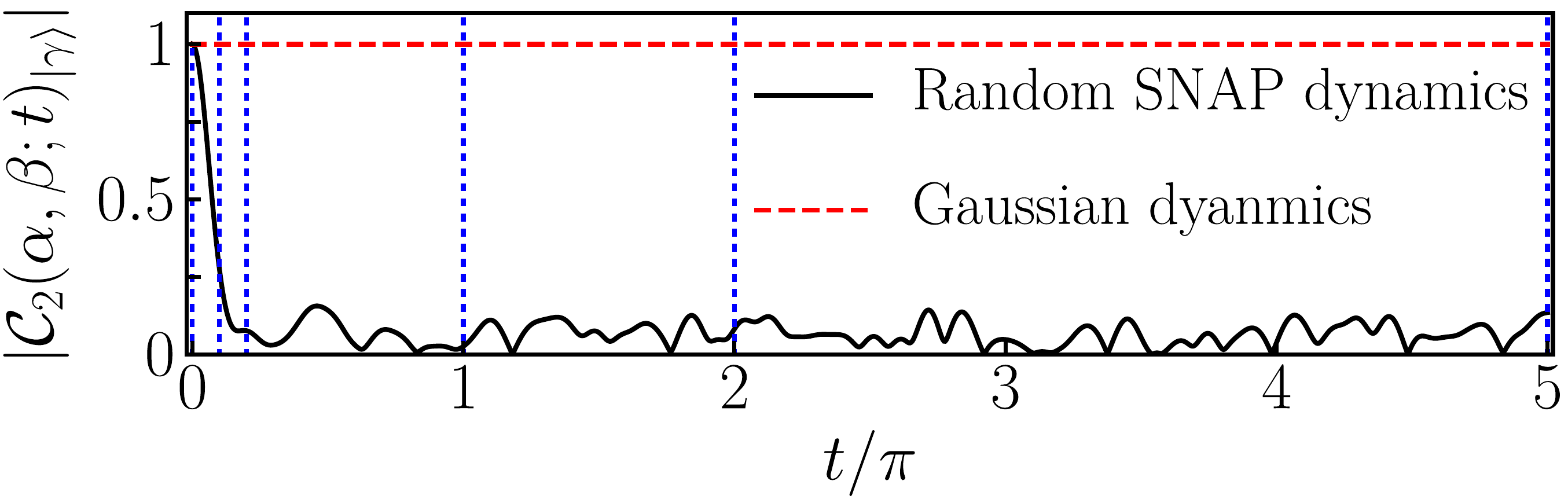}
}
\caption{OTOC amplitude $|\calC_2(\alpha, \beta;t)_{\ket{\gamma}}|$ for a time-dependent single-mode random SNAP gate $S_N(t)$ (black), for $\alpha=2+2i, \beta=2-2i, \gamma=8$. Blue lines indicate the times of the TOC snapshots in Fig.~\ref{TOC_SNAP}. For contrast, the OTOC amplitude does not decay under Gaussian dynamics (red).
\label{OTO_SNAP}
}
\end{figure}

For concreteness, we focus on cavity-QED architectures. 
All Gaussian operations (displacements, beamsplitters, and squeezing operations) can be implemented in these systems~\cite{joshi2017qubit}. 
Furthermore, non-Gaussian effects are typically much stronger than in other platforms.
We focus on a particular non-Gaussian gate that has already been implemented, the so-called Selective Number-dependent Arbitrary
Phase (SNAP) gate~\cite{heeres2015cavity,krastanov2015universal}. Diagonal in the photon number basis, it takes the form
\be
S_N\left(\{\theta_n\}\right)=\sum_{n=0}^\infty e^{i\theta_n} \ket{n}\bra{n},
\ee
where the phase $\theta_n \in [0,2\pi)$ of the $n$-photon number state can in principle be controlled arbitrarily.
The SNAP gate can be experimentally realized using a cavity coupled to a transmon qubit, with the Hamiltonian
\be
H=\omega_c a^\dagger a+\omega_q \state{e}+\chi a^\dagger a \state{e}+\Omega\left(t\right)e^{iw_q t}\ket{e}\bra{g}+c.c.,
\ee
where $a$ is the annihilation operator of the cavity, $\ket{g}$ ($\ket{e}$) is the ground (excited) state of the qubit and c.c. is the complex conjugate of the last term. When the qubit frequency shift $\chi$ is larger than both the qubit and cavity transition line-widths $\omega_q,\omega_c$, and the drive $\Omega\left(t\right)=\Omega_n\left(t\right)e^{-in \chi t}$ is weak, one can apply a phase selectively to the $n$-photon state $\state{n}$, i.e. $\ket{g,n}\to e^{i\theta_n}\ket{g,n}$, while keeping all other states invariant. A drive composed of multiple frequencies $n$ allows the independent implementation of multiple such phases, realizing the SNAP gate. In experiments reported by Ref.~\cite{heeres2015cavity}, phases for up to $n=11$ are precisely controlled. With a cavity lifetime $\sim 50 \, \mu$s, qubit relaxation times $\sim 20-30 \, \mu$s, and $\chi \sim$ MHz, this allowed the implementation of up to 15 SNAP gates with fidelity $> 0.96$ per gate. 


The SNAP gate can simulate a variety of effective cavity Hamiltonians. For example, the Kerr nonlinearity $K_{R}\left(t\right)\equiv\exp\left(-i t H^2\right)$ with $H^2=\left(p^2+q^2\right)^2$ is realized via a single SNAP gate with time-dependent phases $\theta_n=-t \left(4n+2\right)^2$. 
In fact, when combined with Gaussian operations, the SNAP gate is universal for the realization of all Hamiltonians polynomial in quadrature operators~\cite{lloyd1999quantum}.
While in principle this allows one to realize the Henon-Heiles potential and the cubic phase gate of Section~\ref{Sec_genuine_example}, such Hamiltonians generically require long sequences of fast SNAP gates, which may prove less feasible for experiment. 

\begin{figure}
\centering
\includegraphics[width=0.475\textwidth]{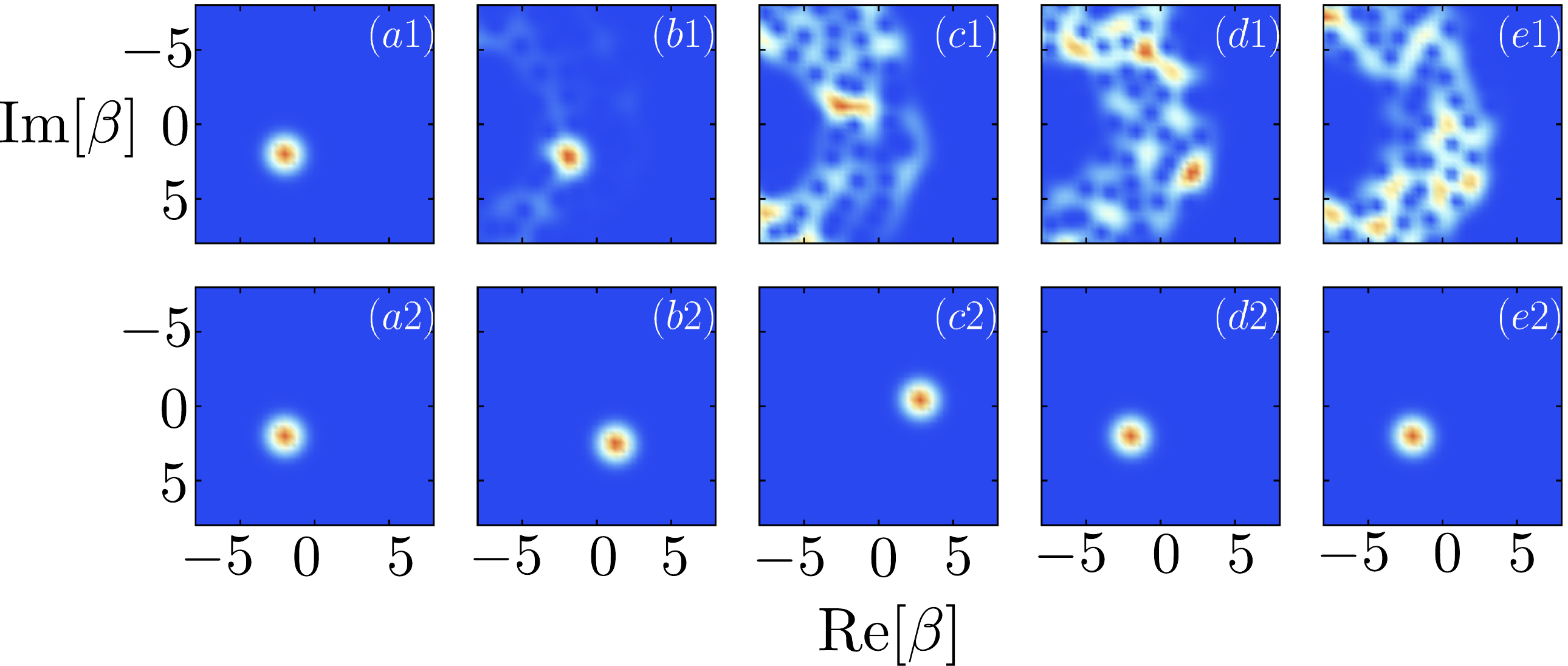}
\caption{(a1)-(e1) Snapshots of the phase space distributions of a displacement operator $\alpha=2+2i$ under the time-dependent random SNAP gate $S_N(t)$, as measured by the TOC $|\calC_1(\alpha, \beta;t)_{\ket{\gamma}}|^2$, with  $\gamma=8$ (blue indicates zero TOC). Time increases from left to right [$t/\pi={0,  0.1, 0.2, 1, 2}$, from (a1)-(e1); see Fig.~\ref{OTO_SNAP}]. (a2)-(e2) For contrast, the distribution remains localized under Gaussian dynamics ($H^1=p^2+q^2$). All plots share the same x- and y-axes, the real and imaginary parts of $\beta$, respectively.
\label{TOC_SNAP}
}
\end{figure}

With near-term experiments in mind, we introduce several genuine scramblers composed of only a moderate number of SNAP gates.
We begin by studying operator spreading in the phase space of a single-mode, under a single time-dependent SNAP gate.
As an example, we consider the random SNAP gate $S_N\left(t\right)=\sum_{n=0}^\infty e^{iw_nt} \ket{n}\bra{n}$, with `energies' $w_n$ distributed uniformly in $[0,2\pi)$~\cite{footnote7}.
Typical of genuine scrambling behavior, we find numerically that OTOCs decay to small values in $O(1/2\pi)$ time and remain small afterwards (see Fig.~\ref{OTO_SNAP}). To visualize the operator spreading responsible for this decay, one can calculate and plot the square of the TOC $|\calC_1(\alpha, \beta;t)_{\ket{\gamma}}|^2$ [see Eq.~(\ref{correlator})], shown in Fig.~\ref{TOC_SNAP}(a). As anticipated, the TOC spreads to occupy a larger phase space volume on a time scale similar to that of the OTOC decay. For contrast, under time-evolution by a \emph{Gaussian} Hamiltonian $H^1=p^2+q^2$, the OTOC amplitude is fixed at unity, and the TOC remains localized in phase space [Fig.~\ref{TOC_SNAP}(b)].


Moving forward, we consider the use of SNAP gates to study genuine scrambling in multi-mode systems.
Such behavior is tremendously difficult to numerically simulate due to the exponential size of the multi-mode Hilbert space, and would benefit greatly from experimental input.
As a first example, the inclusion of SNAP gates in random circuit models like those in Section \ref{Sec_circuit} would break the Gaussianity of the dynamics, and allow one to study operator spreading and entanglement formation of generic locally interacting CV systems.
In addition, SNAP gates conserve photon number, and are thus particularly well-suited for probing CV scrambling in the presence of conservation laws. 
To this end, one might consider a random circuit of only passive linear optics and SNAP gates (no squeezing), organized similarly to Fig.~\ref{random_Gaussian}, with SNAP gates replacing single-mode squeezing operations. 
It would be interesting to observe these circuits' entanglement growth in time: since number conservation seems to forbid the squeezing-induced Hilbert space growth of Section \ref{Sec_circuit}, one might guess that conservation laws cause the system to saturate to DV-like behavior $S \sim t$, in contrast to our previous result $S \sim t^2$.

\begin{figure}
\centering
\subfigure{
\includegraphics[width=0.2\textwidth,valign=t]{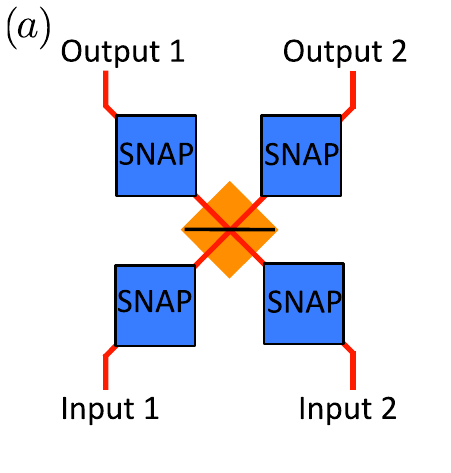}
}\hspace*{-1em}
\subfigure{
\includegraphics[width=0.26\textwidth,valign=t]{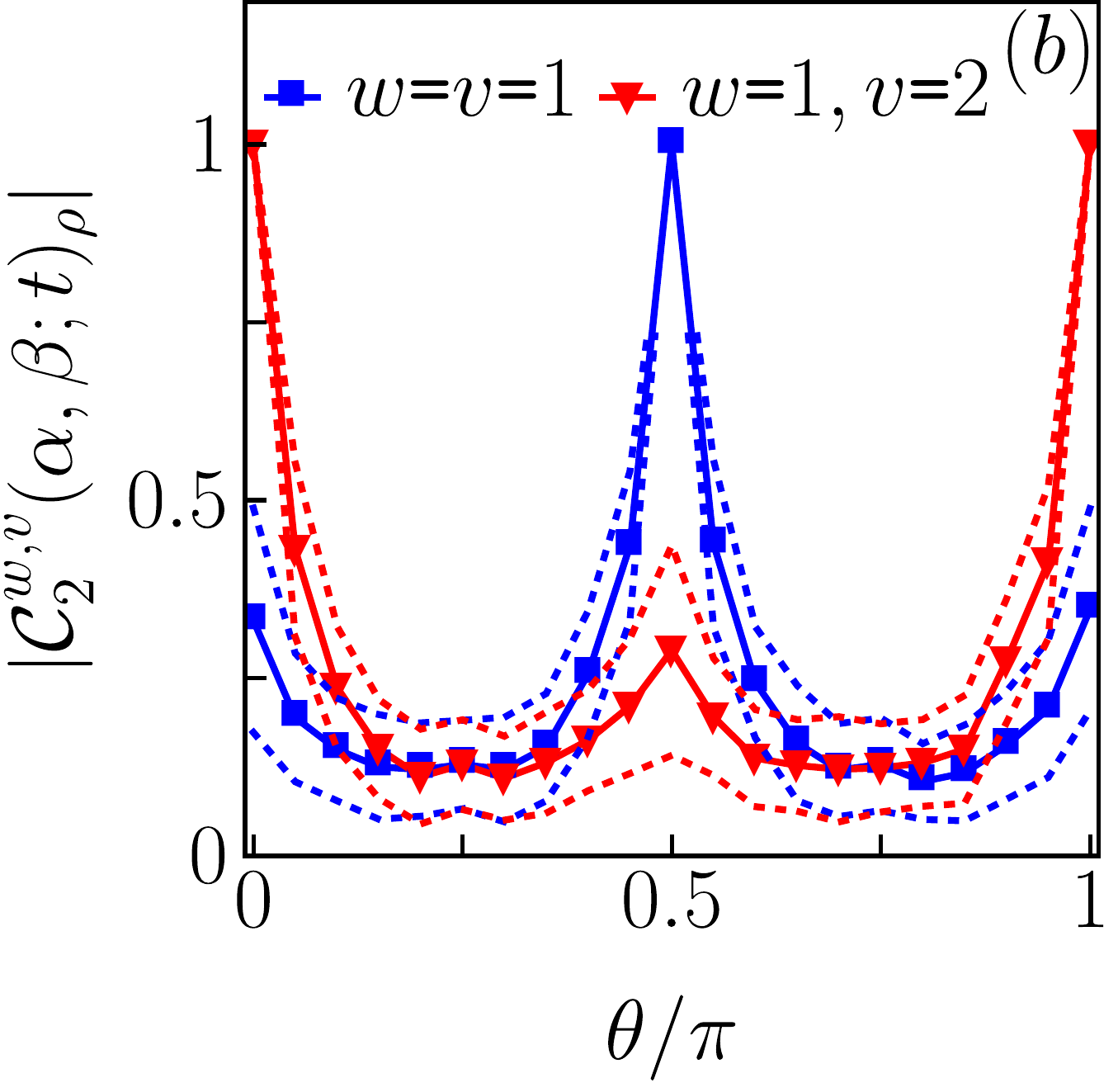}
}
\caption{(a) Simple two-mode model of a SNAP gate random circuit. The center block represents a random beamsplitter. (b) The average OTOC amplitude ($\alpha=(1+i)/2, \beta=(1-i)/2, \gamma_1=\gamma_2=2$) under after the circuit, for both single-mode scrambling ($w=v=1$, blue square) and multi-mode scrambling ($w=1,v=2$, red triangle), as a function of the beamsplitter transmissivity $\cos^2\theta$. Dashed curves indicate standard deviations. Each data point is averaged over 100 samples.
\label{OTO_SNAP_two_mode}
}
\end{figure}

For instructive purposes, we numerically simulate the simplest of such multi-mode, number-conserving circuits: a single-layer circuit on a two-mode system, as shown in Fig.~\ref{OTO_SNAP_two_mode}(a). The circuit consists of a beamsplitter with transmissivity $\cos^2\theta$ sandwiched between four random SNAP gates, each with phases iid uniformly in $[0,2\pi)$. To characterize operator spreading, we compute the amplitude of the OTOC with respect to single-mode displacement operators, averaged over the random SNAP gates. As expected the OTOC strongly depends on $\theta$, which controls the mixing of the two modes [see Fig.~\ref{OTO_SNAP_two_mode}(b)]. At $\theta=0$, there is no mixing between modes, and so $|\calC_2^{1,2}\left(\alpha,\beta;t\right)|=1$. At $\theta=\pi/2$, mode 1 and mode 2 are swapped, giving $|\calC_2^{1,1}\left(\alpha,\beta;t\right)|=1$. The mixing is maximized at $\theta = \pi/4$. Somewhat surprisingly, this value maximizes not only two-mode scrambling (i.e. it minimizes $|\calC_2^{1,2}\left(\alpha,\beta;t\right)|$), but also single-mode scrambling (i.e. it \emph{also} minimizes $|\calC_2^{1,1}\left(\alpha,\beta;t\right)|$).

\begin{figure}
\centering
\subfigure{
\includegraphics[width=0.2\textwidth,valign=t]{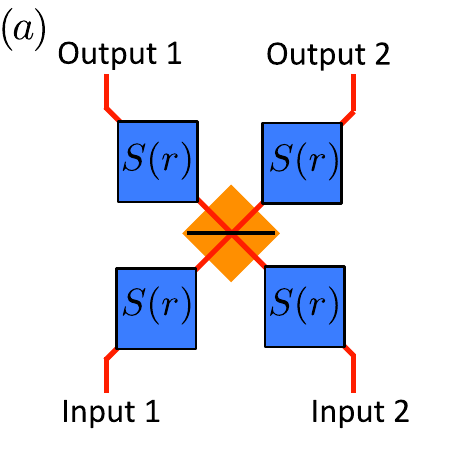}
}\hspace*{-1em}
\subfigure{
\includegraphics[width=0.26\textwidth,valign=t]{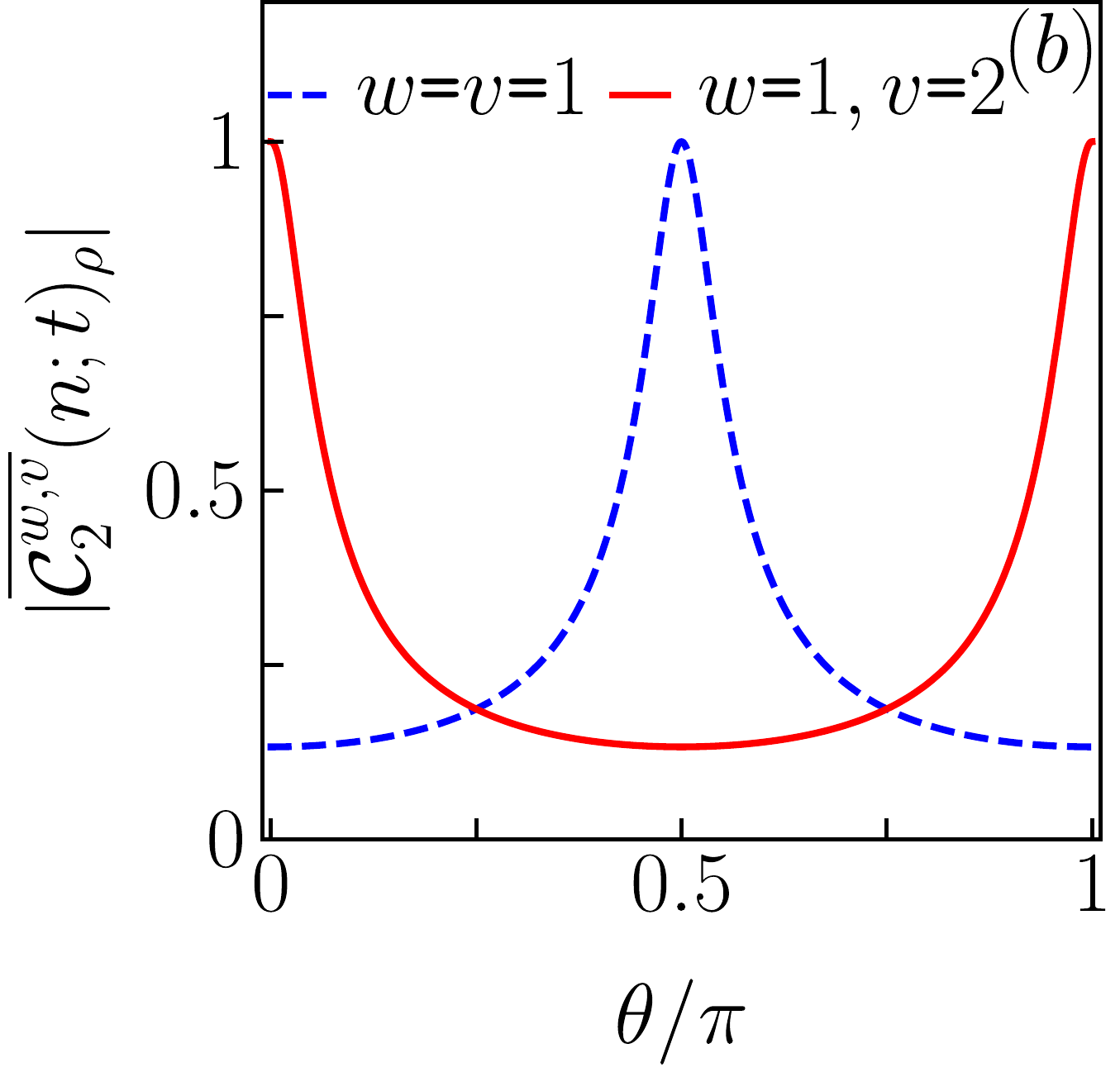}
}
\caption{(a) A two-mode Gaussian analog to the random SNAP gate circuit of Fig.~\ref{OTO_SNAP_two_mode}, where the SNAP gates are replaced with single-mode squeezing of strength $r = 1$. (b) The average OTOC ($n=1$) for single-mode ($w=v=1$, blue dashed) and multi-mode ($w=1,v=2$, red) scrambling, as a function of the beamsplitter transmissivity $\cos^2\theta$. 
\label{OTO_Gaussian_two_mode}
}
\end{figure}

The experimental realization of genuine scrambling circuits would also enable a powerful check on our assumption in Section \ref{Sec_circuit}, that aspects of genuine scrambling could simulated using averages over quasi scrambling systems.
For example, we consider a quasi scrambling analog (Fig.~\ref{OTO_Gaussian_two_mode}) to our previous SNAP gate circuit (Fig.~\ref{OTO_SNAP_two_mode}), with squeezing operations of strength $r$ instead of SNAP gates. In Appendix~\ref{sec:volume_quasi_A}, we compute the circuit's average OTOC $\overline{\mathcal{C}^{\omega,v}_{2}}(n;t)_{\rho} \equiv \overline{\mathcal{C}_{2}} (\mathbb{D}^{\omega}_{n}(t), \mathbb{D}^{v}_{n})$ [Fig.~\ref{OTO_Gaussian_two_mode}(b)]. 
We find qualitatively similar behavior to the (non-averaged) OTOC of the random SNAP gate circuit, supporting our assumption.
Nonetheless, some deviation is observed at the minimally-mixing values $\theta = 0, \pi/2$.
We speculate that this is due to the additional conservation of the individual modes' photon numbers at these values in the SNAP gate circuit, which may inhibit OTOC decay and complicates this particular comparison.



\subsection{Measurement of TOCs and OTOCs}
\label{Sec_experiment_measurement}

\begin{figure}
\centering
\includegraphics[width=0.5\textwidth]{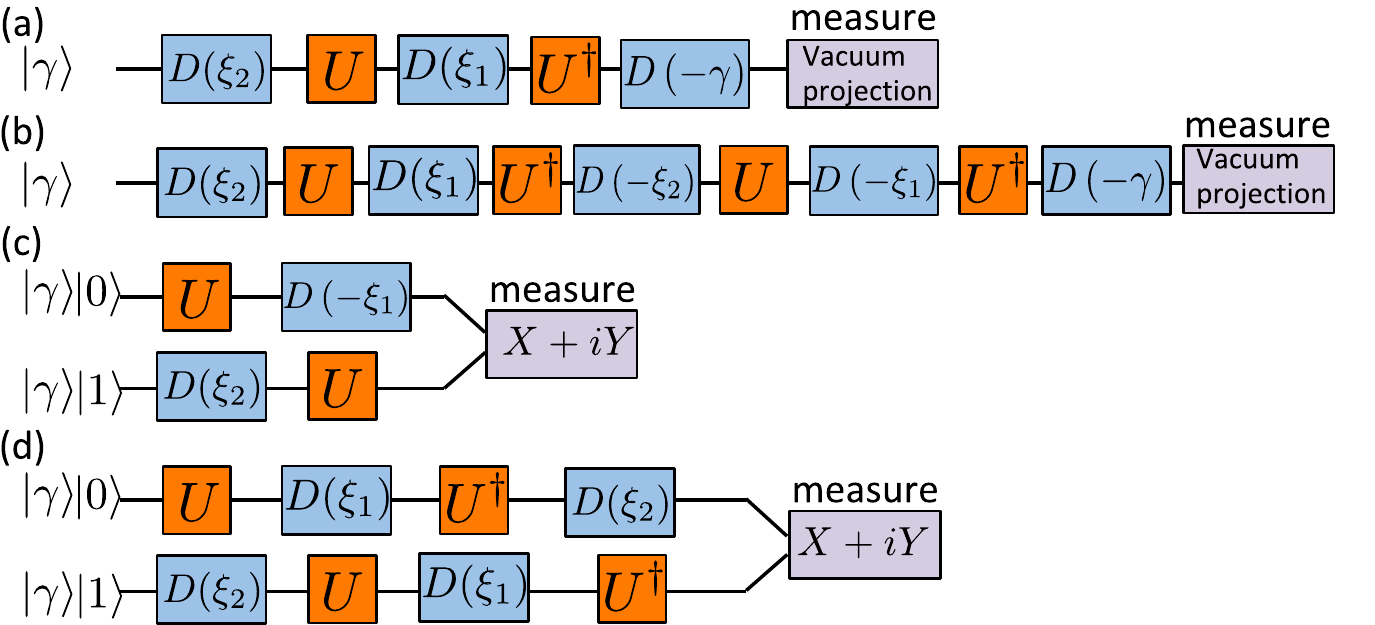}
\caption{
Measurement protocols for (a) the TOC amplitude $|\calC_1(\bm \xi_1,\bm \xi_2;t)_{\ket{\bm \gamma}}|^2$, (b) the TOC $\calC_1(\bm \xi_1,\bm \xi_2;t)_{\ket{\bm \gamma}}$, (c) the OTOC amplitude $|\calC_2(\bm \xi_1,\bm \xi_2;t)_{\ket{\bm \gamma}}|^2$, and (d) the OTOC $\calC_2(\bm \xi_1,\bm \xi_2;t)_{\ket{\bm \gamma}}$.
\label{measurement_scheme_CV}
}
\end{figure}

The most direct way to measure the amplitude of TOCs and OTOCs is to sequentially apply the operators in the correlation function to a state, and measure the probability to remain in that state. For instance, to measure the TOC $|\braket{\bm \gamma| U^\dagger D\left(\bm \xi_1\right) U D\left(\bm \xi_2\right)|\bm \gamma}|^2$, one would apply  $D(\bm \xi_2)$, then $U$, then $D(\bm \xi_1)$, then $U^{\dagger}$, then measure the probability to be in state $\ket{\bm \gamma}$. This is depicted for TOCs and OTOCs in Figs.~\ref{measurement_scheme_CV}(a,c). For simplicity, we consider correlations with respect to a coherent state $\ket{\bm \gamma}$. The probability to be in the coherent state can be measured by performing a displacement $D(-\bm \gamma)$ and measuring the probability to be in the vacuum state.

One can measure TOCs and OTOCs themselves, and not just their amplitudes, using a control qubit and an interferometric scheme similar to Refs.~\cite{swingle2016measuring,yao2016interferometric}. As shown in Fig.~\ref{measurement_scheme_CV}(b,d), here one initializes the control qubit in the state $(\ket{0}+\ket{1})/\sqrt{2}$ and performs different operations on the CV system given different states of the control qubit. The complex-valued TOC/OTOC is found by measuring $X + i Y$, where $X,Y$ are Pauli operators on the control qubit.

Finally, we present a concise scheme to measure average OTOCs, as introduced in Section \ref{Sec_averageOTOC}. 
The only change from our individual OTOC schemes is that we now use an ancillary mode, prepared in the Gaussian state $|\psi_{0}\rangle \sim \int_{-\infty}^{\infty} dq e^{-\frac{1}{2} q^2/\Delta^2 } |q\rangle$ (expressed in the position basis), to perform the `ensemble' of displacements on the CV system of interest~\cite{footnote8}. This is done using the SUM gate,
\begin{align}
\figbox{1.0}{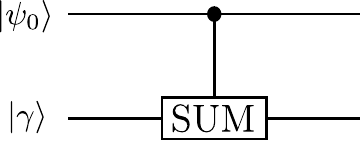},
\end{align}
which acts on quadrature operators as
\begin{align}
&q_{1}\rightarrow q_{1} \qquad \qquad p_{1} \rightarrow p_{1} - p_{2} \\
&q_{2}\rightarrow q_{1}+q_{2} \qquad p_{2} \rightarrow p_{2}.
\end{align}
This implements a displacement $D(q)$ with probability $\sim e^{- q^2/\Delta^2 }$ (shown by tracing out the ancilla). 
Average OTOCs require one to sample pairs of displacement operators $D(\bm \xi)$ and $D(- \bm \xi)$ in a correlated manner, which can be achieved by using the same ancilla for each displacement of the pair~\cite{footnote9}.

\subsection{Robust teleportation-based protocol}
\label{Sec_experiment_teleportation}

In this section, we turn our attention to a theoretical/conceptual question concerning the verification of scrambling. While OTOCs can characterize the phenomena of scrambling when measured perfectly, they are sensitive to experimental noise and decoherence, challenging the experimental measurement of scrambling~\cite{yoshida2018disentangling}. For instance, loss and thermal noise, the most common imperfections in optical systems, both cause the OTOC to decay just as scrambling time-evolution would (see Appendix~\ref{App_OTO_loss}). It is therefore desirable to characterize scrambling in a way which clearly distinguishes scrambling from such errors. 

A robust verification protocol of scrambling has been recently proposed~\cite{yoshida2018disentangling} and experimentally realized~\cite{landsman2018verified} for DV systems. This protocol draws inspiration from ideas in quantum gravity, and can be viewed as a many-body generalization of quantum teleportation~\cite{Yoshida:2017aa}. Here, we briefly describe a similar teleportation-based protocol for CV scrambling. For simplicity, we restrict our attention to measuring scrambling by Gaussian unitaries, where the protocol will succeed with probability unity assuming no experimental error~\cite{footnote10}.

\begin{figure}
\centering
\includegraphics[width=0.4\textwidth]{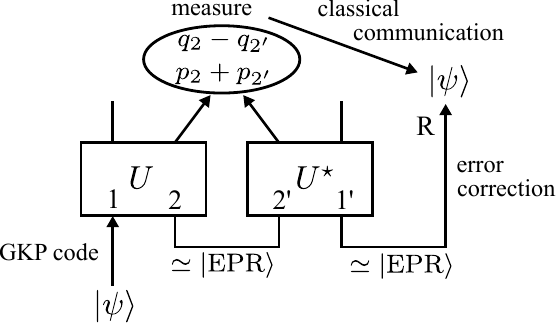}
\caption{
Schematic of the teleportation-based protocol for robustly measuring CV scrambling. An initial state $\ket{\psi}$ and two CV EPR pairs are time-evolved under the unitary of interest $U$ and its conjugate $U^*$. After, a pair of modes $2,2'$ are measured via the operators $q_{2}-q_{2}'$ and $p_{2}+p_{2'}$. The measurement outcome is used to error correct the original state, and the teleportation fidelity serves as a robust measure of information scrambling between modes $1$ and $2$ by $U$. In the presence of measurement uncertainties and imperfection in EPR preparations, the GKP encoding enables error-correction.
\label{fig-teleportation}
}
\end{figure}

We begin our analysis with the ideal, error-free case, and demonstrate our protocol's robustness to error after. We seek to characterize quasi scrambling by a Gaussian unitary $U$ acting on two CV modes. Fig.~\ref{fig-teleportation} displays the set-up for the teleportation-based protocol. Initially, the system is prepared in two EPR pairs on modes $(2,2')$ and $(1',R)$ whereas an arbitrary quantum state $|\psi\rangle$ is prepared on $1$. For a later purpose, it is convenient to use the stabilizer formalism to characterize EPR pairs. For a two mode system with position and momentum operators $(q,p)$, $(q',p')$, the CV EPR pair $|\text{EPR}\rangle$ is defined as the $P= Q= 0$ eigenstate of operators
\begin{align}
P= p + p'  \qquad Q = q - q' 
\end{align}
Note that this corresponds to the infinite squeezing limit of the two-mode squeezed state.

Next, we apply $U$ on $(1,2)$, and its complex conjugate $U^{\star}$ on modes $(2',1')$. While the protocol works with an arbitrary quasi scrambling unitary, we simplify our treatment by considering the following family of quasi scrambling unitaries:
\begin{equation}
\begin{split}
&p_{1}\rightarrow m p_{1} + (m+1) p_{2} \qquad  q_{1} \rightarrow m q_{1} - (m-1) q_{2}\\
&p_{2}\rightarrow (m-1) p_{1} + m p_{2} \qquad q_{2} \rightarrow - (m+1) q_{1} + m q_{2}
\end{split}
\end{equation}
where $m$ is arbitrary real number. When $m\not= 0, -1, 1$, observe that $U$ delocalizes any single-mode displacement operators to a two-mode displacement operator, the criteria for a non-trivial quasi scrambler. It is convenient to write the above transformation in the following, inverted manner:
\begin{equation}
\begin{split}
& m p_{1} - (m+1) p_{2}\rightarrow p_{1}\qquad \ \quad m q_{1}+ (m-1) q_{2} \rightarrow q_{1}\\
&- (m-1)p_{1} + m p_{2}\rightarrow p_{2} \qquad (m+1)q_{1} + mq_{2} \rightarrow q_{2}. \label{eq:inverse}
\end{split}
\end{equation}
The unitary $U$ contains squeezing since the total amplitude of displacements changes. The amount of squeezing $e^r \sim m$ plays an important role in fault-tolerance of the teleportation protocol, as we will see later.

After applying $U$ and $U^{\star}$, we measure $(2,2')$ with the following operators:
\begin{align}
P_{22'}= p_{2} + p_{2'} \qquad Q_{22'} = q_{2} - q_{2'}.
\end{align}
We send the measurement outcomes $Q_{22'}$ and $P_{22'}$ as a classical message to $R$. By applying the inverse transformation from Eq.~\eqref{eq:inverse}, at $t=0$, we have
\begin{align}
&P_{22'}=\big(- (m-1)p_{1} + m p_{2}\big) + \big(- (m-1)p_{1'} + m p_{2'}\big)\\
&Q_{22'}= \big((m+1)q_{1} + mq_{2}\big) - \big((m+1)q_{1'} + mq_{2'}\big).
\end{align}
Since $p_{2}+p_{2}'=0$ and $q_{2}-q_{2'}=0$, we arrive at
\begin{align}
P_{22'} = - (m-1)(p_{1} + p_{1}') \qquad Q_{22'} = (m+1) (q_{1} - q_{1'}).
\end{align}
Hence, the protocol teleports the following state on $R$:
\begin{align}
D\left( \Big( \frac{Q_{22'}}{m+1}, - \frac{P_{22'}}{m-1} \Big) \right)|\psi\rangle
\end{align}
which is just a state shifted from $|\psi\rangle$. Since this shift can be corrected using the classical message, this protocol teleports the quantum state $|\psi\rangle$. 

While we have focused on a particular family of quasi scramblers, our treatment generalizes to generic quasi scrambling unitaries $U$ that delocalize single-mode displacement operators. Namely, we can show that the teleported state is $|\psi\rangle$ up to displacements that can be undone using knowledge of the measurement result. 

So far we have assumed that all implementations of the protocol are perfect. In actuality, one might incur errors due to decoherence, or a mismatch between the experimentally-implemented unitaries $U$, $U^*$. 
This protocol is robust to both of these effects, as discussed in great detail for DV systems in Ref.~\cite{yoshida2018disentangling}.
There are also further imperfections unique to the CV limit.
For example, it is not possible to prepare perfect CV EPR pairs, and one must approximate them with two-mode squeezed states. 
In addition, measurements of $Q_{22'}$ and $P_{22'}$ will involve some uncertainty. 

Both of these imperfections lead to unknown displacement errors on the teleported state, and, at a calculational level, appear similar to inserting some density matrix $\rho$ in the tensor contractions corresponding to the EPR pair and measurement. 
This link to finite-temperature scrambling, as well as experimental relevance, motivate us to qualitatively address when teleportation can succeed despite these imperfections.
We consider the above protocol for a general $N$-mode Gaussian unitary $U_{\b{S}}$ (taking displacements $\b{\xi} \rightarrow \b{S} \b{\xi}$, where $\b{S}$ is a symplectic matrix), in the presence of a measurement error, i.e. one recorded $(Q_{22'}, P_{22'})$ but actually measured $(Q_{22'}, P_{22'}) + \Delta \b{\xi}$.
Imperfections due to an imperfect EPR pair --- a two-mode squeezed state of $\sim n_{\text{EPR}}$ photons --- are treated similarly: they arise from the fact that small displacements $D(\Delta\b{\xi})$ leave the state unchanged for $\Delta \b{\xi} < 1/\sqrt{n_{\text{EPR}}}$.
To see these errors' effect, one can compute the displacement $D(\Delta \b{z})$ on $\ket{\psi}$ that would have given rise to the measurement error: $\Delta \b{\xi} = \b{S}_{1,m} \Delta \b{z}$, where $\b{S}_{1,m}$ is the $2 \times 2$ submatrix of $\b{S}$ between mode $1$ and the measured mode $m$.
Due to the error, the protocol decodes the `wrong' state, off by a displacement $\Delta \b{z}$.
However, as we have seen, small displacements are only distinguishable from the identity operation on subspaces of greater than $1/|\Delta \b{z}|^2$ photons. 
Therefore, if $\ket{\psi}$ has $\lesssim n_\psi$ photons, the error has no effect when $\sqrt{n_\psi} < 1/|\Delta \b{z}|$.
Since an $N$-mode unitary with single-mode squeezing $e^{r}$ has elements of typical magnitude $\sim e^r/\sqrt{N}$, the protocol can only teleport states of photon number $\sqrt{n_\psi} < e^r / (\sqrt{N} |\Delta \b{\xi}|)$~\cite{footnote11}.
This makes sense: a state $\ket{\psi}$ is described by displacement operators separated by distance $\sim 1/\sqrt{n_{\psi}}$ in phase space.
This translates to a distance $\sim (e^r/\sqrt{N}) \times  (1/\sqrt{n_{\psi}})$ on the measured mode after application of $U$.
Our teleportation constraint implies that we can only teleport states when this phase space distance is resolvable despite the imperfections, $(e^r/\sqrt{N}) \times  (1/\sqrt{n_{\psi}}) > \Delta \b{\xi}$.

We also hope to make the teleportation protocol fault-tolerant to these imperfections by using quantum error-correction. Gottesman, Kitaev and Preskill have proposed a way of encoding a finite-dimensional qudit into a single oscillator. This error-correcting code, called the GKP code, enables us to correct small displacement errors. Suppose we prepare a qudit input state and encode it into mode $1$ via the GKP encoding. For simplicity, we consider the error from faulty measurements, and assume that $Q_{22'}$ and $P_{22'}$ suffer from uncertainties $\Delta Q_{22'}$ and  $\Delta P_{22'}$. When the quantum state is teleported to $R$, it suffers from displacement noises $D(q_{R})$ and $D(p_{R})$ of $|q_{R}|\sim \frac{\Delta Q_{22'}}{m+1}$ and $|p_{R}|\sim \frac{\Delta P_{22'}}{m-1}$. By using the GKP code, displacement errors can be corrected if $m$ is sufficiently large, and thus the teleportation protocol can be made fault-tolerant. We remark, however, that the aforementioned protocol does not suppress errors from the imperfect EPR preparations on $(2,2')$ and $(1',R)$. 

Observe that the induced error can only be suppressed when the squeezing $m$ is large. There appears to be an intriguing relation between the amount of squeezing and the amount of teleported information. This leaves an interesting future problem concerning the upper bound on the information capacity under energy constraints. Another relevant future problem concerns the relation between this protocol and the average OTOC. For the quasi scramblers in Eq.~\eqref{eq:inverse}, the average OTOCs becomes small as $m$ increases. Hence, in the presence of measurement uncertainties, the smaller average OTOCs enables larger amount of fault-tolerant teleportation.

\section{Discussions and Conclusions}
\label{Sec_discussion}

We have mainly focused on developing tools and a general framework for characterizing scrambling and complexity in CV systems. The key idea is to introduce a density matrix that represents the Hilbert space of interest, which leads to a smooth regularization of various quantities. Throughout the paper we choose a thermal density matrix $\tilde{\rho}_{n_{\text{th}}}$ --- other choices may lead to different coarse-graining procedures at short-distances and cut-offs at long-distances. Below, we conclude the paper with a few comments and a discussion of open problems.

First, we describe the scenarios where a CV treatment is most suitable. The CV regime is i) large local Hilbert space ii) smoothly cut-off local Hilbert space. The relevant notion of locality is determined by the operators with which we wish to probe the system, and does not necessarily have to agree with the UV lattice cut-off of the system. 
Notably, these conditions can hold even for large but finite dimensional systems, whenever the quantum states under study are associated with a wide range of energy scales. More specifically, consider a Hamiltonian measuring some energy $H_0$, with eigenstates $\ket{E_n}, 1\le n \le N$. When $N$ is large and the relevant states $\rho$ have a distribution $\braket{E_n|\rho|E_n}$ spread out over a range of energies $E_n$, and decaying smoothly towards those states with larger energies, a CV description applies.

Next, we address open problems and future directions in the study of CV scrambling. 
\begin{itemize}

\item[i)]
The first set of open problems relates to our understanding of genuine scrambling. Most importantly, the speed and saturation of operator volume increase characterized by OTOC decay requires further study. For displacement-operator-based OTOCs, the connection between the initial decay and the Lyapunov exponent in classical chaotic systems can be further explored. This may lead to a deeper understanding of different classes of non-Gaussian unitaries. Related to this, it would be interesting to numerically investigate the effects of the small, but finite mass term $m$ in the cubic phase gate model in Sec.~\ref{Sec_genuine_example}. Eventually, one hopes to derive bounds on operators' volume increase when the system is constrained by certain conservation law. 

\item[ii)]
The second set of open questions concerns our understanding of local random Gaussian circuits. In Sec~\ref{Sec_many_mode}, we identified a {\it quadratic} growth of entanglement and linear increase of fluctuation for such systems, arising from an increase in the accessible local Hilbert space. However, a full theoretical model that explains the deviation from the KPZ scaling still needs to be developed. Such a model could deepen our understanding of scrambling dynamics in general CV systems. 

Additionally, extending the study in Sec.~\ref{Sec_experiment_SNAP} on models with photon number conservation law to a larger scale may deepen our understanding of conservation laws' consequences for scrambling dynamics.
Performing the SNAP gate-based experiment proposed in Section~\ref{Sec_experiment_SNAP} would be very instructive on this open problem, as numerical simulation becomes difficult for greater than a few modes. For state-of-art experimental platforms, the realization of this experiment is plausible in the near future. Adding squeezing to such system would also allow the verification of the theoretical results, and assumptions, in Sec~\ref{Sec_many_mode}. Our work lays a solid theoretical foundation for such experimental studies.

\item[iii)]
The third set of open questions regards the construction of CV unitary $k$-designs. We still understand very little about higher designs in the CV case. Evaluating the new frame potential in Eq.~\eqref{frame_potential_new} for $k\ge2$, and more general ensembles, as well as relating it to notions of complexity, are important future directions. Additionally, in light of our findings on Gaussian 2-designs, it remains an open question whether there exists a more appropriate definition of higher CV designs that is compatible with soft energy regularization. Nevertheless, our CV 1-design and 2-design `analog' may still be useful for applications such CV state tomography (1-design) and compressed sensing (2-design)~\cite{roth2018recovering}. 

\item[iv)] 
Another interesting future problem concerns the definition of a `size' for time-evolved operators at finite temperature. In DV systems at infinite temperature, the size of operators $O(t)$ corresponds to the average number of qubit supports in the Pauli decomposition of $O(t)$. As discussed in Section~\ref{sec:genuine}, the size of $O(t)$ can be measured by OTOCs. However, the notion of the size of operators $O(t)$ becomes ambiguous at finite temperature since OTOCs depend on $\rho$. An important question is to how to define the size of operators in the presence of $\rho$ at finite temperature in a physically meaningful manner. In a recent work on the operator growth in the SYK model~\cite{qi2018quantum}, a possible definition of the finite temperature size of $O(t)$ is proposed. Namely, the authors argued that the size should be defined by subtracting the thermal background, i.e. as the difference between the support of $\rho^{1/4}O(t)\rho^{1/4}$ and $\rho^{1/2}$. They showed that the Lyapunov growth of OTOCs corresponds to the exponential growth of this difference.

On the other hand, in Section~\ref{sec:genuine}, we have argued that thermal TOC and OTOC measure coarse-grained volumes of the operator spreading where the resolution of the phase space is set by $\rho$. Specifically, we saw that a thermal state $\tilde{\rho}_{n_{\text{th}}}$ induces a Gaussian blurring of the phase space. One concrete open question concerns the connection between their proposal of subtracting the thermal background and our results developing the notion of coarse-graining. Here we present a heuristic argument while postponing rigorous discussions to the future work. Recall that the size of the operator can be counted by the number of qubits. The coarse-grained volume $V$ in the phase space roughly corresponds to a $\sqrt{V}$-state quantum spin, so it can be embedded in $\frac{1}{2}\log V$ qubits. So, one may assign $\frac{1}{2}\log V$ as the size of the operator. The coarse-grained volume $V$ depends on the scale of the resolution set by $\rho$. Letting an approximate radius of the spreading be $R$ and the resolution be $\delta$, the volume is $V = R^2/\delta^2$, the size will be given by $\frac{1}{2} (\log R - \log \delta) $. Hence we may interpret the second term as the thermal background.

\item[v)]
A final set of open questions is on the connection to resource theory.
It is well known that universal quantum computation requires non-Clifford operations, since Clifford operations admit efficient classical simulations according to the Gottessman-Knill theorem. It would therefore be interesting to assign a resource theoretic interpretation of $\sum_{Q \in \text{Pauli}} |f\left[Q;P(t)\right]|^4$, thereby relating quantum computational power and scrambling/decay of OTOCs. Similarly in CV systems, the preservation of OTOC amplitude by Gaussian unitaries (quasi scramblers) and the decay of OTOC amplitude caused by non-Gaussian unitaries (genuine scramblers) might lead to a new resource theory framework for non-Gaussianity~\cite{marian2013relative,genoni2008quantifying,genoni2010quantifying,zhuang2018resource,takagi2018convex,albarelli2018resource}. Finally, there has also been some recent interest in characterizing the complexity of Gaussian CV states relevant for quantum field theory~\cite{chapman2018towards}; we speculate that our general approach, based upon frame potentials and OTOC decays, can be applied towards a more broad characterization of the complexity of states in field theories.

\end{itemize}


We conclude by summarizing our work and clarifying a number of distinctions between our work and prior studies.
Scrambling in CV systems has been studied in a number of  seminal previous works~\cite{rozenbaum2017lyapunov,hashimoto2017out,rammensee2018many,chavez2018quantum,borgonovi2018emergence,cotler2018out}, where OTOCs for quadrature operators and number operators are explored in the context of specific example Hamiltonians. 
These studies revealed that the OTOC with quadratures operators can enable a quantum-classical correspondence within the Ehrenfest time~\cite{rozenbaum2017lyapunov,cotler2018out}. 
In this paper, we attempt to understand scrambling in CV systems from a quantum information theoretic perspective: (i) We give general interpretations of TOCs and OTOCs in terms of operator spreading. (ii) We investigate the scrambling dynamics of generic local circuits based upon a CV analog of 2-designs. (iii) Our choice of displacement operators in OTOCs not only enables a quantum-classical correspondence in phase space, but also enables OTOCs to be measurable in quantum optical experiments. (iv) We provide an experimental blueprint for probing scrambling in cavity QED systems.

\let\oldaddcontentsline\addcontentsline
\renewcommand{\addcontentsline}[3]{}
\begin{acknowledgements}

We thank useful discussions with Vinay Ramasesh, Leigh Martin, Raphael Bousso, Irfan Siddiqi, Jeffrey H. Shapiro and Brian Swingle. 
This work was supported by the GeoFlow grant: DE-SC0019380, the Office of Science, Office of High Energy Physics, of the U.S. Department of Energy under Contract No. DEAC02-05CH11231 through the COMPHEP pilot “Probing information scrambling,” the AP Sloan foundation and the David and Lucille Packard Foundation.
Research at the Perimeter Institute is supported by the Government of Canada through Industry Canada and by the Province of Ontario through the Ministry of Research
and Innovation. T.S. acknowledges support from the National Science Foundation Graduate Research Fellowship Program under Grant No. DGE 1752814. 
\end{acknowledgements}

\appendix

\section{Basic Gaussian unitaries}\label{App_Gaussian}

We begin by giving some examples of symplectic matrices corresponding to Gaussian unitary operations~\cite{Weedbrook_2012}. 
\begin{enumerate}
\item 
Single-mode phase rotation:
\begin{align}
& 
\bm R\left(\theta\right) =
\left(
\begin{array}{cccc}
\cos\theta&\sin\theta\\
-\sin\theta&\cos\theta
\end{array} 
\right).
\label{R_mat}
\end{align}

\item 
Single-mode squeezing:
\begin{align}
& 
\bm S\left(r\right) =
\left(
\begin{array}{cccc}
e^{-r}&0\\
0&e^{r}
\end{array} 
\right).
\label{S_mat}
\end{align}

\item 
Two-mode beamsplitter:
\begin{align}
& 
\bm B\left(\eta\right) =
\left(
\begin{array}{cccc}
\sqrt{\eta} \, {\mathbf I}&\sqrt{1-\eta}\, {\mathbf I}\\
-\sqrt{1-\eta} \, {\mathbf I}&\sqrt{\eta} \, {\mathbf I}
\end{array} 
\right),
\label{B_mat}
\end{align}
where ${\mathbf I}$ is the $2 \times 2$ identity matrix.

\item 
Two-mode squeezing:
\begin{align}
& 
\bm S_2\left(r\right) =
\left(
\begin{array}{cccc}
\cosh(r)\,{\mathbf I}&\sinh(r)\,{\mathbf Z}\\
\sinh(r)\,{\mathbf Z}&\cosh(r)\,{\mathbf I}
\end{array} 
\right),
\label{S2_mat}
\end{align}
where ${\mathbf Z}$ is the $2 \times 2$ Pauli matrix.

\end{enumerate}

Next, we list some useful identities for displacement operators and Gaussian unitaries, all of which have analogs for Pauli operators and Clifford unitaries in DV systems~\cite{veitch2014resource}. The completeness of displacement operators follows from,
\be 
\tr\left( D\left({\bm \xi}\right) D\left({\bm \xi}^\prime\right)\right)=\pi^N \delta\left(\bm \xi+\bm \xi^\prime \right),
\nonumber
\ee
and
\be
A={1}/{\pi^N}\int d^{2N} {\bm \xi} \ \chi\left({\bm \xi};A\right)D\left(-{\bm \xi}\right),
\nonumber
\ee 
when $\chi\left({\bm \xi};A\right)$ exists. These properties give rise to the identity
\begin{align}
&\frac{1}{\pi^N}\int d^{2N} {\bm \xi} \ \tr \left[D\left({\bm \xi}\right) A \right] \tr \left[D^\dagger\left({\bm \xi}\right) B
\right]
 =\tr\left(AB\right),
\label{lemma4_eq2}
\end{align}
proven in Ref.~\cite{cahill1969ordered}. Finally, we note that a Gaussian operations corresponds to a linear coordinate transform of the Wigner characteristic function,
\be 
\chi \left({\bm \xi};{U}_{\bm S,\bm d} A {U}_{\bm S,\bm d}^\dagger\right)=\chi \left({\bm S}^{-1}{\bm \xi};A\right) \exp\left(i {\bm d}^T {\bm \Omega \bm \xi }\right),
\ee 
which follows from the action of Gaussian unitaries on displacement operators, and is proved in Ref.~\cite{takagi2018convex}.


\section{Operator distributions and OTOCs in quasi scramblers}\label{sec:volume_quasi}

\subsection{Volume and average OTOCs}\label{sec:volume_quasi_A}

Consider an initial ensemble $\mathcal{E}$ of displacement operators. Under quasi scrambling time-evolution by a Gaussian unitary $U(t)$ these displacement operators will evolve into different displacements, which form a new ensemble $\mathcal{E}(t)$. Our goal is to develop a probe of $\mathcal{E}(t)$ via OTOCs. For this purpose, let us introduce a formal definition of the average OTOC for quasi scramblers. Given a pair of displacement operator ensembles $\calE_1$ and $\calE_2$, we define the average quasi scrambling `OTOC'
\be
\overline{\calC_2}\left(\calE_1,\calE_2\right)_\rho \equiv
\mathbb{E}_{V\sim \calE_1, W\sim \calE_2}
\tr\left[\rho V^\dagger W^\dagger V W\right].
\label{OTO_ave_quasi}
\ee
When characterizing time-evolution, $\mathcal{E}_{1}$ may be regarded as the time-evolved distribution of interest, e.g., $\mathcal{E}_{1} = \mathcal{E}(t)$, while $\mathcal{E}_{2}$ is a probing distribution. The necessity of considering an average OTOC is understood from the fact that the amplitude of individual OTOCs for Gaussian time-evolution is always unity. 

Although this average quasi scrambling OTOC bears great similarity to that of genuine scramblers in Eq.~\eqref{OTO_ave_general} of the main text, there is a subtle difference. For genuine scramblers, $D^{\dagger}(\bm \xi_1;t)$ and $D(\bm \xi_1;t)$ are usefully decomposed as sums of displacement operators using the characteristic functions $\chi^\star[\bm \xi]$ and $\chi[\bm \xi^\prime]$. Due to the density matrix $\rho$, the average genuine scrambling OTOC receives contributions for $\bm \xi \not = \bm \xi^\prime$. This is not the case for quasi scramblers, where we sample the same displacement operator for $U^{\dagger}$ and $U$. This implies that the average quasi scrambling OTOCs does \emph{not depend on} $\rho$. The $\rho$ dependence is recovered when considering thermally regulated OTOCs, but detailed discussions of this are beyond the scope of this paper.

As with genuine scramblers, the average OTOC is closely related to operator distributions. For instance, consider an arbitrary ensemble of displacement operators $\mathbb{D}_{P\left(\cdot\right)}$ [with probability distribution $P\left(\cdot\right)$] and a Gaussian probe ensemble $\calE_{2} = \mathbb{D}_n$ (defined in Eq.~(\ref{Dn_def}) in the main paper):
\begin{align}
\overline{\calC_2}\left(\mathbb{D}_n,\mathbb{D}_{P\left(\cdot\right)}\right)_\rho&=
\mathbb{E}_{V\sim \mathbb{D}_n, W\sim \mathbb{D}_{P\left(\cdot\right)}}
\tr\left[\rho V^\dagger W^\dagger V W\right] \\
 &= \mathbb{E}_{\bm \xi\sim P\left(\cdot \right)}
\left[
\exp\left(-n|\bm \xi|^2\right)
\right].
\end{align}
This average quasi scrambling OTOC therefore measures the extent of the operator spreading with a Gaussian coarse-graining, of familiar width $1/\sqrt{n}$ in phase space. 

In fact, an explicit correspondence between these average OTOCs and the frame potential can be derived when the distributions of interest are Gaussian distributions with zero mean. For example, consider the zero mean ensemble $
\mathbb{D}_{\bm0,\bm V}
$ in Eq.~\eqref{D_Gaussian_ensemble} in the main paper to be our `time-evolved' ensemble $\calE_1$.
One can show that
\be 
\overline{\calC_2}\left(\mathbb{D}_{\bm 0,\bm V},\mathbb{D}_n\right)_\rho=\prod_{\ell=1}^{2N}\frac{1}{\sqrt{1+2\lambda_\ell n}},
\ee
where $\lambda_\ell$'s are the eigenvalues of the matrix $\bm V$. In particular, if $n=2\left(2n_{\text{th}}+1\right)$, comparing with Eq.~\eqref{frame_Gauss} in the main text we have
\be
\overline{\calC_2}\left(\mathbb{D}_{\bm 0,\bm V},\mathbb{D}_n\right)_\rho=\calF_{\mathbb{D}_{\bm \xi_0,\bm V}}\left(\tilde{\rho}_{n_{\text{th}}}\right).
\label{OTO_ave_volume}
\ee

One important advantage of using average OTOCs over the frame potential is that OTOCs can measure not only the ensemble volume, but also volumes of ensembles projected onto subspaces of the $2N$-dimensional phase space (Fig.~\ref{weak_scrambling} in the main text). For instance, if one sets $\calE_2$ to be a Gaussian ensemble of displacement operators localized on certain mode $w$, $
\mathbb{D}_n^w=\left\{D^w\left(\bm \xi\right)| \bm \xi\sim P_D^G\left(\bm \xi; n \right)\equiv 
\exp\left(- |\bm \xi|^2/n\right)
/ \left(\pi n\right)  \right\}
$, we have
\be
\overline{\calC_2}\left(\mathbb{D}_{\bm 0,\bm V},\mathbb{D}_n^w\right)_\rho=\frac{1}{\sqrt{\left(1+2\lambda_1^w n\right)\left(1+2\lambda_2^wn\right)}},
\ee 
where $\lambda_1^w,\lambda_2^w$ are the eigenvalues of the covariance matrix projected onto the mode $w$ subspace, $\bm V_{2w,2w+1}$. This enables us to define the projected volume on mode $w$ as
\be
{\rm vol}^w\left(\calE\right)=\lim_{n\to\infty}
\left(\frac{1}{2n}\right)
\frac{1}{\overline{\calC_2}\left(\calE,\mathbb{D}_n^w\right)_\rho},
\label{vol_project}
\ee
in analog to Eqs.~(\ref{def_vol}, \ref{Volume_TOC}) in the main text. Note that the limit $n\rightarrow \infty$ only applies to the probe ensemble.

It is worth looking at a few simple examples. Consider average OTOCs of the form
\begin{align}
\overline{\mathcal{C}^{w,v}_{2}}(n;t)_{\rho} \equiv \overline{\mathcal{C}_{2}} (\mathbb{D}^{w}_{n}(t), \mathbb{D}^{v}_{n}),
\label{Average OTOC}
\end{align}
where displacement operators are chosen to be Gaussian distributed on modes $w,v$ at $t=0$, and $\mathbb{D}^{w}_{n}$ evolves to $\mathbb{D}^{w}_{n}(t)$ under a Gaussian unitary $U$. First, we look at single-mode cases $w=v$.

i) For $U$ composed only of displacements and phase rotations, we have 
\be
\overline{\calC^{w,v}_2}(n;t)_\rho =\frac{1}{1+n^2}.
\ee 
This includes the case where $U$ is an identity operator, and indicates the fact that displacements and phase rotations leave the mean-zero ensemble $\mathbb{D}^{w}_{n}(t)$ invariant.

ii)
For single-mode squeezing of strength $r$, we compute
\be
\overline{\calC^{w,v}_2}(n;t)_\rho =\frac{1}{\sqrt{1+n^4+2n^2\cosh(2r)}},
\ee 
so that the average OTOC decays as the squeezing increases. This demonstrates the increase of the coarse-grained volume measured by finite temperature OTOCs under Gaussian time-evolution.

Next we consider the multi-mode cases. 

iii) Passive linear optics. Let $\eta_{w,v}$ be the transmissivity between the two modes $w,v$. We have 
\begin{align}
\overline{\calC^{w,v}_2}(n;t)_\rho= \frac{1}{(1+\eta_{w,v}^2n^2)}.
\end{align}
When $\eta_{w,v}=1$, the passive linear optics act as a swap gate and $\overline{\calC^{w,v}_2}(n;t)_\rho={1}/{\left(1+n^2\right)}$. For generic passive linear optics, a typical transmissivity between the two modes $\eta_{w,v}\sim 1/\sqrt{N}$ decreases with the system size, and so the decay of the average OTOCs becomes less significant. This is specific to the case of the passive Gaussian (i.e. free boson) dynamics.

iv) For two-mode squeezing of strength $r$ between modes $w,v$, we find 
\begin{align}
\overline{\calC^{w,v}_2}(n;t)_\rho={1}/{(1+\sinh^2 \left(r\right)n^2)}.
\end{align}
As the amount of squeezing increases, $\overline{\calC^{w,v}_2}(n;t)_\rho\sim e^{-2|r|}$ decreases exponentially.

\subsection{Quantum Liouville's theorem}\label{App_Louiville}

While genuine scrambling (via non-Gaussian unitaries) is necessarily of a quantum nature, quasi scrambling (via Gaussian unitaries) admits a description by a classical Hamiltonian dynamics, since a single phase space point stays localized. This observation motivates us to generalize the classical Liouville's theorem to Gaussian quantum dynamics. 

In classical mechanics, Liouville's theorem asserts that phase space volume is preserved under Hamiltonian dynamics. (Although conserved, this volume may nonetheless stretch and distort over time --- a key feature classical chaos.) To formulate an analog to the classical theorem, let us consider an arbitrary ensemble $\mathbb{D}_{P\left(\cdot\right)}$ of displacement operators, with probability distribution $P\left(\cdot\right)$. We are interested in the volume of the time-evolved ensemble $
\mathbb{D}_{P\left(\cdot\right)}\left(t\right)$. To be rigorous, we consider the volume of operator distributions as defined in Eq.~\eqref{def_vol} in the main text. Noticing that volume is given by an integration over $\bm \xi$, of a function of $\bm S^{-1} \bm \xi$ (see Eq.~\eqref{vol_general_displacement} in the main text), and using $\det\left(\bm S\right)=1$, we immediately have the following theorem:

\emph{Quantum Liouville's theorem} --- The phase space volume of a general ensemble of displacement operators is preserved if $U\left(t\right)$ is a Gaussian unitary.

Despite its simplicity, this theorem provides interesting lessons. First, since the theorem indicates that there is no volume increase for Gaussian dynamics, one might wonder why Gaussian quasi scramblers can transform displacements into larger displacements. Indeed, given the fact that OTOCs are sensitive to volume growth, the decay of the average OTOC appears to contradict the theorem. The resolution is that average OTOCs with respect to local modes measure \emph{projected volumes}. The projected volumes on local modes may indeed increase in the presence of large squeezing (Fig.~\ref{weak_scrambling} in the main paper), even if the volume in the entire $2N$-dimensional space is fixed. 

Second, this theorem only applies to the infinite temperature limit; as seen in the main text, the `coarse-grained' volumes measured at finite temperature may not be preserved. The coarse-grained volume also has a nice correspondence to the Kolmogorov-Sinai (KS) entropy in classical Hamiltonian dynamics. Similar to the coarse-grained volume, the KS entropy counts the number of coarse-grained phase space boxes of the phase space volume, which increases for generic chaotic dynamics. 

Another interesting implication of this theorem concerns the characterization of quantum chaos in CV systems. The theorem implies that any non-Gaussian effect will lead to some change in the volume. Although the volume may either decrease or increase depending on $U$ and the initial state, we expect that the volume will generically increase for chaotic non-Gaussian dynamics. We speculate that if the volume of operator distributions is a monotonically increasing quantity, it may possess a similar intuition to the second-law of entropies.  

\section{OTOC in presence of loss}
\label{App_OTO_loss}

In this section, we quickly demonstrate that OTOCs decay in the presence of a loss channel, i.e. decoherence (for a detailed discussion of this in DV systems, see Ref.~\cite{yoshida2018disentangling}. Consider a unitary channel $\rho \to U^\dagger\left(t\right) \rho U\left(t\right)$ combined with a thermal loss channel. Suppose the loss happens before the unitary, so that
\be
D^w\left(\alpha;t\right)=U\left(t\right)^\dagger \calN^{N_E}_\eta \left(D^w\left(\alpha\right)\right) U\left(t\right).
\ee 
This channel maps $x\to \sqrt{\eta}x+\sqrt{1-\eta} x_e$, where $x_e$ are quadrature operators for an ancilla in thermal state $\rho_E$ with mean photon number $N_E$. To obtain how operators evolve under the loss channel we consider the trace
\begin{align}
&\tr_A\left( D_A\left(\alpha \right) \calN^{N_E}_\eta \left( \rho_A \right) \right)
\nonumber
\\
&=\tr_A\left( D_A\left(\alpha \right) \tr_E U_{AE,\eta} \left( \rho_A \otimes \rho_E \right) U_{AE,\eta}^\dagger \right)
\nonumber
\\
&=\tr_E\tr_A\left( U_{AE,\eta}^\dagger D_A\left(\alpha \right)  U_{AE,\eta} \rho_A \otimes \rho_E   \right)
\nonumber
\\
&=\tr_E\tr_A\left( D_A\left(\sqrt{\eta} \alpha\right) D_E\left(\sqrt{1-\eta}\alpha\right)    \rho_A \otimes \rho_E  \right)
\nonumber
\\
&=\chi \left({\rm Re}\sqrt{1-\eta}\alpha, {\rm Im}\sqrt{1-\eta}\alpha ;\rho_E\right) \tr_A\left( D_A\left(\sqrt{\eta} \alpha\right)    \rho_A  \right)
\nonumber
\\
&=\exp\left(-\left(1-\eta\right)|\alpha|^2(2N_E+1)/2\right)\tr_A\left( D_A\left(\sqrt{\eta} \alpha\right)    \rho_A  \right),
\end{align}
where we have used the Stinespring dilation of a pure loss channel. Thus for a single loss channel
\be
\calN^{N_E}_\eta \left(D\left(\alpha \right)\right)=\exp\left(-\left(1-\eta\right)|\alpha|^2(2N_E+1)/2\right) D\left(\sqrt{\eta}\alpha\right).
\ee
From this, we expect that the OTOC decreases exponentially in presence of loss and noise. The actual decay depends on the specific scheme used to measure the OTOC, but a rule-of-thumb estimation by considering two lossy noisy channel gives
$
\calC^{w,v}_2(\alpha,\beta;t)_T|_{\eta, N_E}\sim \exp\left(-\left(1-\eta\right)|\alpha|^2(2N_E+1)\right).
$

\section{Twice-regulated frame potential}\label{App_FP}

In the main text, we employed a generalization of a `once-regulated' finite-temperature frame potential from DV systems to characterize ensembles of CV unitaries. Despite the fact that it provides us with a quantitative understanding of volumes of distributions, there were two drawbacks to this frame potential. First, it is lower bounded by zero even though $\rho$ is a quantum state with finite entropy. This implies that no normalized distribution of unitaries can saturate the lower bound, despite the fact that our Hilbert space of interest is regulated by $\rho$ and may be finite. This concern also relates to a second drawback: when seeking an appropriate definition for CV designs, it may be more physical to consider ensembles of unitaries that approximately preserve a chosen density matrix $\rho$. In this Section, we construct a new, twice-regulated frame potential, which solves both of these drawbacks by weighting the contributions from unitaries based on their preservation of $\rho$ .

Our guiding principle in constructing the new frame potential is the definition in the main text of CV unitary $k$-designs, as the $d\to \infty $, limit of Eq.~\eqref{design_def_2}. For general $k$, this gives:
\begin{equation}\label{DV Haar}
\, \lim_{d\to\infty}
 d^k \, 
\mathbb{E}_{\text{Haar}}
\left\{ U^{\otimes k} \otimes (U^{\dagger})^{\otimes k} \right\}
= \sum_{\pi} S_{\leftrightarrow} ( W_{\pi} \otimes W^{-1}_{\pi} ),
\end{equation}
where lower-order contributions in $d$ are neglected (this can be derived from equations in Ref.~\cite{roberts2017chaos}). Here $W_{\pi}$ performs the permutation $\pi$ on the $k$-copied Hilbert space $\mathcal{H}^{\otimes k}$ and $S_{\leftrightarrow}$ is the swap operator between the two copies of $\mathcal{H}^{\otimes k}$ (one acted on by $U^{\otimes k}$ and one acted on by $(U^{\dagger})^{\otimes k}$). 

The new frame potential seeks to characterize how closely the above expectation value over an ensemble $\mathcal{E}$ resembles the Haar value. To do so we will view the CV operators, and integrals over them, as `vectors' in a higher-dimensional space, and measure the distance between them with a prescribed inner product. Define the following:
\begin{align}
| \mathcal{H} ) \equiv  \,
\mathbb{E}_{\text{Haar}}
\, \left\{ U^{\otimes k} \otimes (U^{\dagger})^{\otimes k}  \right\} 
\\
| \mathcal{E} ) \equiv \mathcal{N} \, 
\mathbb{E}_{\calE}
\, \left\{ U^{\otimes k} \otimes (U^{\dagger})^{\otimes k}  \right\},
\end{align}
with some appropriate normalization constant $\mathcal{N}$ which will be specified later. We take the inner product between operators $\mathcal{O}_1,\mathcal{O}_2$ on the $2k$-copied Hilbert space $\mathcal{H}^{\otimes 2k}$ to be
\begin{equation}
\ip{\mathcal{O}_1}{\mathcal{O}_2} = \tr\left( \mathcal{O}_1^{\dagger} \,P_k\, \mathcal{O}_2 \,P_k\,\right),
\label{inner_product}
\end{equation}
with respect to positive Hermitian operators (e.g. density matrices) $P_k = [{\left(\rho^{\frac{1}{2k}}\right)}^{\otimes k} \otimes \left(\rho^{\frac{1}{2k}}\right)^{\otimes k}]$. One can show that this satisfies all properties of an inner product if $\rho$ is full rank. If $\rho$ is not full rank, it satisfies the properties of a semi-definite inner product. It can also be viewed as the usual matrix inner product of the `low-energy weighted' operators $P_k^{1/2} \, \mathcal{O}_1 \, P_k^{1/2}$ and $P_k^{1/2} \, \mathcal{O}_2 \, P_k^{1/2}$.


The inner product provides a natural distance measure with which to compare $\mathcal{E}$ and $\mathcal{H}$,
$\min_\calN
\norm{| \mathcal{E} ) - | \mathcal{H} )}^2  \equiv \min_\calN\big[ ( \mathcal{E} | - ( \mathcal{H} | \big] \, \big[ | \mathcal{E} ) - | \mathcal{H} ) \big] \ge 0.
$
Here we minimize over the normalization constant $\calN$ such that the distance is minimum. After some calculations, we arrive at
\begin{align}
\min_{N} \norm{| \mathcal{E} ) - | \mathcal{H} )}^2  = \mathcal{H}^{(k)}(\rho ) - \frac{1}{\calJ_\calE^{(k)}\left(\rho\right) }\geq 0,
\end{align}
where 
\begin{align}
\mathcal{H}^{(k)}(\rho) 
= k! \sum_{\pi} |\tr
\left( 
W_{\pi} \left(\rho^{\frac{1}{2k}}\right)^{\otimes k} \left(\rho^{\frac{1}{2k}}\right)^{\otimes k}
\right)
|^2,
\label{H2 def_main}\\
\calJ_\calE^{(k)}\left(\rho\right) 
= 
\frac{
\mathbb{E}_{U, V\in \calE}
\left[  
|\tr\left( U^{\dagger} \rho^{\frac{1}{2k}} V \rho^{\frac{1}{2k}}\right)|^{2k}
\right]
}{\left(k!\right)^2
\bigg[  \mathbb{E}_{U\in \calE}  \big[ \tr\left( U^{\dagger} \rho^{1/k} U \rho^{1/k}\right) \big]^k \bigg]^2}.
\label{frame_potential_new}
\end{align}
Note that $\mathcal{H}^{(k)}(\rho )$ depends only on $\rho$. Observing
\begin{align}
\calJ_\calE^{(k)}\left(\rho\right) 
\ge \frac{1}{\mathcal{H}^{(k)}(\rho )}, 
\end{align}
we recognize $\calJ_\calE^{(k)}\left(\rho\right)$ as the frame potential. For $k=1,2$, we find the following lower bounds 
\begin{align}
\mathcal{H}^{(1)}(\rho) =  1,\qquad
\mathcal{H}^{(2)}(\rho) =  2  + 2 \,\tr{\left( \rho^{1/2}\right)}^4.
\end{align}

We would like to emphasize that the density matrix $\rho$ reflects the Hilbert space of interest. Namely, when the frame potential achieves its lower bound, the ensemble $\mathcal{E}$ is CV $k$-design with respect to $\rho$. As a trivial example, for a pure state $\rho = |\psi\rangle \langle \psi |$, the lower bound is saturated for any ensemble (the only operator on a single-state Hilbert space is the identity, and so all unitary designs are trivial).

The new frame potential can be used to verify CV $k$-designs. As an illuminating example, we calculate the $k = 1$ CV frame potential for our 1-design candidate $\mathbb{D}_n$ on a thermal state $\tilde{\rho}_{n_{\text{th}}}$ of mean photon number $n_{\text{th}}$. We obtain
\begin{align}
\calJ_{\mathbb{D}_n}^{(1)}\left(\tilde{\rho}_{n_{\text{th}}}\right) 
=\left(\frac{(1+n+2n_{\text{th}})^2}{(1+n)^2+4nn_{\text{th}}}\right)^N
\ge 1=\frac{1}{\calH^{(1)}},
\end{align}
where $N$ is the number of modes. It is illuminating to examine two limits: $n \gg n_{\text{th}}$, where the ensemble $\mathbb{D}_n$ contains an approximate basis for all operators acting on the subspace of $\lesssim n_{\text{th}}$ photons, and $n \ll n_{\text{th}}$, where it does not. For the latter, the frame potential scales as $( n_{\text{th}} / n )^N$ and continues decreasing as we increase the size $n$ of the ensemble. Once $n\gtrsim n_{\text{th}}$, the frame potential algebraically saturates to its lower bound of $1$, up to terms of order $(1/n)$ and $(n_{\text{th}}/n)^2$, confirming again that $\text{lim}_{n\to\infty}  \mathbb{D}_n$ is a proper CV 1-design.

\let\addcontentsline\oldaddcontentsline

\let\oldaddcontentsline\addcontentsline
\renewcommand{\addcontentsline}[3]{}
\let\addcontentsline\oldaddcontentsline




\begin{thebibliography}{102}
\bibitem{footnote1}In DV systems, our current understanding of quantum chaos and random matrix theory has been deeply influenced by ideas and tools from Haar randomness.
\bibitem{footnote3}
We observe systematic overestimation of the velocity by theory near the point of saturation $e^{c_R} = 2$. We speculate that this arises from finite-size effects and discrepancy between the average OTOC, $ \overline{\exp\left(-n f\left(x,t\right)\right)}$, and the quantity we theoretically model, $\exp\left(-n \overline{f}\left(x,t\right)\right)$. These effects can only decrease the numerical estimate of the velocity due to the strict upper bound $v_B \leq 1$.
\bibitem{footnote7}
We choose a random SNAP gate, instead of the better-known Kerr gate, due to a periodicity revival in the Kerr gate: $K_R\left(\pi/16\right)=\exp\left(-i\pi/4\right)\bm I$.
\bibitem{footnote8}
A Gaussian thermal state, instead of a Gaussian pure state, could also be used as the ancilla.
\bibitem{footnote9}
We note that this procedure for taking averages over Gaussian distributions of displacement operators is performed using only Gaussian operations. From Section \ref{Sec_nonGaussianity}, this implies that one can measure the non-Gaussianity of an operation $U\otimes U \otimes U^{\dagger}\otimes U^{\dagger}$ via only Gaussian operations, suggesting a possible resource theoretic interpretation of the quantity.
\bibitem{footnote10}
In DV systems, the analogous protocol succeeds with probability unity for Clifford unitaries, but also with lesser probability $1/d^2$ for non-Clifford unitaries. This probability goes to zero in the CV limit, hence our avoidance of non-Gaussian unitaries. In reality, imperfections in EPR preparation and measurement will `cut-off' this CV limit, and one might expect that the protocol can also succeed with finite probability for non-Gaussian unitaries.
\bibitem{footnote11}
For an imperfect EPR pair, an additional constraint arises from squeezing's effect on the two-mode squeezed state. Although squeezing increases the state's typical photon number, it vastly reduces its width in certain, typically many-mode, quadratures. This implies that there exist many-mode displacements of magnitude $|\Delta \b{\xi}| \sim e^r/\sqrt{n_{\text{EPR}}}$ that leave the many-body state (after the application of $U$, $U^*$) unchanged. Projecting to the measured mode, these are equivalent to measurement errors $ \sim e^r/(\sqrt{N}\sqrt{n_{\text{EPR}}})$. The previous arguments then give the constraint $\sqrt{n_\psi} > \sqrt{n_{EPR}}$, which is a more powerful constraint at high squeezing $e^r > \sqrt{N}$. We see that squeezing can only assist teleportation up to a certain limit, after which the teleportation is bounded solely by the strength of the imperfections.
\end{thebibliography}
\end{document}